\documentclass[twocolumn]{aastex631}

\begin{document}

\defcitealias{Yang+etal_2020}{Y20}

\title{Early Planet Formation in Embedded Disks (eDisk). XI. \par A high-resolution view toward the BHR 71 Class 0 protostellar wide binary}


\correspondingauthor{Sacha Gavino}
\email{sacha.gavino@nbi.ku.dk}

\author[0000-0001-5782-915X]{Sacha Gavino}, 
\affiliation{Niels Bohr Institute, University of Copenhagen, Øster Voldgade 5-7, 1350, Copenhagen K, Denmark}

\author[0000-0001-9133-8047]{Jes K. Jørgensen}
\affiliation{Niels Bohr Institute, University of Copenhagen, Øster Voldgade 5-7, 1350, Copenhagen K, Denmark}

\author[0000-0002-0549-544X]{Rajeeb Sharma}
\affiliation{Niels Bohr Institute, University of Copenhagen, Øster Voldgade 5-7, 1350, Copenhagen K, Denmark}

\author[0000-0001-8227-2816]{Yao-Lun Yang}
\affiliation{Star and Planet Formation Laboratory, RIKEN Cluster for Pioneering Research, Wako, Saitama 351-0198, Japan}

\author[0000-0002-7402-6487]{Zhi-Yun Li}
\affiliation{University of Virginia, 530 McCormick Rd., Charlottesville, Virginia 22904, USA}

\author[0000-0002-6195-0152]{John J. Tobin}
\affil{National Radio Astronomy Observatory, 520 Edgemont Rd., Charlottesville, VA 22903 USA} 

\author[0000-0003-0998-5064]{Nagayoshi Ohashi}
\affiliation{Academia Sinica Institute of Astronomy \& Astrophysics \\
11F of Astronomy-Mathematics Building, AS/NTU, No.1, Sec. 4, Roosevelt Rd \\
Taipei 10617, Taiwan, R.O.C.}

\author[0000-0003-0845-128X]{Shigehisa Takakuwa}
\affiliation{Department of Physics and Astronomy, Graduate School of Science and Engineering, Kagoshima University, 1-21-35 Korimoto, Kagoshima,Kagoshima 890-0065, Japan}
\affiliation{Academia Sinica Institute of Astronomy and Astrophysics, 11F of Astro-Math Bldg, 1, Sec. 4, Roosevelt Rd, Taipei 10617, Taiwan}

\author[0000-0002-9912-5705]{Adele L.\ Plunkett}
\affiliation{National Radio Astronomy Observatory, 520 Edgemont Rd, Charlottesville, VA, 22903, USA}

\author[0000-0003-4022-4132]{Woojin Kwon}
\affiliation{Department of Earth Science Education, Seoul National University, 1 Gwanak-ro, Gwanak-gu, Seoul 08826, Republic of Korea}
\affiliation{SNU Astronomy Research Center, Seoul National University, 1 Gwanak-ro, Gwanak-gu, Seoul 08826, Republic of Korea}

\author[0000-0003-4518-407X]{Itziar de Gregorio-Monsalvo}
\affiliation{European Southern Observatory, Alonso de Cordova 3107, Casilla 19, Vitacura, Santiago, Chile}

\author[0000-0001-7233-4171]{Zhe-Yu Daniel Lin}
\affiliation{University of Virginia, 530 McCormick Rd., Charlottesville, Virginia 22904, USA}

\author[0000-0001-6267-2820]{Alejandro Santamaría-Miranda}
\affiliation{European Southern Observatory, Alonso de Cordova 3107, Casilla 19, Vitacura, Santiago, Chile}

\author[0000-0002-8238-7709]{Yusuke Aso}
\affiliation{Korea Astronomy and Space Science Institute, 776 Daedeok-daero, Yuseong-gu, Daejeon 34055, Republic of Korea}

\author[0000-0003-4361-5577]{Jinshi Sai (Insa Choi)}
\affil{Academia Sinica Institute of Astronomy and Astrophysics, 11F of Astronomy-Mathematics Building, AS/NTU, No.\ 1, Sec.\ 4, Roosevelt Rd, Taipei 10617, Taiwan}

\author[0000-0003-3283-6884]{Yuri Aikawa}
\affiliation{Department of Astronomy, Graduate School of Science, The University of Tokyo, 7-3-1 Hongo, Bunkyo-ku, Tokyo 113-0033, Japan}

\author[0000-0001-8105-8113]{Kengo Tomida}
\affiliation{Astronomical Institute, Graduate School of Science, Tohoku University, Sendai 980-8578, Japan}

\author[0000-0003-2777-5861]{Patrick M. Koch}
\affil{Academia Sinica Institute of Astronomy and Astrophysics, 11F of Astronomy-Mathematics Building, AS/NTU, No.1, Sec. 4, Roosevelt Rd, Taipei 10617, Taiwan, R.O.C.}

\author[0000-0003-3119-2087]{Jeong-Eun Lee}
\affiliation{Department of Physics and Astronomy, Seoul National University, 1 Gwanak-ro, Gwanak-gu, Seoul 08826, Korea}

\author[0000-0002-3179-6334]{Chang Won Lee}
\affiliation{Korea Astronomy and Space Science Institute, 776 Daedeok-daero Yuseong-gu, Daejeon 34055, Republic of Korea}
\affiliation{University of Science and Technology, 217 Gajeong-ro Yuseong-gu, Daejeon 34113, Republic of Korea}

\author[0000-0001-5522-486X]{Shih-Ping Lai}
\affiliation{Institute of Astronomy, National Tsing Hua University, No. 101, Section 2, Kuang-Fu Road, Hsinchu 30013, Taiwan}
\affiliation{Center for Informatics and Computation in Astronomy, National Tsing Hua University, No. 101, Section 2, Kuang-Fu Road, Hsinchu 30013, Taiwan}
\affiliation{Department of Physics, National Tsing Hua University, No. 101, Section 2, Kuang-Fu Road, Hsinchu 30013, Taiwan}
\affiliation{Academia Sinica Institute of Astronomy and Astrophysics, P.O. Box 23-141, 10617 Taipei, Taiwan}

\author[0000-0002-4540-6587]{Leslie W. Looney}
\affiliation{Department of Astronomy, University of Illinois, 1002 West Green St, Urbana, IL 61801, USA}

\author[0000-0002-0244-6650]{Suchitra Narayanan}
\affiliation{Institute for Astronomy, University of Hawai`i at Mānoa, 2680 Woodlawn Dr., Honolulu, HI 96822, USA}

\author[0000-0002-4372-5509]{Nguyen Thi Phuong}
\affiliation{Korea Astronomy and Space Science Institute, 776 Daedeok-daero, Yuseong-gu, Daejeon 34055, Republic of Korea}
\affiliation{Department of Astrophysics, Vietnam National Space Center, Vietnam Academy of Science and Techonology, 18 Hoang Quoc Viet, Cau Giay, Hanoi, Vietnam}

\author[0000-0003-0334-1583]{Travis J. Thieme}
\affiliation{Institute of Astronomy, National Tsing Hua University, No. 101, Section 2, Kuang-Fu Road, Hsinchu 30013, Taiwan}
\affiliation{Center for Informatics and Computation in Astronomy, National Tsing Hua University, No. 101, Section 2, Kuang-Fu Road, Hsinchu 30013, Taiwan}
\affiliation{Department of Physics, National Tsing Hua University, No. 101, Section 2, Kuang-Fu Road, Hsinchu 30013, Taiwan}

\author[0000-0002-2555-9869]{Merel L.R. van 't Hoff}
\affil{Department of Astronomy, University of Michigan, 1085 S. University Ave., Ann Arbor, MI 48109-1107, USA}

\author[0000-0001-5058-695X]{Jonathan P. Williams}
\affiliation{Institute for Astronomy, University of Hawai‘i at Mānoa, 2680 Woodlawn Dr., Honolulu, HI 96822, USA}

\author[0000-0003-1412-893X]{Hsi-Wei Yen}
\affiliation{Academia Sinica Institute of Astronomy and Astrophysics, 11F of Astro-Math Bldg, 1, Sec. 4, Roosevelt Rd, Taipei 10617, Taiwan}





\begin{abstract}
We present Atacama Large Millimeter/submillimeter Array (ALMA) observations of the binary Class 0 protostellar system BHR 71 IRS1 and IRS2 as part of the Early Planet Formation in Embedded Disks (eDisk) ALMA Large Program. We describe the $^{12}$CO ($J$=2--1),  $^{13}$CO ($J$=2--1), C$^{18}$O ($J$=2--1), H$_2$CO ($J=3_{2,1}$--$2_{2,0}$), and SiO ($J$=5--4) molecular lines along with the 1.3 mm continuum at high spatial resolution ($\sim$0\farcs08 or $\sim$5~au). Dust continuum emission is detected toward BHR 71 IRS1 and IRS2, with a central compact component and extended continuum emission. The compact components are smooth and show no sign of substructures such as spirals, rings or gaps. However, there is a brightness asymmetry along the minor axis of the presumed disk in IRS1, possibly indicative of an inclined geometrically and optically thick disk-like component. Using a position-velocity diagram analysis of the C$^{18}$O line, clear Keplerian motions were not detected toward either source. If Keplerian rotationally-supported disks are present, they are likely deeply embedded in their envelope. However, we can set upper limits of the central protostellar mass of 0.46 M$_\odot$ and 0.26 M$_\odot$ for BHR 71 IRS1 and BHR 71 IRS2, respectively. Outflows traced by $^{12}$CO and SiO are detected in both sources. The outflows can be divided into two components, a wide-angle outflow and a jet. In IRS1, the jet exhibits a double helical structure, reflecting the removal of angular momentum from the system. In IRS2, the jet is very collimated and shows a chain of knots, suggesting episodic accretion events.

\end{abstract}


 \section{Introduction} \label{sec:intro}
 
Circumstellar disks are the result of the conservation of angular momentum during the gravitational collapse of a molecular cloud core composed of dust and gas \citep[e.g.][]{Terebey+etal_1984, Shu+Adam_1987, Ohashi+etal_1997, Momose+etal_1998}. These disks are ultimately the formation site of planets \citep{Armitage_2011, Testi+etal_2014}. It is expected that the dust and gas distributions in the disk evolve quickly after the collapse and grain growth can occur at very early stages \citep[e.g.][]{Kwon+etal_2009, Harsono+etal_2018}. Although other processes could be at play, theoretical models and observations show that planet forming disks can generate substructures such as spirals and gaps, where material has been cleared from their orbit around the protostar \citep[e.g.][]{Johansen+etal_2010, Kley+Nelson_2012, Dong+etal_2015, Zhang+etal_2018}. 

Substructures in the dust continuum of Class II disks are found to be common \citep[e.g.][]{Partnership+etal_2015, Andrews+etal_2018, Cieza+etal_2019}, and gas substructures have also been observed in multiple disks \citep[e.g.][]{Yen+etal_2016, Law+etal_2021, Zhang+etal_2021}. Many of the line emission substructures are spatially coincident with the substructures in the dust millimeter continuum in the inner regions of the disks where planets are expected to form \citep{Law+etal_2021}. 

The stage at which these substructures emerge and the timescale of planet formation are still open questions. The Early Planet Formation in Embedded Disks (eDisk) ALMA Large Program aims to provide new insights by investigating at high spatial resolution ($\sim$7~au) the 1.3 mm dust continuum as well as a number of selected lines observable with a single setting toward embedded disks around 12 Class 0 and 7 Class I protostars \citep{Ohashi+edisk_2023}. In this study, we report the observations of the protostellar binary BHR 71 as part of the eDisk Large Program.

BHR 71 is a Bok globule nearby the Southern Coalsack. The distance to the Southern Coalsack is 176 pc as derived by \citet{Voirin+etal_2018} from measurements of the Chamaleon clouds. We adopt this value as the distance to the BHR 71 protostellar system. The BHR 71 system hosts two protostars, a primary (IRS1) and a secondary (IRS2) with a measured separation of $\sim$16$\arcsec$ \citep{Tobin+etal_2019}, or $\sim$2800 au at 176 pc.

The bolometric temperatures and bolometric luminosities, derived from the newly compiled spectral energy densities (SEDs) made as part of the eDisk Large Program \citep[see][for the details]{Ohashi+edisk_2023}, are $T_{\rm bol} = 66$~K and $L_{\rm bol} = 10$~L$_{\rm \odot}$ for BHR 71 IRS1, and $T_{\rm bol} = 39$~K and $L_{\rm bol} = 1.1$~L$_{\rm \odot}$ for BHR 71 IRS2. The characteristics of both SEDs are consistent with those of Class 0 sources \citep{Ohashi+edisk_2023}.

To date, no substructures such as rings and spirals, nor Keplerian motion have been detected in BHR 71. However, \citet{Yang+etal_2020} (hereafter Y20) have demonstrated that IRS1 presents obvious signatures of infall motion and their infalling envelope model underestimates the observed high-velocity emission of HCN. They argue that a Keplerian disk may contribute to the velocity excess but that their observations were not able to put constraints on the presence of a disk.

Each source has prominent and spectacular outflows. They were first discovered by \citet{Bourke+etal_1997} and have been extensively studied \citep[e.g.][]{Garay+etal_1998, Bourke_2001, Parise+etal_2006, Yang+etal_2017, Tobin+etal_2019,Yang+etal_2020}. Furthermore, there are strong indications of the existence of high-velocity components. Using the Infrared Spectrograph (IRS) of the \textit{Spitzer Space Telescope}, \citet{Neufeld+etal_2009} carried out spectroscopic mapping observations and measured energetic properties of the outflows. Observations of CO, H$_2$, and SiO combining SOFIA, APEX, and the \textit{Spitzer} telescopes, \citet{Gusdorf+etal_2011} and \citet{Gusdorf+etal_2015} analyzed shocks in the northern lobe of the BHR 71 IRS1 outflow and measured shock velocities of around 20–25 km~s$^{-1}$. \citet{Mottram+etal_2014} identified shock components via the observation of multiple water transitions as part of the WISH survey, showing H$_2$O spectra with velocities $>$~20~km~s$^{-1}$. \citet{Benedettini+etal_2017} performed observations in the far infrared domain with \textit{Herschel} and revealed the presence of knots of shocked gas at high velocities.

The main goals of this study are to address the main questions of the eDisk Large Program by investigating (i) the presence of disks in the BHR 71 system as well as their sizes and mass, and (ii) possible substructures in the dust continuum in each source. We also perform a simple kinematic analysis of the outflows in order to better characterize their structure. 

The study is organized as follows: Section\,\ref{sec:obs} describes the details about the observations and data reduction; Section\,\ref{sec:results} introduces the 1.3 mm dust continuum and molecular line results; Section\,\ref{sec:analysis} provides an analysis of the dust mass, a kinematic analysis at disk scale to derive stellar masses as well as a kinematic analysis of the outflows; Section\,\ref{sec:discussion} discusses the asymmetry visible in the mm component of BHR 71 IRS1 and then the stellar to disk mass ratio of both sources. Finally, Section\,\ref{sec:summary} gives a brief summary.

\section{Observations} \label{sec:obs}
Observations of BHR 71 IRS1 and IRS2 (IRS1 and IRS2 in the following) were performed as part of the eDisk ALMA Large Program 2019.1.00261.L (PI: N. Ohashi). An overview of the program and the details of the overall data reduction procedure are described in \cite{Ohashi+edisk_2023}. Here we summarize the key points concerning the observations. The two sources were observed separately in two different configurations in ALMA's Band 6 at 1.3 mm (correlator setup centered at $\sim225$ GHz). The observations were performed in the C-5 configuration of ALMA (sampling baselines between 15 to 2517~m) on 2021 May 04 and 09 with 44 antennas available in the array on both days. They were likewise observed in the C-8 configuration (baselines of 91 to 8983~m) on 2021 October 15, 16, 19 and 20 with 41 to 43 antennas available. Additional observations in the C-8 configuration were performed on 2021 October 14: those were rated ``semi-pass'' during the quality assurance (QA2). This was due to slight phase decorrelation that was easily fixed with self-calibration and those observations were therefore included here. The most compact configuration of our observations has a maximum recoverable scale, ($\theta_\mathrm{MRS}$), of 2$\farcs$91. The pointing center for the two sources were $12^\mathrm{h}01^\mathrm{m}36\fs8$, $-65\arcdeg08^\mathrm{m}49\fs2$ for IRS1 and $12^\mathrm{h}01^\mathrm{m}34\fs1$, $-65\arcdeg08^\mathrm{m}47\fs4$ for IRS2. While the two sources could easily be observed as part of the same scheduling block with shared calibrators, their separation $\sim$16$\arcsec$ is too large relative to the ALMA primary beam at 1.3~mm (25$''$) for them to be optimally observed within a single pointing. The data for the two sources were therefore also reduced and imaged separately. 

The ALMA short and long baseline data, with the C-5 and C-8 configurations, respectively, were reduced using Common Astronomy Software Application (CASA version 6.2.1 \citep{Mcmullin+etal_2007}. The self-calibration and imaging of the short+long baseline (SB+LB) data were performed using standardized eDisk reduction scripts \citep{john_tobin_2023_7986682}. The continuum was self-calibrated first. For the SB-only data, seven iterations of phase-only self-calibration were performed for IRS1 and IRS2, separately. Similarly, for the SB+LB data, we performed seven iterations of phase-only self-calibration on the SB-only visibilities and then seven iterations of phase-only self-calibration on the combined SB+LB visibilities for each source separately. The line self-calibration used the self-calibration solutions from the continuum. Overall, the reduction directly follows the description in \cite{Ohashi+edisk_2023}.

A range of robust parameters from -2.0 to 2.0 were explored for the imaging. We adopted the robust value of 0.5, determined by a balance between sensitivity and resolution, to create the final images of both the continuum and the spectral lines. For the continuum, the resulting beam size was $0\farcs073 \times 0\farcs053$ with a position angle ($P.A.$) of 21$\fdg$55, and a noise level of 0.045~mJy~beam$^{-1}$ in root mean square (RMS) (see summary in Tab.~\ref{tab:obsdetails}). 

As described in \cite{Ohashi+edisk_2023} the observing setups of the eDisk data cover several different spectral windows around 220 and 230~GHz including the $J=2-1$ transitions of the main CO isotopologues ($^{12}$CO, $^{13}$CO, C$^{18}$O), the 217--218~GHz transitions of H$_2$CO, c-C$_3$H$_2$, SiO 5--4, DCN 3--2 and other species. In this study, we focus on the species listed in Table~\ref{tab:obsdetails}. All moment images presented in the following sections were created using the CASA task \texttt{immoments}. The systemic velocity is taken as $V_\mathrm{sys} = -4.45$ km~s$^{-1}$ as measured by \citet{Bourke+etal_1997} from NH$_3$. \citet{Tobin+etal_2019} also found a similar value from C$^{18}$O, although slightly more negative ($\sim -4.6$ km~s$^{-1}$). We follow the conclusion of \citet{Tobin+etal_2019} that both IRS1 and IRS2 have the same systemic velocity along the line of sight.

\begin{table*}
\caption{Overview of the presented ALMA maps.}
\label{tab:obsdetails}
\begin{tabular}{lccccc}\hline\hline
Continuum / line & robust & Frequency &   Beam size (P.A.) & Spectral resolution & RMS \\ 
\colhead{}  & \colhead{}  & \colhead{[GHz]} & \colhead{[$\arcsec$]} & \colhead{[km~s$^{-1}$]} & \colhead{[mJy~beam$^{-1}$]}  \\
\hline
1.3~mm continuum & 0.5 & 225.000 &  $0.073 \times 0.053$ (21.55$\arcdeg$) & $-$ & 0.045 \\
$^{12}$CO $J=2-1$   & 0.5 & 230.53800  & $0.107 \times 0.096$ (7.59$\arcdeg$) & 0.63     & 1.32 \\ 
$^{13}$CO $J=2-1$   & 0.5 & 220.39868  & $0.110 \times 0.100$ (5.19$\arcdeg$) & 0.17     & 2.90 \\
C$^{18}$O $J=2-1$   & 0.5 & 219.56035  & $0.110 \times 0.100$ (10.73$\arcdeg$) & 0.17     & 1.96 \\
H$_2$CO $J=3_{2,1}-2_{2,0}$  & 0.5 & 218.76007  & $0.108 \times 0.097$ (11.20$\arcdeg$) & 0.17     & 2.21 \\ 
SiO $J=5-4$  & 0.5 & 217.10498  & $0.108 \times 0.097$ (13.74$\arcdeg$) & 1.35   & 0.65 \\ \hline
\end{tabular}
\end{table*}

\section{Observational results} \label{sec:results}
\subsection{1.3 mm continuum} \label{sec:res_cont}

\begin{figure*}
\plotone{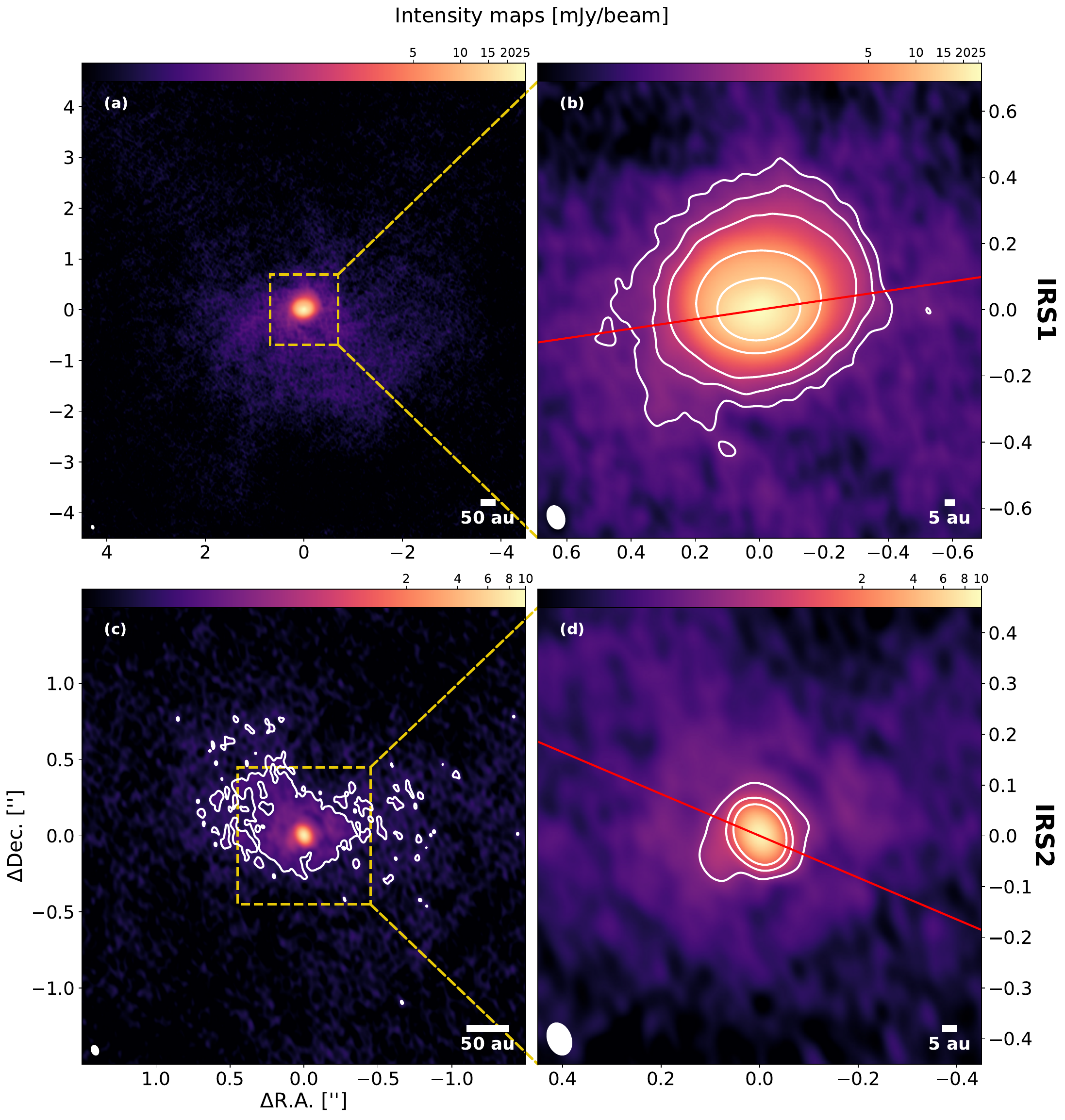}
 \caption{Continuum emission at 1.3 mm with robust parameter $r$ = 0.5 for IRS1 (top row) and IRS2 (bottom row). The right panels are zoomed-in views of the left panels. The white contours in the right panels represent increasing selected values of $\sigma$ [7$\sigma$, 14$\sigma$, 70$\sigma$, 140$\sigma$, 700$\sigma$] and [15$\sigma$, 40$\sigma$, 80$\sigma$] for IRS1 and IRS2, respectively. The 4$\sigma$ contour level for IRS2 is also shown in panel c. The white ellipse represents the beam size. The red lines follow the position angle of the semi-major axis as derived from the best-fit model (in IRS2, the direction of the red line does not follow the apparent semi-major axis because the  image is marginally resolved). Note that the color scales are not identical but depend on the peak intensity in each panel.}
\label{fig:cont_emission}
\end{figure*}

Figure \,\ref{fig:cont_emission} reveals the ALMA observation of the 1.3 mm dust continuum emission of IRS1 (top) and IRS2 (bottom), both with robust=0.5. The left column shows a large-scale view of the sources and the right column presents zoomed-in views. The angular resolution allows us to measure a projected separation of 15$\farcs$6 ($\sim$2756~au at a distance of 176 pc) from IRS1 toward IRS2, consistent with previous measurements. There is extended emission in both sources, extending up to a radius of $\sim$3$\arcsec$ in IRS1 and up to a radius of 1$\arcsec$ in IRS2, which is likely tracing the inner envelope as noted by \citetalias{Yang+etal_2020}. However, this extended emission is very faint as most of it lies below 4$\sigma$ in both sources. The zoomed-in view shows compact continuum emissions with elongated structures, suggesting that they may trace a disk around IRS1 and IRS2. The view reveals a marginally resolved image of IRS2's continuum emission, which appears smooth and fully symmetric along both axes inside 50$\sigma$. The 15$\sigma$ contour has a ``finger" toward the south-east direction, in the direction of the red-shifted outflow (see Section\,\ref{sec:res_fast}).

We perform a simple 2-D Gaussian fitting using the CASA tool \texttt{imfit}. The best-fit 2-D Gaussian model parameters are provided in Table\,\ref{tab:bestfit}. The emission of IRS1 is composed of a compact object, with a measured major and minor axes FWHM of 0$\farcs$28 ($\sim$50~au) and 0$\farcs$22 ($\sim$35~au), respectively, from the 2-D Gaussian fit to emission above 10$\sigma$. We estimate an inclination angle of 39$\fdg$4 from the deconvolved aspect ratio\footnote{$i$ = arccos($\theta_\mathrm{min}$/$\theta_\mathrm{maj}$) where $\theta_\mathrm{min}$ and $\theta_\mathrm{max}$ are the FWHM of the minor and major axes, respectively.}, assuming an infinitesimally-thin disk. This inclination angle should be most likely a lower limit because of the flared structure (see Section\,\ref{sec:discus_asymmetry}). We derive a P.A. value of the semi-major axis of 98$\fdg$2 $\pm$ 0$\fdg$4. For comparison, \citetalias{Yang+etal_2020} derived a P.A. of 113$\fdg$7$\pm$ 2$\fdg$0 at the deconvolved size, which differs by $\sim$~15$\arcdeg$ from the value measured in this study. This difference may be a result of the emission not being well-resolved by the beam of 0$\farcs$39$\times$0$\farcs$27 from the data of \citetalias{Yang+etal_2020}.

From the Gaussian fit, we find that IRS1 has a peak of 50.95 $\pm$ 0.39 mJy~beam$^{-1}$ and a flux density of 413.3 $\pm$ 3.5 mJy measured by integrating the emission above 4$\sigma$. The zoomed-in view Fig.\,\ref{fig:cont_emission}b reveals a resolved disk-like morphology without the presence of visible substructures (ring, gap, spiral). The component appears to be symmetric across the major axis ($P.A. \sim98\arcdeg$) but shows an asymmetry across the minor axis.

For IRS2, we measure major and minor axes FWHM of 0$\farcs$049 ($\sim$10~au) and 0$\farcs$042 ($\sim$7~au), respectively, using the best-fit 2-D Gaussian model performed above 15$\sigma$. We derive a deconvolved P.A. of 67.6° $\pm$ 3.7° and estimate an inclination angle of 30.9° from the deconvolved aspect ratio. The peak intensity is 8.34 $\pm$ 0.03 mJy~beam$^{-1}$ and the integrated flux density is 14.05 $\pm$ 0.07 mJy (above 4$\sigma$), about 30 times smaller than IRS1. Note that because of the marginally resolved image, the apparent direction of the semi-major axis of the compact emission is likely to follow the beam direction, hence the discrepancy between the apparent semi-major axis and the red line in Fig.\,\ref{fig:cont_emission}d, which shows the deconvolved position angle.

\begin{figure*}
\plotone{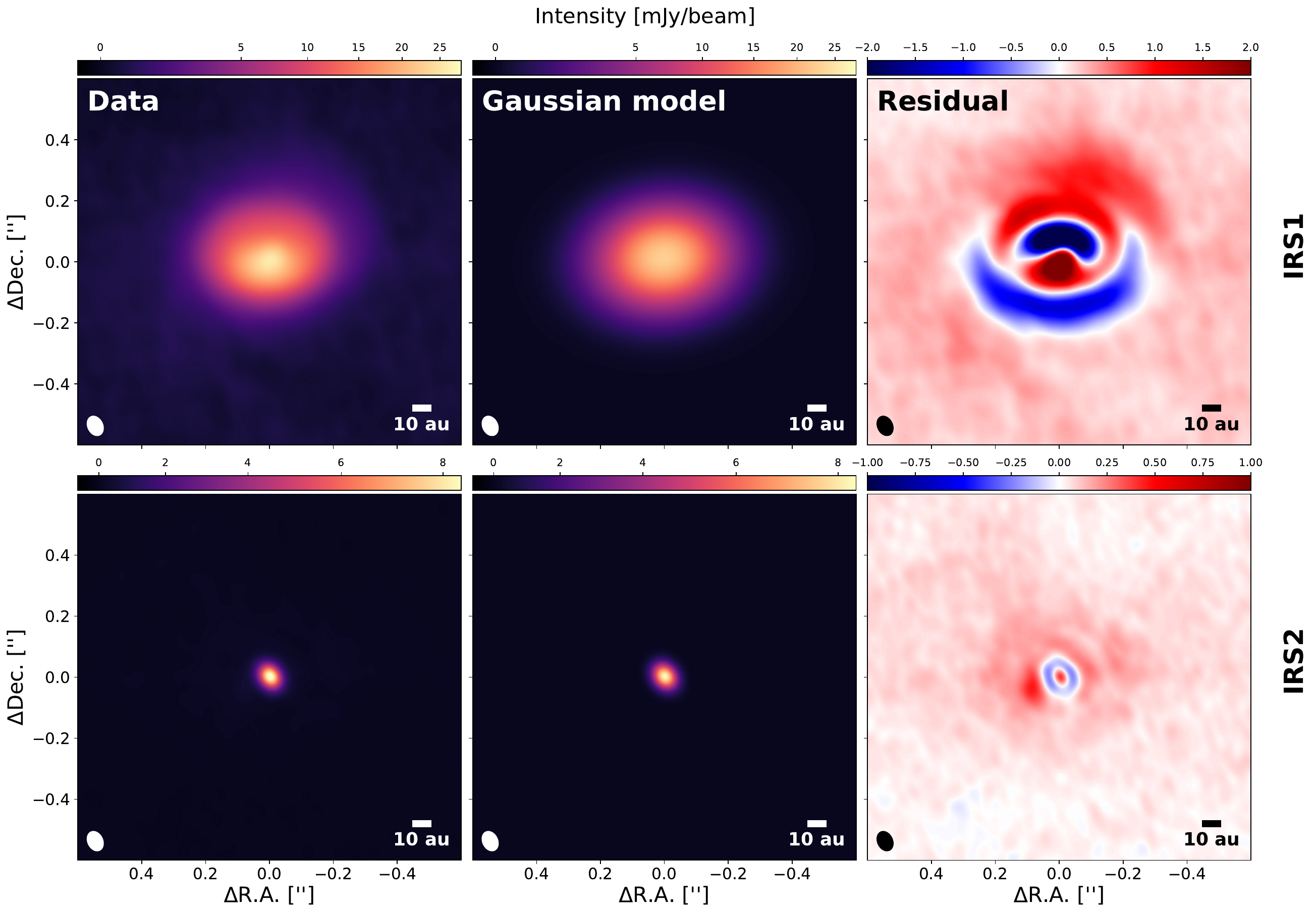}
 \caption{Continuum emission at 1.3 mm, Gaussian model, and residuals (the observed emission subtracted by the model) from left to right for IRS 1 (upper row) and for IRS 2 (bottom row). The ellipse in the bottom-left corner represents the beam size.}
\label{fig:cont_res}
\end{figure*}

\begin{figure}
\plotone{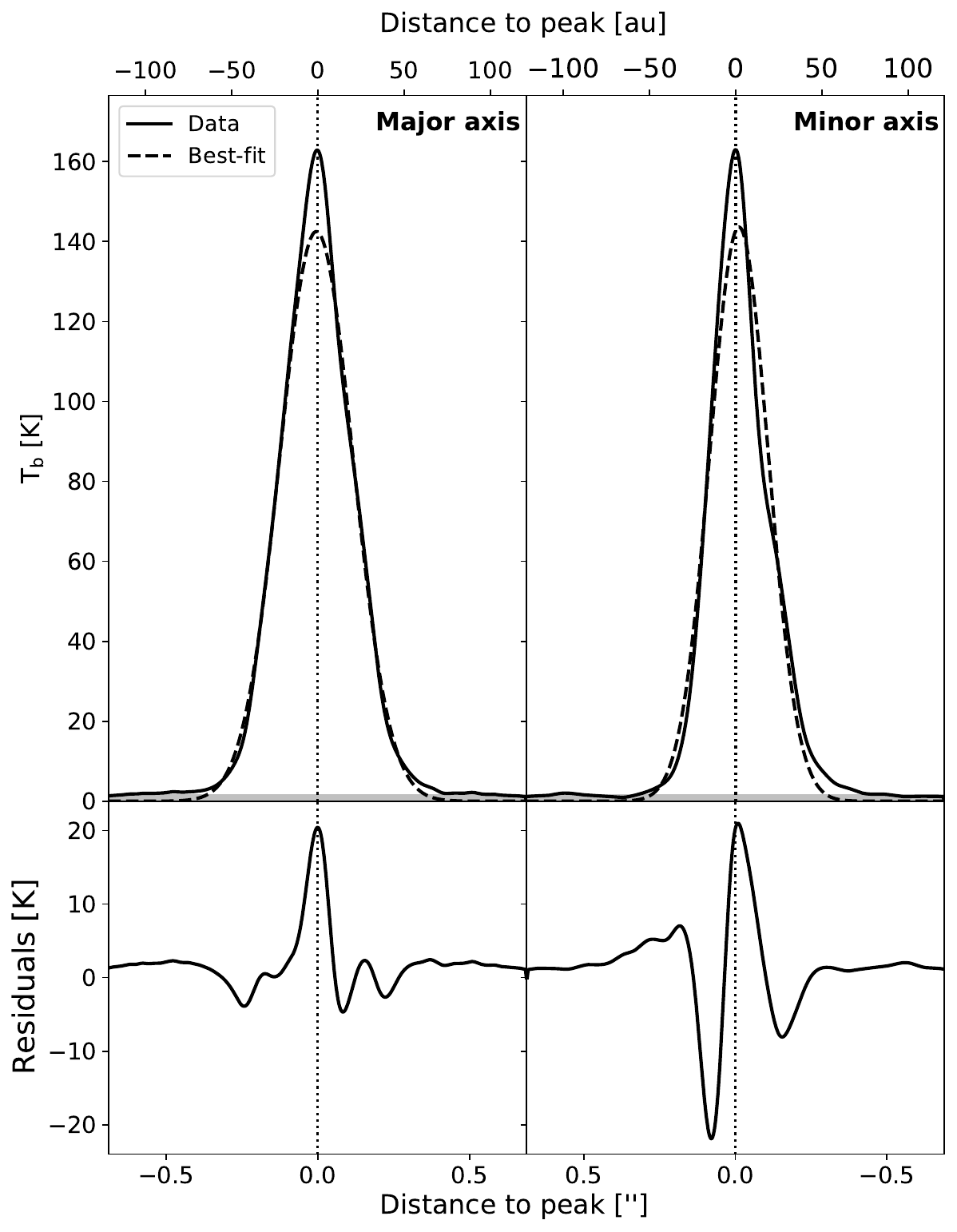}
 \caption{Brightness temperature cuts averaged over widths of 10 pixels along the major (left) and minor (right) axes of the continuum emission of IRS1 centered on the peak emission of the observed image. The solid lines represent the cuts in the observed continuum. The dashed lines show the cuts of the best-fit 2-D Gaussian model. The shaded region is the Brightness temperature below 4$\sigma$.The bottom panel shows the residuals.}
\label{fig:cont_cut1}
\end{figure}

In Figure\,\ref{fig:cont_res} we present the comparison between the observed 1.3 continuum emission in the image plane and the best 2-D Gaussian fitting. The residuals (Fig.\,\ref{fig:cont_res}, upper right) of IRS1 exhibit four (two inner and two outer) nested lobes that originate from the fact that the brightest emission, which is shifted toward the south along the minor axis, does not correspond to the geometrical center in the observed image. This deviation is well highlighted by the single Gaussian fitting. The residuals also show a ’cap’ toward NW above 0$\farcs$2 corresponding to an excess of emission.  Figure\,\ref{fig:cont_cut1} provides a more detailed comparison of the emission along the major and minor axes. The major axis has minimal deviation, meaning that the source is roughly symmetric along this axis. On the other hand, the deviation along the minor axis is more prominent and there is a slight offset of the peak due to the asymmetry along the minor axis. The origin of this asymmetry is discussed in Section\,\ref{sec:discus_asymmetry}.

We note that the residuals do not reach zero even at large distances. This is because the profile of the observed images has 'tails' at larger distances ($> 0\farcs3$). In fact, the observed structure is more likely best fitted with a power-law profile than with a single 2-D Gaussian \citep[see][]{Long+etal_2018, Sheehan+etal_2022}. However, a Gaussian fit provides valuable information about emission skewness and asymmetry.

Unlike IRS1, IRS2 shows no asymmetry. There are concentric ring-like structures visible in the residuals in Fig.\,\ref{fig:cont_res} (bottom right panel) but it is only due to the Gaussian not being properly adapted to fit the structure. However, a strong and relatively extended deviation appears in the southeastern red part of the residual image, which is due to the finger structure seen in Fig.~\ref{fig:cont_emission}d.

\begin{table*}
\centering
\caption{Best-fit parameters of the continuum and beam sizes using a 2-D Gaussian model.}
\label{tab:bestfit}
\begin{tabular}{lll}
\hline
& BHR 71 IRS1 & BHR 71 IRS2 \\
\hline
\hline
  Parameter [units] & Value &  Value\\
\hline
    R.A. [h:m:s] &  12:01:36.476 & 12:01:34.008 \\
    Dec. [d:m:s] &  -65:08:49.372 & -65:08:48.080 \\
    Beam size [$\arcsec$] & 0.09 $\times$ 0.08 (P.A. = 95.9 $\pm$ 1.3°)  & 0.07$\times$ 0.05 (P.A. = 30.88 $\pm$ 0.56°)\\
\hline
    Flux density [mJy] & 413.3 $\pm$ 3.5 & 14.048 $\pm$ 0.066 \\
    Peak intensity [mJy.beam$^{-1}$] & 50.95 $\pm$ 0.39   & 8.342 $\pm$ 0.026  \\
    semi-major axis FWHM [mas] & 278.94 $\pm$ 0.69 (49.30 $\pm$ 0.12 au)  & 48.72 $\pm$ 0.59 (8.61 $\pm$ 0.10 au) \\
    semi-minor axis FWHM [mas] & 215.53 $\pm$ 0.59 (38.08 $\pm$ 0.10 au) & 41.79 $\pm$ 0.63 (7.38 $\pm$ 0.11 au) \\
    Position angle [°] & 98.15 $\pm$ 0.41  & 67.6 $\pm$ 3.7 \\
    Estimated inclination$^{\ast}$ [°] & 39.42 & 30.93 \\
\hline
\multicolumn{3}{l}{$\ast$ At zero angle the object is seen face-on.}\\
\end{tabular} 
\end{table*}


\subsection{Outflows}\label{sec:res_outflow}
\begin{figure*}
\plotone{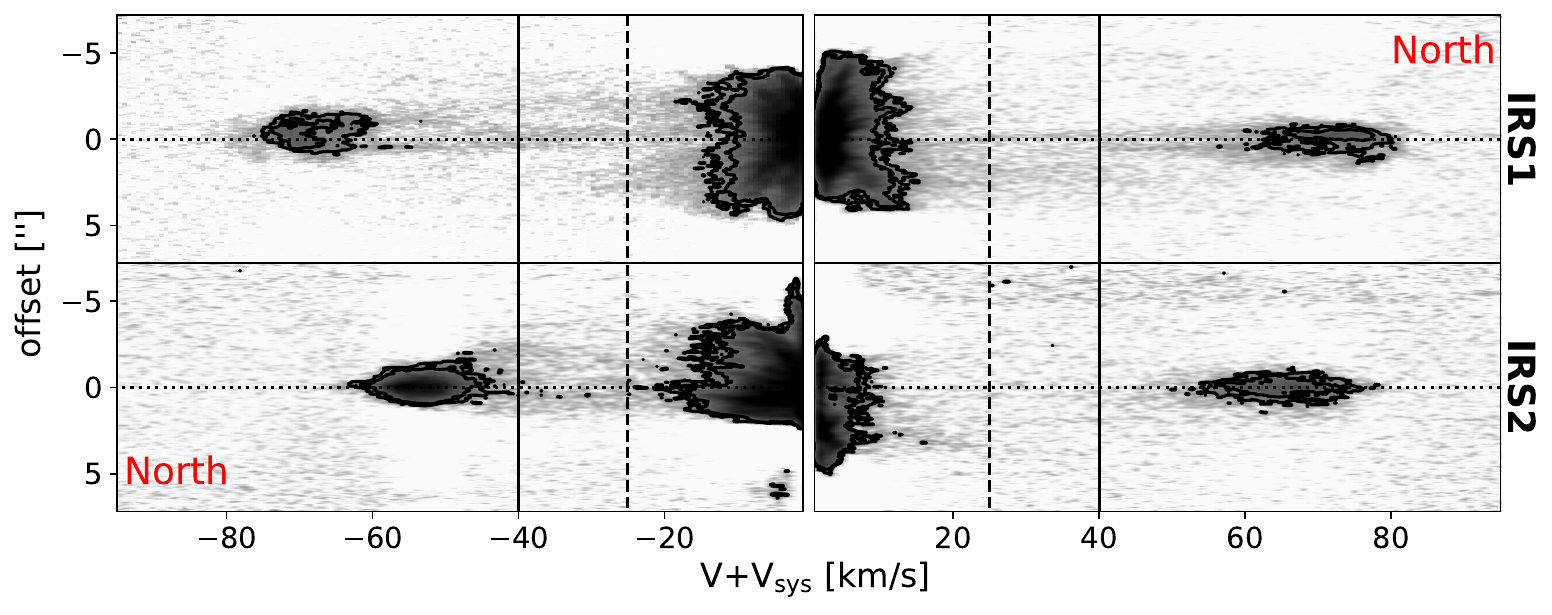}
 \caption{$^{12}$CO P-V diagrams made perpendicular to the outflows of each source averaged from multiple P-V diagrams performed every 0.01$\arcsec$ between a distance of 1$\arcsec$ and 13$\arcsec$ from the star in each pole. The upper row shows the P-V diagrams made in the northern outflow (right) and southern outflow (left) of IRS1, the bottom row shows the same P-V diagrams for IRS2. The black contours mark the 3$\sigma$ and 6$\sigma$ values. The vertical dashed and solid lines mark the standard high-velocity and extremely high-velocity, respectively.}
\label{fig:pvoutflows}
\end{figure*}

BHR 71 exhibits well-known prominent bipolar outflows powered by both IRS1 and IRS2. Figure\,\ref{fig:pvoutflows} shows $^{12}$CO P-V diagrams made perpendicular to the outflows of each source. There is a discontinuity in emission that clearly separates two components of the outflows. This pattern is observed in both sources. Based on these P-V diagrams, we can define two velocity regimes without ambiguity, a standard high-velocity (SHV) regime ($\lessapprox 25$ km~s$^{-1}$), corresponding to the wide-angle outflows, and an extremely high-velocity (EHV) regime ($\gtrapprox 40$ km~s$^{-1}$), corresponding to the collimated jets.

\subsubsection{Wide-angle outflows} \label{sec:res_wide}

\begin{figure*}
\plotone{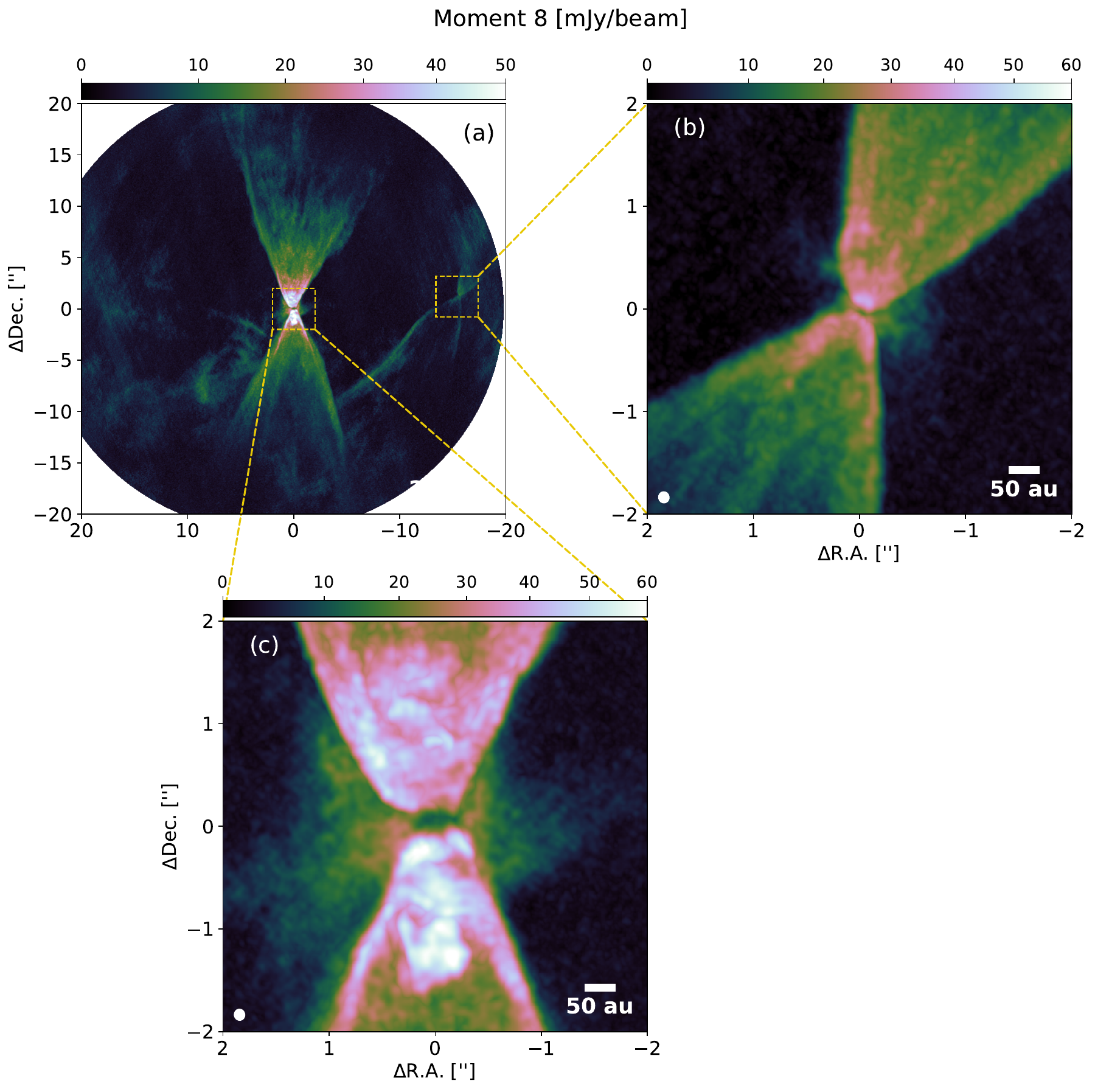}
 \caption{Maximum intensity for $^{12}$CO. Panel (a) shows a composite view of the system centered on IRS1. Panels (b) and (c) shows the outflows at a smaller scale of IRS2 and IRS1, respectively. The maps are generated using a velocity range of -29 km~s$^{-1}$ $< V + V_\mathrm{sys} <$ +29~km~s$^{-1}$. This range includes the envelope and excludes the high-velocity jet. The white and black ellipses in the lower-left corner of each panel mark the beam shape.}
\label{fig:composite}
\end{figure*}

In order to compare the brightness and structure of the two outflows, we show in Figure\,\ref{fig:composite}a a global view of the two wide-angle outflows associated with IRS1 and IRS2. Figure\,\ref{fig:composite}b and c shows a zoomed-in image ($2\arcsec \times2\arcsec$) toward IRS2 and IRS1, respectively. The moments were generated over a velocity range of $\pm$29 km~s$^{-1}$ with respect to the systemic velocity, $V_{\rm sys}$. The outflow emission from IRS2 is fainter than that of IRS1.

\begin{figure*}
\plotone{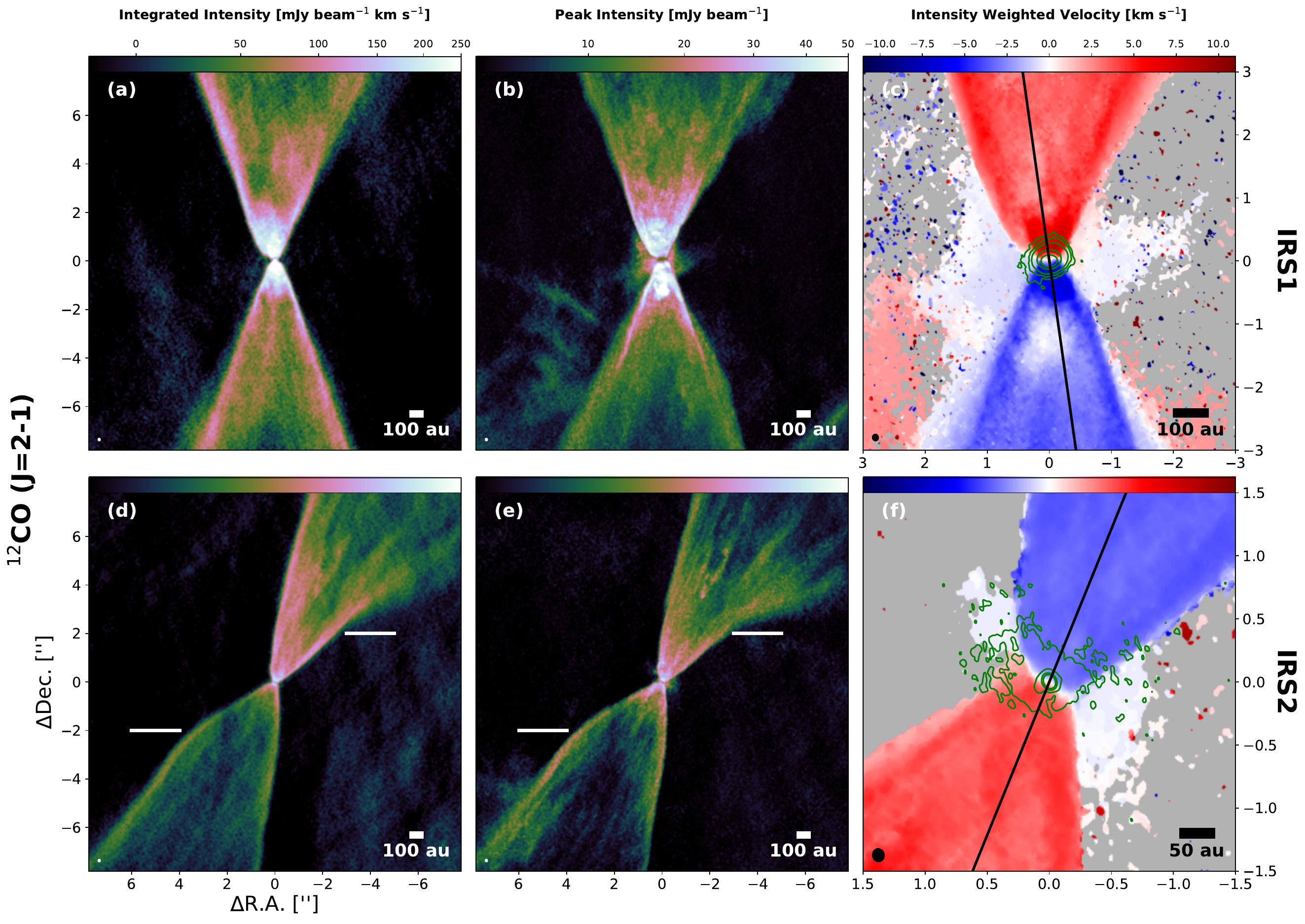}
 \caption{SHV outflow emission in $^{12}$CO in IRS1 (upper row) and IRS2 (bottom row). The left column shows the integrated intensity (moment 0) maps, the middle column the peak intensity maps, and the right column the velocity-weighted intensity (moment 1) maps. The ellipse in the bottom-left corner is the beam size. All maps are generated over a velocity range of -29 km~s$^{-1}$ $< V + V_\mathrm{sys} <$ +29~km~s$^{-1}$. The green contours mark the continuum emission with increasing values of $\sigma$ [4$\sigma$, 7$\sigma$, 14$\sigma$, 70$\sigma$, 140$\sigma$] and [4$\sigma$, 15$\sigma$, 50$\sigma$, 100$\sigma$] for IRS1 and IRS2, respectively. The black lines in the moment 1 maps correspond to the direction orthogonal to the semi-major axis $P.A.$ value of the 1.3-mm continuum emission derived from the best-fit model. The horizontal white lines in panels (d) and (e) show the location where the left side of the southern lobe and right side of the northern lobe are bent.}
\label{fig:12CO_maps}
\end{figure*}

To examine each outflow individually, Figure \,\ref{fig:12CO_maps} shows the wide-angle outflows of IRS1 and IRS2. Panels (a) and (d) present their integrated intensity (moment 0) maps integrated over the same velocity range as in Fig.~\ref{fig:composite}, and within a $8\arcsec \times 8\arcsec$ region centered on the protostars. Similarly, panels (b) and (e) present their peak intensity maps. In both sources, the outflows extend a distance larger than $\sim$2600~au across the protostellar position, corresponding to the limit of the ALMA primary beam (or field of view). Yet, previous observations of the outflows \citep[e.g.][]{Bourke+etal_1997, Garay+etal_1998} showed that the jet's length is extending up to $\sim$0.14~pc.

While the brightness of each side seems approximately symmetric in IRS1, there is an evident brightness asymmetry in IRS2, the southern outflow of IRS2 is much fainter than the northern one. Moreover, the outflows in IRS2 are geometrically asymmetric. The left side of the southern lobe and right side of the northern lobe bend at approximately equal distances from the protostar ($\sim3\arcsec$ in projected distance), marked by the horizontal white lines (Fig.\,\ref{fig:12CO_maps}d and e). This feature can be a sign of a change in launch direction from the source. 

Fig.\,\ref{fig:12CO_maps}c and f show the intensity-weighted
mean velocity (moment 1) maps. The maps reveal redshifted and blueshifted components. While the estimated inclinations from the disk-like structures are similar (39.42° and 30.93° for IRS1 and IRS2, respectively), the kinematics orientation of their outflows are opposite: The northern side of IRS1 is redshifted (far-side) while the one of IRS2 is blueshifted (near-side).  

On the assumption that the axes of the outflows are orthogonal to the major axes derived from the 2-D Gaussian fitting of the continuum (Section~\ref{sec:res_cont}), the inferred P.A. values of the outflow axes are 188$\fdg$2 and 337$\fdg$6 for IRS1 and IRS2, respectively. The black lines in Fig.~\ref{fig:12CO_maps}c and f mark these directions. We see a clear but small misalignment between the $P.A.$ values and the outflows themselves.

\begin{figure*}
\plotone{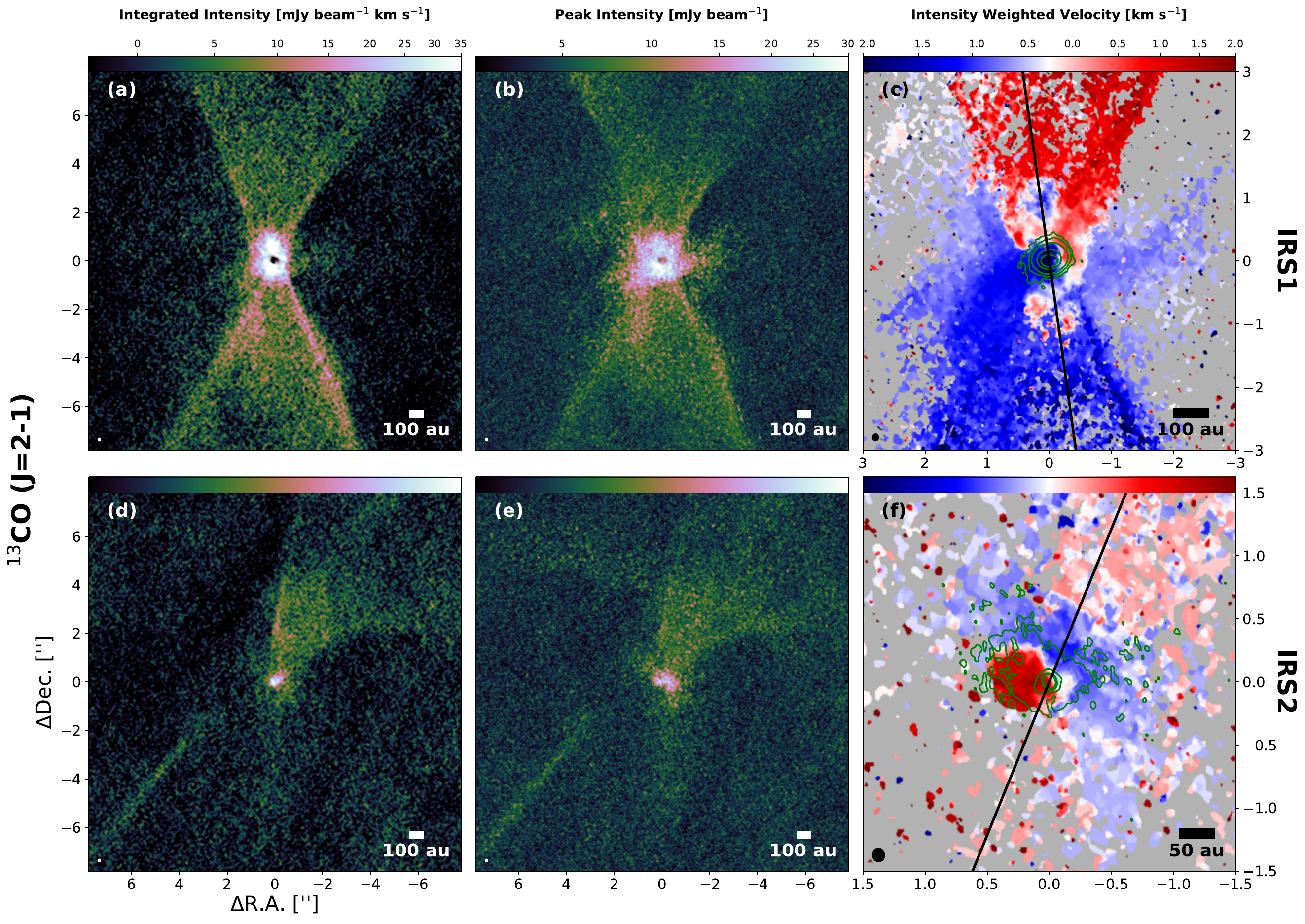}
 \caption{Emission from $^{13}$CO in IRS1 (upper row) and IRS2 (bottom row). The left column shows the integrated intensity (moment 0) maps, the middle column the peak intensity maps, and the right column the mean velocity (moment 1) maps. All maps are generated over a velocity range of -7.55~km~s$^{-1}$ $< V + V_\mathrm{sys} <$ +7.55~km~s$^{-1}$. The green contours mark the continuum emission with increasing values of $\sigma$ [4$\sigma$, 7$\sigma$, 14$\sigma$, 70$\sigma$, 140$\sigma$] and [4$\sigma$, 15$\sigma$, 50$\sigma$, 100$\sigma$] for IRS1 and IRS2, respectively. The ellipse in the bottom-left corner is the beam size. The black lines in the moment 1 maps correspond to the direction orthogonal to the semi-major axis $P.A.$ value derived from the best-fit models.}
\label{fig:13CO}
\end{figure*}

The outflows are also traced in the $^{13}$CO emission. Figure\,\ref{fig:13CO}a and d show the maps of the $^{13}$CO (J=2$-$1) emission detected above 3$\sigma$ levels within a velocity range of $\pm$7.55 km~s$^{-1}$ with respect to the systemic velocity, $V_{\rm sys}$, for IRS1 (top row) and IRS2 (bottom row), respectively. The outflow cavity is clearly visible in both IRS1 and IRS2, forming a distinct X-shape spatially coincident with the emission observed in $^{12}$CO. IRS2's southern outflow appears much fainter than its northern one, following the pattern observed in $^{12}$CO. The moment 1 maps (Fig.\,\ref{fig:13CO}c and f) provide the kinematic structure. In the case of IRS1, the map unambiguously delineates the redshifted northern and blueshifted southern parts of the outflows. 

It is notable that the moment 1 map of IRS1 (Fig.\,\ref{fig:13CO}c) shows two small regions in the outflow direction with velocities opposite to the kinematics of the outflow, one redshifted at -1$\arcsec$ and one blueshifted at 1$\arcsec$ (Fig.\,\ref{fig:13CO}c). Similar features traced by HCN was seen by \citetalias{Yang+etal_2020} who suggested that it either was associated with the wide opening angle of the outflow or infalling material from the larger scale envelope.

\subsubsection{High-velocity jets}\label{sec:res_fast}
\begin{figure*}
\plotone{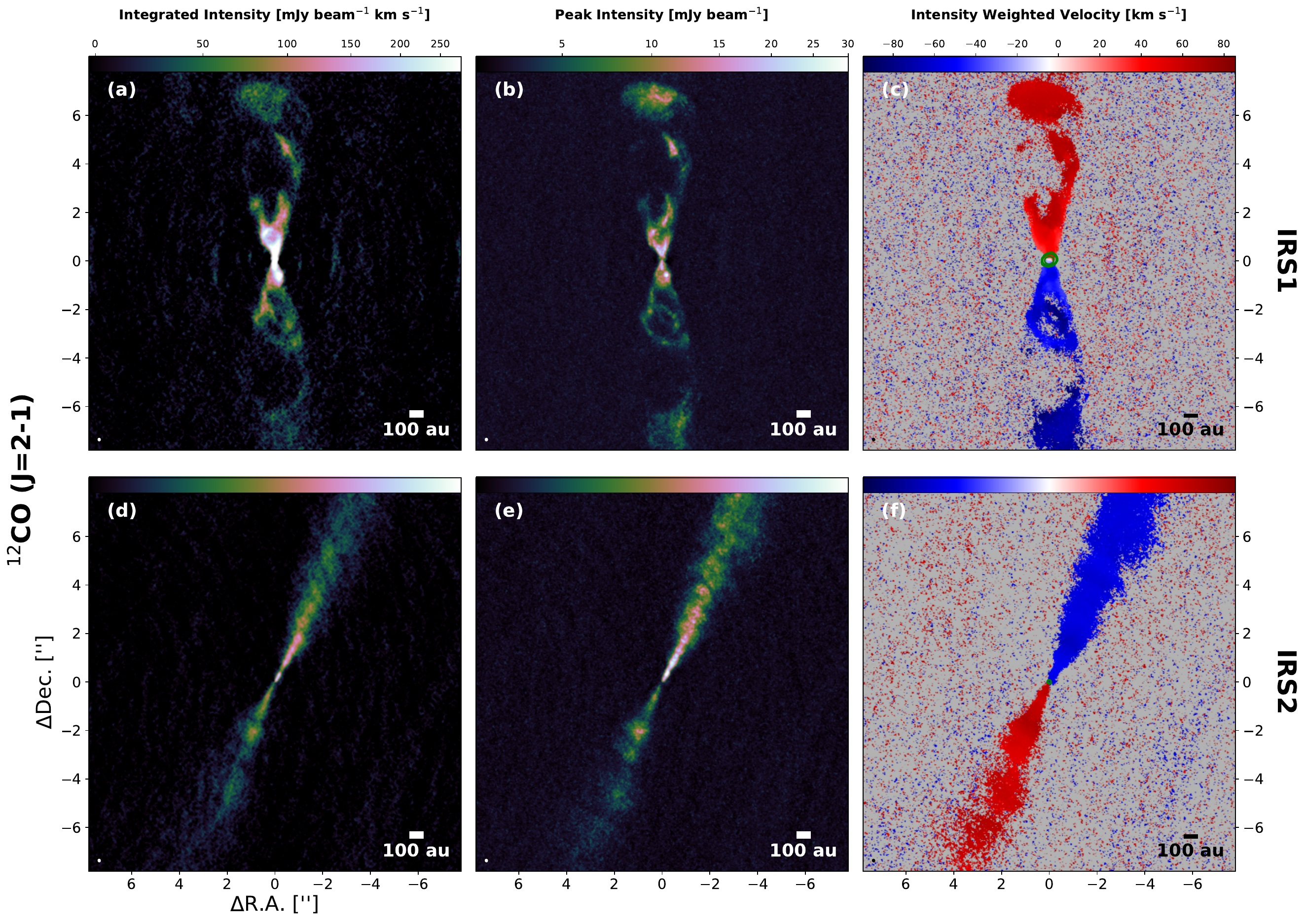}
 \caption{EHV emission from $^{12}$CO in IRS1 (upper row) and IRS2 (bottom row). The left column shows the integrated intensity (moment 0) maps, the middle column the peak intensity maps, and the right column the velocity-weighted intensity (moment 1) maps. The ellipse in the bottom-left corner is the beam size. The maps are generated covering velocities $|V + V_\mathrm{sys}| >$ 30~km~s$^{-1}$.}
\label{fig:12CO_jet}
\end{figure*}

Figure \,\ref{fig:12CO_jet} shows the collimated jet in the EHV regime in $^{12}$CO. Panels (a) and (d) present the integrated intensity (moment 0) of IRS1 and IRS2, respectively, and show a region of $8\arcsec \times 8\arcsec$ centered on the protostars. We see that although the IRS1 and IRS2's jets resemble each other in width and velocity, there is a clear difference in morphology, which is possibly indicative of how angular momentum is carried away in each source, or indicative of different evolutionary stages. IRS1's jet exhibits a striking and prominent double helical structure on each pole. The individual channel maps displayed in Appendix\,\ref{app:chanmap_IRS1} give a better appreciation of the helical structure.

In contrast, IRS2's jets are narrower and seem to show compact chains of knots and bow shocks which are signs of episodic accretions, a more commonly observed characteristic among Class 0 protostars \citep[e.g.][]{Santiago-Garcia+etal_2009, Tafalla+etal_2015,Tafalla+etal_2017, Lee+etal_2022}. Also, the southern jet is fainter than the northern one, following the trend also visible in the wide-angle outflow.

\begin{figure*}
\plotone{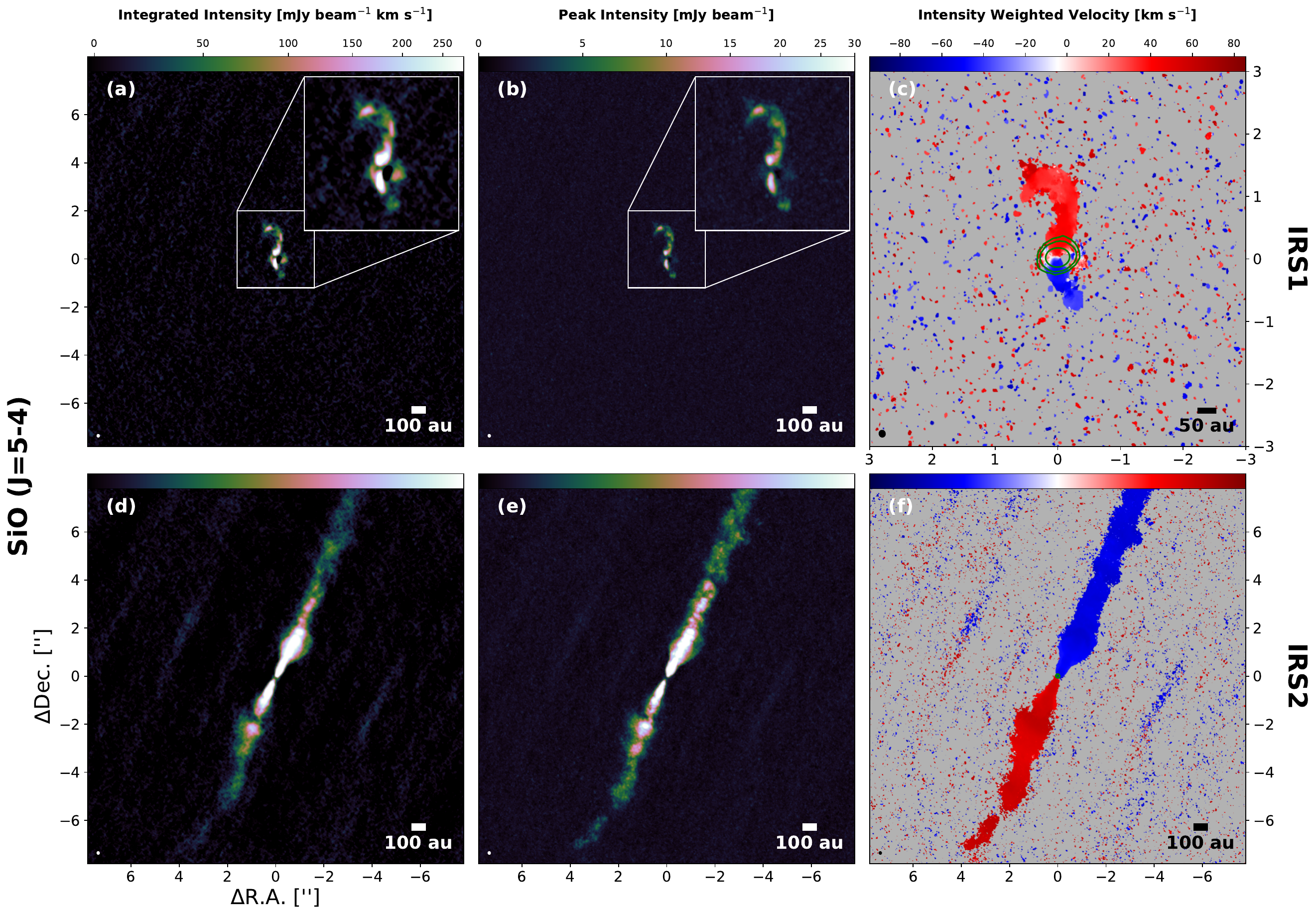}
 \caption{EHV emission from SiO in IRS1 (upper row) and IRS2 (bottom row). The left column shows the integrated intensity (moment 0) maps, the middle column the peak intensity maps, and the right column the velocity-weighted intensity (moment 1) maps. The ellipse in the bottom-left corner is the beam size. The maps are generated covering velocities $|V + V_\mathrm{sys}| >$ 30~km~s$^{-1}$. The green contours mark the continuum emission with increasing values of $\sigma$ [4$\sigma$, 7$\sigma$, 14$\sigma$, 70$\sigma$, 140$\sigma$] and [4$\sigma$, 15$\sigma$, 50$\sigma$, 100$\sigma$] for IRS1 and IRS2, respectively.}
\label{fig:sio_jet}
\end{figure*}

Fig.\,\ref{fig:sio_jet} shows the emission from SiO in the same velocity range (EHV) as in Fig.\,\ref{fig:12CO_jet}. SiO is generally assumed to be a good tracer of shocked gas \citep{Schilke+etal_1997, Gusdorf+etal_2008, Guillet+etal_2008} associated with protostellar jets \citep[e.g.][]{Codella+etal_2007, Cabrit+etal_2012, Tafalla+etal_2017}. In IRS1, although clearly belonging to the jet, the SiO emission is not prominent and is confined within 2$\arcsec$ from the protostar with an south-north asymmetry: the northern redshifted lobe is more extended than the southern lobe. In IRS2, on the other hand, the SiO emission traces the jet over the same distance to the protostar as $^{12}$CO and both are unambiguously spatially coincident, although the SiO jet appears slightly narrower than in $^{12}$CO. We note that at velocities $<$~20~km~s$^{-1}$, SiO shows an northern blueshifted 'tail' extending up to 10$\arcsec$ from the protostar and shows strictly no extended emission in the southern redshifted part (see the channel maps in the Appendix). This seems to contradict the observations made by \citet{Garay+etal_1998}, who reported the presence of SiO tracing the outflows in a very large spatial scale in a velocity range $|v| < 10$~km~s$^{-1}$. A possible explanation may be due to their larger beam. With a beam size of $40\arcsec$ for SiO, their data did not allow the resolution of emission within distances to the protostar $< 20\arcsec$, which roughly represents our field of view. There may be a spatial gap of up to $40\arcsec$ between the protostar and the rise of the SiO low-velocity outflow emission. Alternatively, with a larger beam, they could be picking up larger-scale, low-surface brightness emission that the ALMA data are not sensitive to. On another hand, the outflow observed by \citet{Garay+etal_1998}, both in $^{12}$CO and SiO, seems to lean toward the NW-SE direction, similarly to IRS2's outflow and unlike IRS1's outflow (which is toward the S-N direction) from our dataset. This should suggest that the SiO (and $^{12}$CO) large-scale emission detected by \citet{Garay+etal_1998} is mostly arising from IRS2.

\begin{figure*}
\plotone{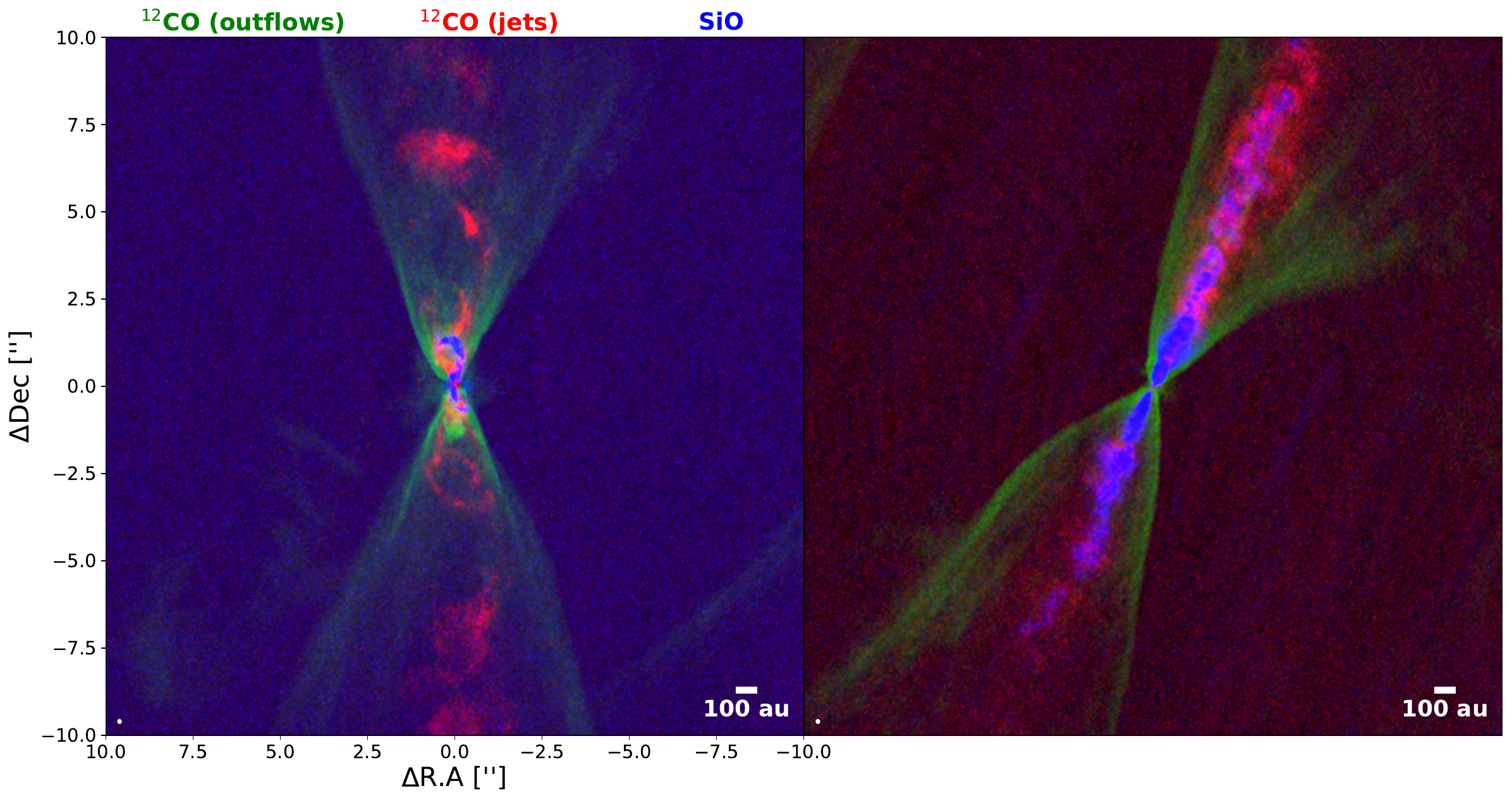}
 \caption{Overlay of the moment 0 maps of $^{12}$CO and SiO. The $^{12}$CO emission is integrated over the SHV regime (green) and over the EHV regime (red). The SiO emission (blue) is integrated over all velocities.}
\label{fig:overlay}
\end{figure*}

To summarize and provide a complete view on the different components (outflows and jets) around IRS1 and IRS2, images are created from an overlay of the $^{12}$CO and SiO intensity maps, integrated at velocities tracing the SHV components and the EHV collimated jets separately. The images are presented in Figure\,\ref{fig:overlay}, where the green and red represent the outflows (SHV regime) and the jets (EHV regime), respectively, in $^{12}$CO, and the blue emission is SiO. This overlay reveals the difference of nature of IRS1 and IRS2's jets: while the outflow appears similar in both sources, their respective jet shows different features.

\subsection{Gas material on envelope and disk scales}\label{sec:res_lowvel}

In addition to the outflow, $^{12}$CO also shows weak, blueshifted emission on the east and west side of the protostar in the moment 1 map of IRS1 (Fig.\,\ref{fig:12CO_maps}c). Similarly, there is weak, blueshifted $^{12}$CO emission on the northeast and southwest side of IRS2 in its moment 1 map (Fig.\,\ref{fig:12CO_maps}f). These blueshifted emissions perpendicular to the outflows are likely associated with infalling envelopes around the protostars.

The integrated and peak intensity maps of the $^{13}$CO emission (Fig.\,\ref{fig:13CO}a, b, d, e), unlike $^{12}$CO, show a strong compact emission around the central protostars which seems to trace material on envelope and disk scales rather than the outflows. There is a central depression in IRS1 inside $\sim$10~au, which is probably due to absorption by optically thick dust. There is no such visible depression in IRS2. 

The $^{13}$CO moment 1 maps (Fig.\,\ref{fig:13CO}c and f) shows that the $^{13}$CO compact emission exhibits velocity gradients along the major axes of the continuum emission. In IRS1, down to the scale of the 1.3-mm continuum emission ($\lesssim$ 0$\farcs$5), there appears to be a velocity gradient along the major axis of the continuum emission, possibly indicating disk rotation. However, it is noticeable that the center of the mm component does not coincide with a velocity near the systemic velocity (-4.45 km~s$^{-1}$). Indeed, most of the mm continuum is overlaid by strong blueshifted emission. This is likely because the emission at the line center and the continuum are optically thick, which skews the velocities in the intensity-weighted velocity maps. 





\begin{figure*}
\plotone{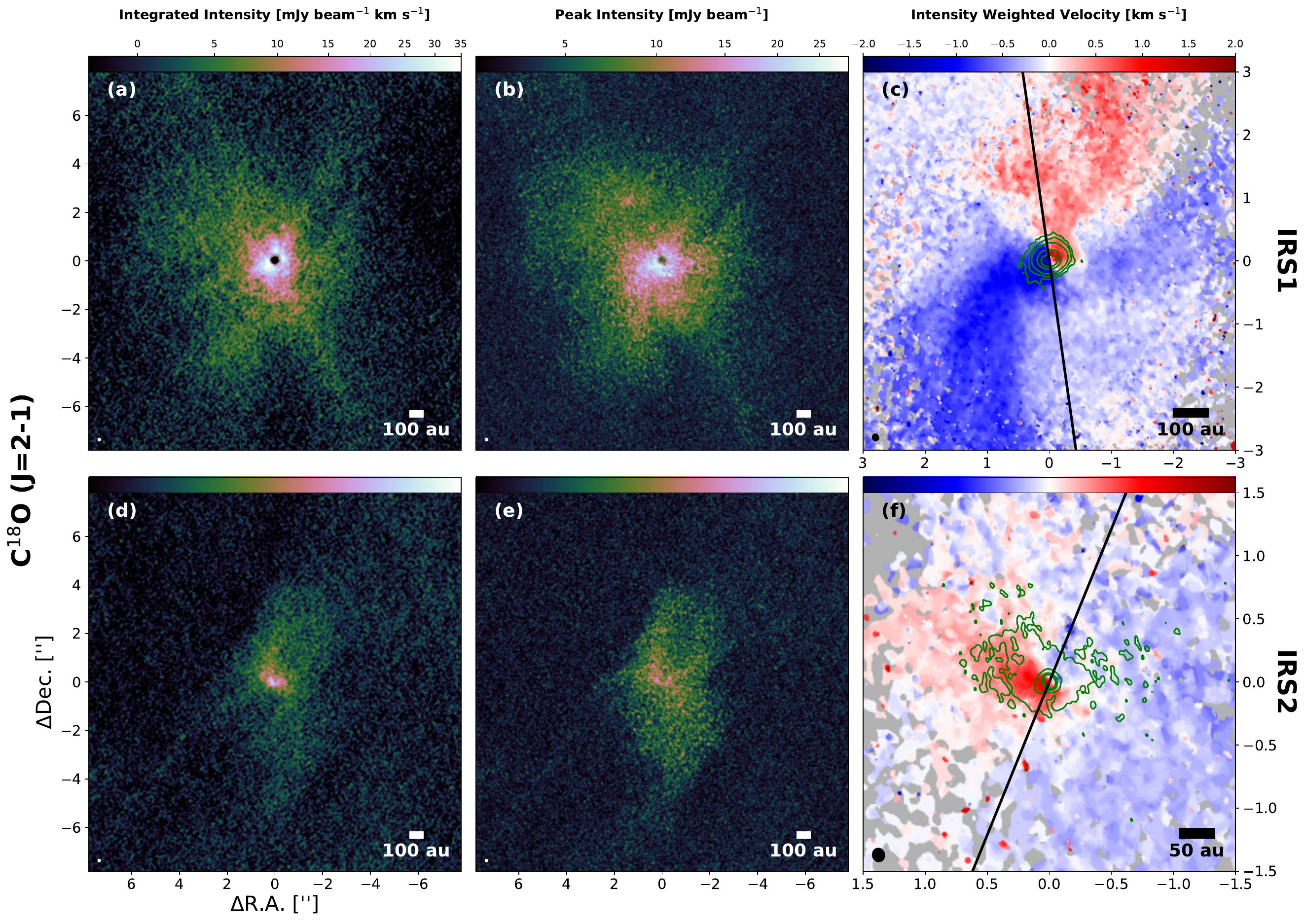}
 \caption{Emission from C$^{18}$O in IRS1 (upper row) and IRS2 (bottom row). The left column shows the integrated intensity (moment 0) maps, the middle column the peak intensity maps, and the right column the mean velocity (moment 1) maps. The maps are generated using a velocity range of -7.55~km~s$^{-1}$ $< V + V_\mathrm{sys} <$ +7.55~km~s$^{-1}$. The green contours mark the continuum emission with increasing values of $\sigma$ [4$\sigma$, 7$\sigma$, 14$\sigma$, 70$\sigma$, 140$\sigma$] and [4$\sigma$, 15$\sigma$, 50$\sigma$, 100$\sigma$] for IRS1 and IRS2, respectively. The ellipse in the bottom-left corner is the beam size. The black lines in the moment 1 maps correspond to the direction orthogonal to the semi-major axis $P.A.$ value derived from the best-fit models.}
\label{fig:C18O}
\end{figure*}

The C$^{18}$O ($J$=2--1) emission shows even more compact structures than the $^{13}$CO, as demonstrated in Figure\,\ref{fig:C18O} presenting maps of the C$^{18}$O detected above 3 sigma levels for IRS1 (top row) and IRS2 (bottom row). This compact emission is considered to trace envelopes and possibly disks as well. Note that there is very faint C$^{18}$O emission showing similar extended structures to the $^{12}$CO outflow, suggesting that the C$^{18}$O also weakly traces the outflows. We also note that the C$^{18}$O is depressed at the central position of IRS1, as also seen in the $^{13}$CO. As explained in the case of the $^{13}$CO above, this is probably due to absorption by optically thick dust.

The C$^{18}$O moment  1 map of IRS1 (Fig. 9c) shows that the overall velocity gradient of the compact C$^{18}$O emission around IRS1 is from southeast (blueshifted) to northwest (redshifted). Unlike in $^{13}$CO, the C$^{18}$O velocity gradient appears more consistent with the elliptic continuum structure, with the isovelocity line at systemic velocity crossing the protostellar position. This velocity gradient can be explained as disk rotation.

The C$^{18}$O moment 1 map of IRS2 (Fig. 9f) shows that the overall velocity gradient of the compact C$^{18}$O emission around IRS2 is from northwest (blueshifted) to southeast (redsshifted), which is consistent with that seen in the $^{13}$CO. Note that the direction of the velocity gradient is significantly tilted from the major axis of the continuum emission, which is also consistent with the $^{13}$CO, suggesting that the velocity gradient may not be explained as pure disk rotation and an infalling and rotating envelope may be also required to interpret the velocity gradient \citep[e.g.][]{Momose+etal_1998,Flores+edisk_2023}. More details of the kinematics of the compact C$^{18}$O emission, particular its possible rotating motions, is discussed in Section~\ref{sec:ana_starmass}.


\section{Analysis} \label{sec:analysis}

\subsection{Dust mass}\label{sec:ana_mass}
Under the assumption that the dust emission is optically thin and isothermal, we can estimate the dust mass using the following equation,

\begin{equation}\label{eq:surdensg}
M_{\rm dust} =  \frac{F_\nu D^2} { \kappa_\mathrm{225\mathrm{GHz}}B_\nu(T_\mathrm{dust}) },
\end{equation}	

\noindent where $F_\mathrm{\nu}$ is the continuum flux density, $D$ is the distance to the source, $B_\nu$ is Planck's law of black-body radiation, and $\kappa_\mathrm{225GHz}$ is the opacity of the dust at the observation frequency 225 GHz (1.3 mm). We use $\kappa_\mathrm{225GHz} = 2.30$ cm$^2$/g (per unit dust mass) from \citet{Beckwith+etal_1990} and assume a dust temperature $T_\mathrm{dust} = 20$ K for both sources, which is the median value from Taurus disks \citep{Andrews+Williams_2005,Ansdell+etal_2016} (a value of $T_\mathrm{dust} = 30$ K is also commonly used for embedded disks, e.g. \citet{Tobin+etal_2015}, which would result in a smaller derived mass value). The values of $F_\nu$ are derived from the best-fit 2-D Gaussian models with parameters given in Table\,\ref{tab:bestfit}. We find a total mass $M_\mathrm{tot} = 0.114$~M$_\odot$ for IRS1 and $M_\mathrm{tot} = 3.86\times10^{-3}$~M$_\odot$ for IRS2, assuming a gas-to-dust mass ratio of 100.

A different way to estimate the dust mass is to scale the dust temperature with the bolometric luminosity of the source. We adopt the average dust temperature 

\begin{equation}\label{eq:Tbol}
    T_{\rm dust} = T_\mathrm{0} \bigg(\frac{L_{\text{bol}}}{L_{\odot}}\bigg)^{1/4}
\end{equation}

\noindent calculated using a grid of radiative transfer models \citep{Tobin+etal_2020}, where $T_0 = 43$~K and $L_{\rm bol} = 10$~L$_\odot$ for IRS1 and $L_{\rm bol} = 1.1$~L$_\odot$ for IRS2, computed from the SEDs in \citet{Ohashi+edisk_2023}. This gives a T$_{dust}$ value of 76 K for IRS1 and 44 K for IRS2. Using the same parameters as above we derive a total mass $M_\mathrm{tot} = 2.40\times10^{-2}$~M$_\odot$ and $M_\mathrm{tot} = 1.50\times10^{-3}$~M$_\odot$ for IRS1 and IRS2, respectively.

The \citet{Beckwith+etal_1990} opacity is commonly used for the analysis of dust in circumstellar disks and adopted in the other studies of eDisk sources. Other previous studies, such as \citet{Tobin+etal_2019} and \citetalias{Yang+etal_2020}, adopted dust opacities from \citet{Ossenkopf+Henning_1994} to derive the dust mass in BHR 71, corresponding to grains with thin ice mantles, typically adopted for modeling the dust in protostellar envelopes. The value at 230 GHz for those is 0.899 cm$^2$~g$^{-1}$ \citep{Tobin+etal_2019} i.e. about a factor 2.5 lower than the value adopted here. This opacity would result in a dust mass larger by a factor 2.5 (Eq.~\ref{eq:surdensg}). For IRS1, \citet{Tobin+etal_2019} found a total (dust+gas) mass of 1.13 M$_\odot$ and \citetalias{Yang+etal_2020} found a total mass of 2.1$\times10^{-2}$~M$_\odot$. Both values deviate from a factor 2.5 compared to our mass estimates. The other differences should be due to either a different measured flux density and/or a different adopted dust temperature (and to a lesser extent a different assumed distance to the source). Using the same distance to the sources, opacity value, and temperature as in \citet{Tobin+etal_2019}, we measure mass values of about a factor $\sim$~3 and $\sim$~8.5 lower than their values for IRS1 and IRS2, respectively. These factors correspond exactly to those of the respective measured flux density values between our study and their study. Their larger flux density values are probably due to the spatial scales probed by the observations compared to the different spatial scales probed by the respective observations: the larger numbers of the estimated dust masses from \citet{Tobin+etal_2019} stem from observations using the ALMA including the Compact Array with a beam of $\sim1\arcsec$ and sensitive to the extended emission in the larger scale envelope, which might have led to include envelope emission in the flux density. \citetalias{Yang+etal_2020} assumed a mass-weighted dust temperature of 148 K, which in turn results in a smaller inferred mass for the same flux than our initial estimates. These derived masses are in reasonable agreement with the estimates we obtain using the dust temperature scaled according to the source luminosity with the remaining differences due to the different beam sizes and dust opacities.


We note that Equation\,\ref{eq:surdensg} comes along with uncertainties. In addition to the uncertainties due to the temperature and dust opacities described above, assuming an optically thin emission underestimates the value of the mass if the source is optically thick. In the case of IRS1, the observed asymmetry across the minor axis and its peak brightness temperature (Fig.\,\ref{fig:cont_cut1}) strongly suggest that the source is optically thick \citep{Takakuwa+edisk_2024}. As previously mentioned the fact that the dust emission is brighter on the lower part indicates that the dust is not settled and the object is geometrically thick. Given its relatively low measured inclination (39°), the presumed disk is expected to have a very steep flaring index and/or a mass significantly larger than the one derived with Eq.\,\ref{eq:surdensg}. 


\subsection{Stellar mass}\label{sec:ana_starmass}

In this section, we analyze details of the kinematics of the compact emission around IRS1 and IRS2 to search for signature of Keplerian rotation with a goal to constrain the dynamical stellar mass of IRS1 and IRS2. Among the three CO isotopologues, C$^{18}$O is the best candidate to search for Keplerian motion. $^{12}$CO is too optically thick at small scales (a few tens of au) to detect disk material and the emission is mostly dominated by the outflows. Both the $^{13}$CO and C$^{18}$O emissions are optically thinner than $^{12}$CO and should be dominated by the disk and/or inner envelope (although both also trace the outflow walls) but the optical depth of $^{13}$CO is larger due to its higher abundance. This common characteristic has been reported in most other sources among the eDisk sample \citep[see, for instance, ][]{vanHoff+edisk_2023}. 

\begin{figure*}
\plotone{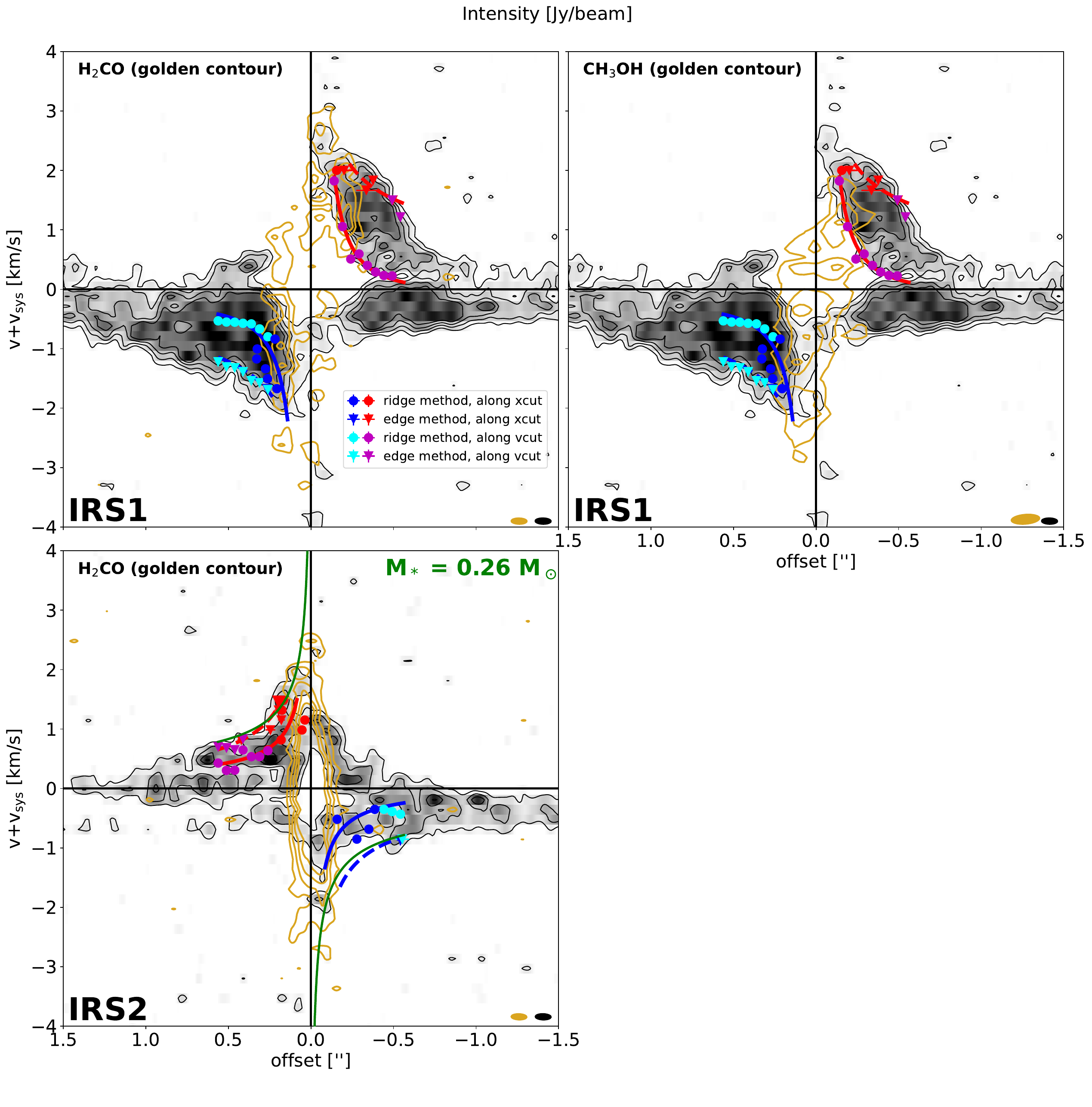}
 \caption{C$^{18}$O P-V diagrams (gray-scale maps and black contours) in IRS1 (top) and IRS2 (bottom). Contours mark the emission at 3$\sigma$, 5$\sigma$, 7$\sigma$, 9$\sigma$ for both IRS1 and IRS2. The cuts were made with a width of 1 beam ($0\farcs09$ along $P.A. = 98\fdg15$ for IRS1 and $0\farcs07$  along $P.A. = 67\fdg6$ for IRS2). In IRS1, the golden contours mark the emission of H$_2$CO (left column) and CH$_3$OH (right column) at their 3$\sigma$, 5$\sigma$, 7$\sigma$, 9$\sigma$ values. In IRS2, the golden contours mark the emission of H$_2$CO at the same $\sigma$ values. The velocity resolution is 0.17 km s$^{-1}$ for both C$^{18}$O and H$_2$CO, and 0.15 km s$^{-1}$ for CH$_3$OH. The golden ellipses in the bottom-right corners are the beam sizes of H$_2$CO and CH$_3$OH and the black ones are that of C$^{18}$O. The upside down triangles and circles are the data points derived from the edge and ridge methods, respectively, and their fitted rotation curves are shown in dashed and solid lines, respectively. The blue and red data points are obtained from a fit along the position axis (xcut), while the cyan and magenta points are obtained from a fit along the velocity axis (vcut). The green lines show the line-of-sight Keplerian profiles with the stellar mass indicated in the top right corner.}
\label{fig:pv_full}
\end{figure*}

Figure\,\ref{fig:pv_full} shows the Position-velocity (P-V) diagrams in C$^{18}$O (gray-scale map) for IRS1 and IRS2. The cuts were made along $P.A. = 98.15\arcdeg$ and $P.A. = 67.6\arcdeg$ for IRS1 and IRS2, respectively, corresponding to the $P.A.$ values of the dust continuum major axis derived from the best-fit models. Although both sources exhibit evidence of differential rotation, the diagrams also reveal that a significant fraction of the emission is visible in all non-Keplerian quadrants (top left and bottom right for IRS1, top right and bottom left for IRS2). This indicates the presence of rotating infalling motion \citep[e.g.][]{Ohashi+etal_1997, Momose+etal_1998}. Moreover, there is a quadrant that systematically shows less emission (top left for IRS1, lower left for IRS2) compared to the opposite quadrant, which should suggest that the infall is asymmetric. Both sources also show emission at large distances $> 1.5\arcsec$ ($> 260$~au) close to systemic velocity, indicative of envelope material. More importantly, the diagrams confirm the previous observation made by \citet{Tobin+etal_2019} which revealed that the two objects rotate in opposite directions. The diagrams (in IRS1 in particular) have an inner depression at small distances ($<0\farcs5$), suggesting that the dust emission is optically thick in the center where not much line emission is received because of the continuum oversubtraction. Note that the depression of the C$^{18}$O in IRS1 is also seen in its moment 0 and 8 maps (Fig.~\ref{fig:C18O}a and b). 

For further investigations of the kinematics in this region, we also show the emission of H$_2$CO (218.76 GHz) and CH$_3$OH (250.291 GHz)\footnote{The CH$_3$OH datacube is provided by the ALMA program 2021.1.00262.S (PI: Y.-L. Yang) focused on IRS1 only and with a better spectral resolution.} (golden contours) on top of the C$^{18}$O emission. Interestingly, both H$_2$CO and CH$_3$OH show up at the inner rim of C$^{18}$O, filling in the central region where the continuum emission is optically thick. These lines are likely to trace the warm and dense gas closer to the star whereas C$^{18}$O traces the colder outer part of the disk and/or the infalling/rotating envelope.



We fit the C$^{18}$O P-V diagram with the emission ridge method \citep{Ohashi+etal_2014, Aso+etal_2015, Aso+etal_2017, Yen+etal_2017,Sai+etal_2020} and the outer emission edge method \citep{Seifried+etal_2016, Alves+etal_2017, Reynolds+etal_2021} using the Python package \texttt{pvanalysis} included in the Spectral Line Analysis/Modeling code \texttt{SLAM}\footnote{\url{https://github.com/jinshisai/SLAM}} \citep{Aso_2023}. The fitting process is detailed in \citet{Ohashi+edisk_2023}. First, the code computes the edge and ridge points of coordinates ($r$, $v$) in the diagram, using a threshold $> 5\sigma$ (with $\sigma = 1.54\times10^{-3}$ Jy beam$^{-1}$ for IRS1 and $\sigma = 1.72\times10^{-3}$ Jy beam$^{-1}$ for IRS2). Then, using the Markov-Chain Monte Carlo (MCMC) method, the code fits the ridge and edge points separately with the single power-law function,

\begin{equation}\label{eq:fit}
    V = V_m\left(\frac{R}{R_m}\right)^{-p_{in}} + V_{\rm sys},
\end{equation}

\noindent  where $R_{\rm m}$ is the characteristic radius, $V_{\rm m}$ is the velocity at radius $R_{\rm m}$, $V_{\rm sys}$ is the systemic velocity, and $p_{in}$ is the power-law index. The fitting was made with emission detected within 0.6" from the central stars to avoid possible contamination of envelope emission. The results are shown in Fig.\,\ref{fig:pv_full}, on top of the P-V diagrams. The fit provides the pairs ($R_{\rm m}$, $V_{\rm m}$) from which we can derive the dynamical stellar mass $M_\star = (R_{\rm m} V_{\rm m}^2)/G/{\rm sin}^2 i$, provided that the index $p_{in}$ is close enough to the Keplerian value of 0.5. The fitting values are summarized in Table\,\ref{tab:fitting}.

For IRS1,  the edge method gives a power-law index $p_\mathrm{in}$ of 0.47$\pm$ 0.03, which is consistent with a Keplerian rotation. From the results obtained with the edge method we can estimate an upper limit on the stellar mass of $0.46 \pm 0.01$~M$_\odot$, because the edge method tends to overestimate the mass \citep{Maret+etal_2020}. The ridge method gives a $p_\mathrm{in}$ value of 1.61, which is not physical. A possible reason why the ridge method does not provide a reasonable result might be because the C$^{18}$O emission is reduced at the center due to the optically thick continuum emission, and as a result, the redshifted C$^{18}$O emission measured by the SLAM code does not arise from the ridge of the emission, but from the inner edge. For IRS2, we derive a value of p$_\mathrm{in} = 0.66 \pm 0.05$ and $0.78 \pm 0.03$ for the edge and ridge method, respectively. These values are not Keplerian, but remain between 0.5 and 1, meaning that the observed rotation is between what is expected from the emission of a rotating envelope and a Keplerian disk. We are therefore limited in what can be quantitatively done with regard to confirming Keplerian motion, because the kinematics is likely contaminated by the emission of the envelope. Nevertheless, the results may still suggest that the outer edge of the C$^{18}$O P-V diagram can be fitted with a Keplerian rotation curve which could provide a reliable estimation of an upper limit of stellar mass. Instead of using p$_\mathrm{in}$ as a free parameter, we can fix it at 0.5 by assuming that there is a Keplerian rotation and then fit the P-V diagram. We derive of stellar mass of 0.26 $\pm$ 0.01 using this method. This sets a reasonable upper limit on the stellar mass of IRS2. The Keplerian profile for this stellar mass is showed in the C$^{18}$O P-V diagram of IRS2 (Fig.\,\ref{fig:pv_full}).

\begin{deluxetable*}{llcc|ccc}
\tablecaption{Results of the Best Fitting to the C$^{18}$O P-V Diagram. \label{tab:fitting}}
\tablehead{
\colhead{Parameter [units]} & \multicolumn{1}{l}{Description} & \multicolumn{2}{c}{IRS1} & \multicolumn{3}{c}{IRS2} \\
\colhead{} & \colhead{} & \colhead{edge} & \colhead{ridge} & \colhead{edge} & \colhead{edge ($p_\mathrm{in}$ fixed)} & \colhead{ridge}
}
\startdata
$R_\mathrm{m}$ [au] & characteristic radius & 85.81 $\pm$ 2.200 & 98.76 $\pm$ 0.240 & 55.78 $\pm$ 2.080 &  52.32 $\pm$ 1.760 &  47.50 $\pm$ 0.710 \\
$V_\mathrm{m}$ [km~s$^{-1}$] & velocity at $R_\mathrm{m}$ & 1.386 $\pm$ 0.000 & 0.219 $\pm$ 0.000 & 1.072 $\pm$ 0.000 & 1.072 $\pm$ 0.000 & 0.574 $\pm$ 0.000 \\
$p_\mathrm{in}$  & Power of the fitting function & 0.466 $\pm$ 0.031 & 1.605 $\pm$ 0.008 & 0.658 $\pm$ 0.046 & 0.5 & 0.781 $\pm$ 0.026 \\
$V_{\rm sys}$  [km~s$^{-1}$] & systemic velocity & $-$4.533 $\pm$ 0.013 & $-$4.836 $\pm$ 0.006 & -4.403 $\pm$ 0.021 & -4.384 $\pm$ 0.015 & -4.244 $\pm$ 0.005 \\
$M_\mathrm{*}$ [$M_{\sun}$] & Mass of the protostar & 0.461 $\pm$ 0.011 & 0.226 $\pm$ 0.002$^{\dag}$ & 0.273 $\pm$ 0.011 & 0.257 $\pm$ 0.009 & 0.129 $\pm$ 0.005 \\
\enddata
\tablecomments{$^{\dag}$ If $p_\mathrm{in}$ is significantly deviated from 0.5, the derived stellar mass is not reliable.}
\label{tab:pv}
\end{deluxetable*}

\subsection{Kinematics of the jets}\label{sec:ana_outflowkin}

\begin{figure*}
\plotone{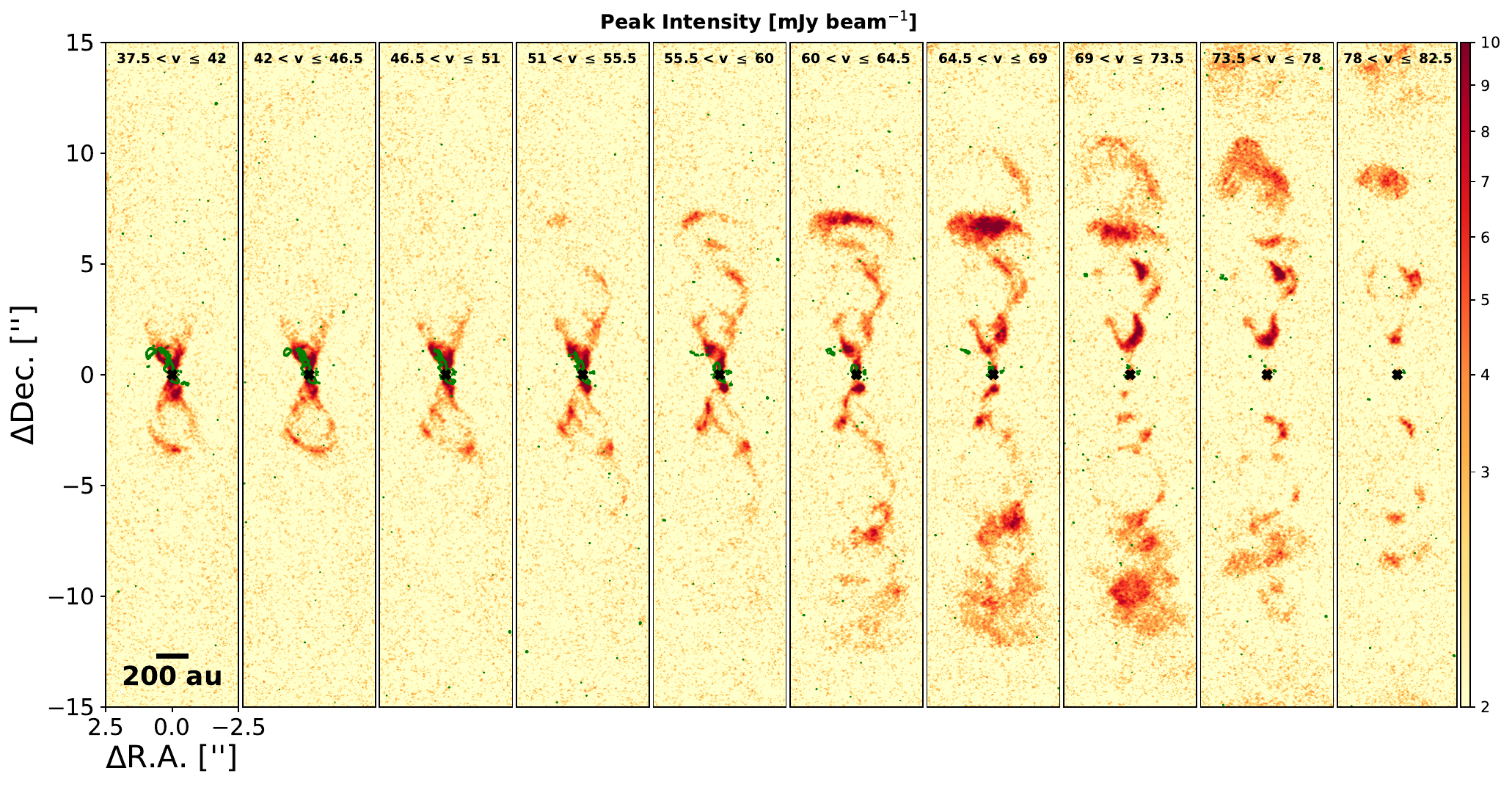}
 \caption{Peak intensity maps of the protostellar jet from the $^{12}$CO emission line in IRS1. The velocity increases from left to right panels from $v \sim$37~km~s$^{-1}$ to $v \sim$83~km~$s^{-1}$ where $v = |V - V_{sys}|$. Each panel represents a velocity range of about 4.5~km~s$^{-1}$ with the lowest velocity range in the leftmost panel and the largest range in the rightmost panel. The whole range spans 45~km~s$^{-1}$. The green contours mark the SiO peak intensity maps at 3$\sigma$, 4$\sigma$, 5$\sigma$, and 6$\sigma$ in the same velocity ranges as the respective panel. The images are rotated so that the jets are shown vertically. The black cross marks the position of the protostar.}
\label{fig:irs1_12CO_mapjet}
\end{figure*}

We present selected peak intensity maps of the $^{12}$CO and SiO jet of IRS1 in Figure
\,\ref{fig:irs1_12CO_mapjet}, showing successive increasing velocity intervals relative to the systemic velocity, from left to right.

The EHV jet presents a clear helical shape which is evidence of rotation and is indicative of the mechanisms carrying away the angular momentum from small scales. \citet{Kwon+etal_2015} reported the presence of a similar helical structure in the outflow of the Class 0 protostellar system L1157, which is interpreted as two precessing jets. The morphology of IRS1's EHV jets could agree with this two-jet scenario. In particular, Figures.\,\ref{fig:irs1_12CO_mapjet}, \ref{fig:chan_irs1_vhv_12co_blue}, and \ref{fig:chan_irs1_vhv_12co_red} seem to show two apparent collimated jets extending up to 5$\arcsec$. However, similar morphologies can also be explained by other scenarios. A wide-angle disk wind can drive a rotating outflow \citep[][]{deValon+2022, lopez+2023}, and \citet{machida+etal_2007} showed with numerical simulations that the magnetic field can naturally be at the origin of a twisted jet.

The mechanism of precession itself is not well-understood and various theoretical models have been proposed to explain its origin. Most models involve the presence of a companion protostar \citep[e.g.][]{Terquem+etal_1999, Masciadri+etal_2002, Montgomery_2009}. Other models explain the precession by the misalignment between the disk rotation and ejection axes \citep[e.g.][]{Frank+etal_2014}, and IRS1 shows such misalignments. 

\begin{figure*}
\plotone{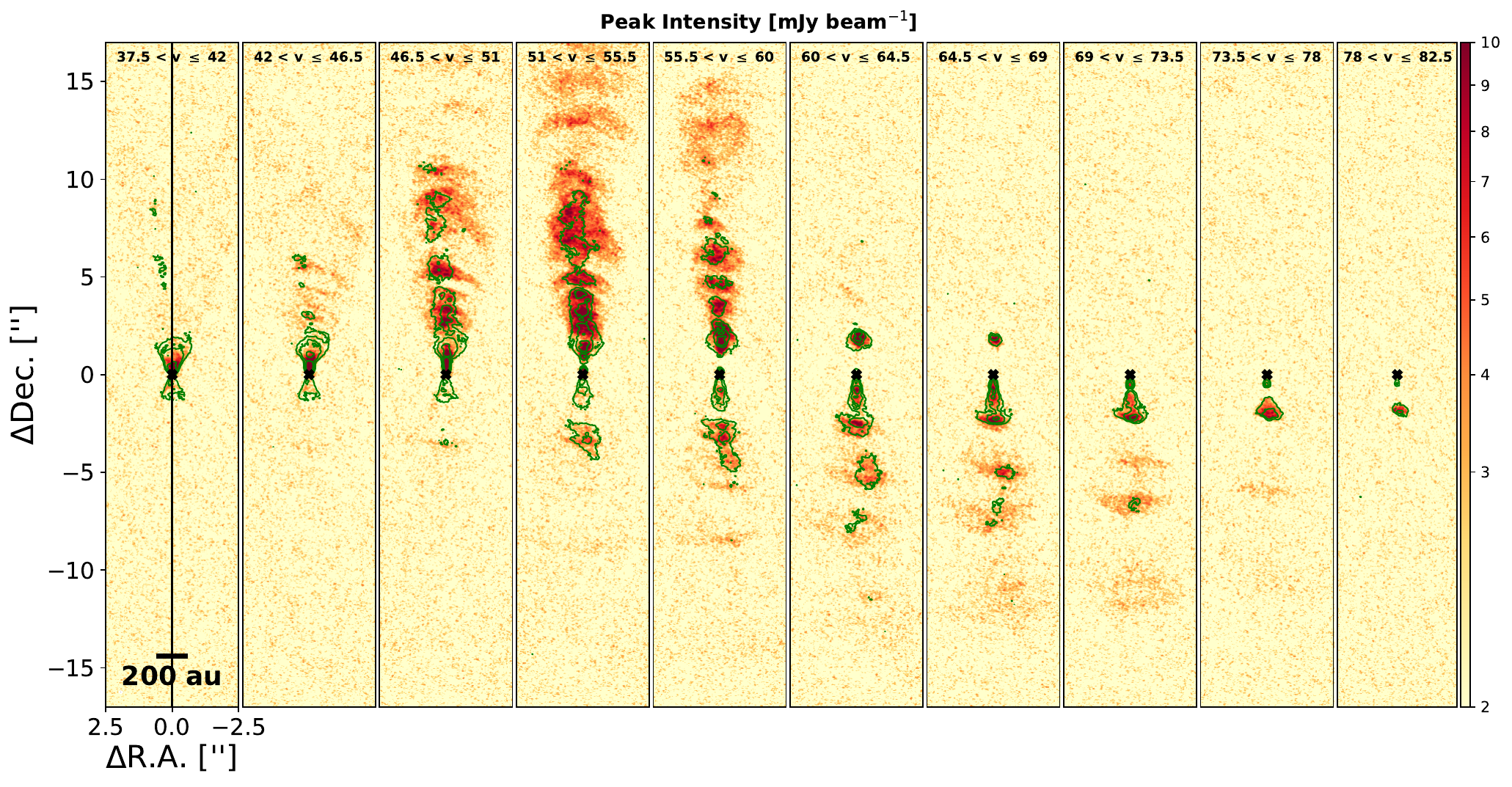}
 \caption{Peak intensity maps of the protostellar jet from the $^{12}$CO emission line in IRS2. The velocity increases from left to right panels from $v \sim$37~km~s$^{-1}$ to $v \sim$83~km~$s^{-1}$ where $v = |V - V_{sys}|$. Each panel represents a velocity range of about 4.5 km~s$^{-1}$ with the lowest velocity range in the leftmost panel and the largest range in the rightmost panel. The green contours mark the SiO peak intensity maps at 8$\sigma$, 24$\sigma$, and 80$\sigma$ in the same velocity ranges as the respective panel. The images are rotated so that the jets are shown vertically. The vertical black line in the first panel indicates the axis along which the P-V diagrams in Fig.\,\ref{fig:pvjet} are made. The black cross marks the position of the protostar.}
\label{fig:irs2_12CO_mapjet}
\end{figure*}

Figure \ref{fig:irs2_12CO_mapjet} shows selected peak intensity maps of the $^{12}$CO jet in IRS2 covering the same velocity ranges as in Fig.\,\ref{fig:irs1_12CO_mapjet}. The SiO emission (green contours) is also presented on top of the CO emission. Most of the SiO emission is confined within a narrower region of the jet than the $^{12}$CO emission and appears to depart from the jet axis (black line). The southern jet deviates toward the west at a projected distance of about 5$\arcsec$ from the star, then comes back to the vertical direction of the figure at higher distances (This is particularly evident in the sixth and seventh panels of Fig.\,\ref{fig:irs2_12CO_mapjet} as well as in the velocity maps of SiO shown in the appendix). The northern jet, on the other hand, appears to slightly deviate toward the east. This behavior is a strong indication that the jet is precessing.

\begin{figure*}
\plotone{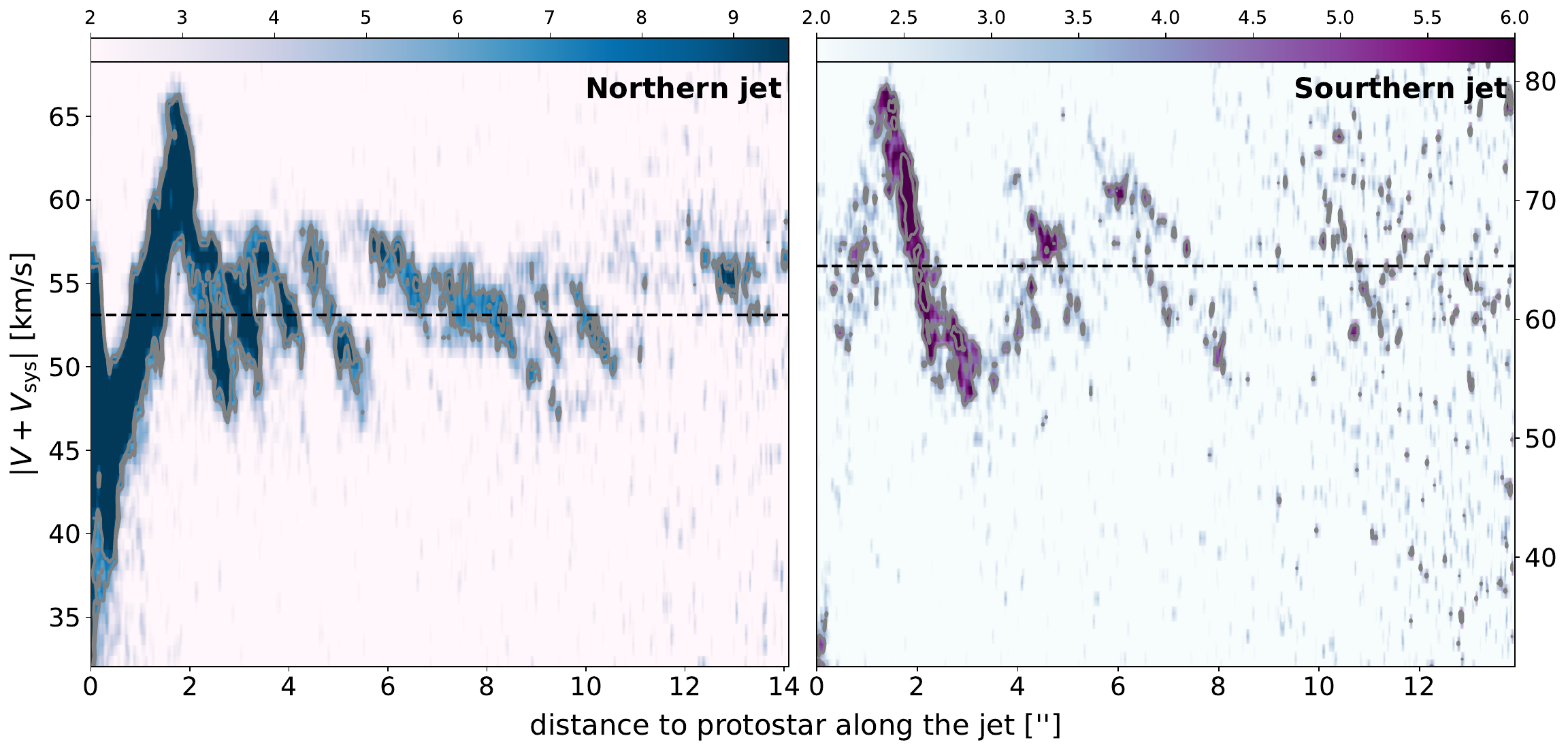}
 \caption{$^{12}$CO P-V cuts across the jet's axis in IRS2 with a width of 1 beam. Left: northern blueshifted jet. Right: southern redshifted jet. The protostar position is at a distance of zero. The contours mark the emission at 3$\sigma$ and 5$\sigma$. The horizontal dashed lines show the mean velocity of the southern (65~km~s$^{-1}$) and northern jets (53~km~s$^{-1}$).}
\label{fig:pvjet}
\end{figure*}

The jet in IRS2 is also composed of internal velocity gradients, as illustrated in Figure\,\ref{fig:pvjet}, which shows the P-V diagrams in $^{12}$CO, made across the jet's axis (vertical black line in the first panel of Fig.\,\ref{fig:irs2_12CO_mapjet}), with a width of 1 beam. The velocity structure is a succession of slow and fast emission features along the jet's axis, producing a characteristic 'sawtooth' structure. In each of these features, the gas closer to the source (tail) moves faster than the gas further away from the source (head). The difference in velocity between the head and the tail is around 10-20 km~s$^{-1}$ in both the southern and northern jets. The velocity field of the jet is only affected locally by these velocity gradients, so the gas does not accelerate or decelerate along the stream, hence the horizontal 'sawtooth' structure. This indicates that the gas in the jet should not be affected by its environment. The same patterns have been observed in IRAS04166+2706 jet \citep[][]{Santiago-Garcia+etal_2009}. Theoretical and numerical models have predicted that this sawtooth pattern can be the result of internal working surfaces in a pulsed jet, where internal material compression generates lateral ejections in a knot, making the upstream gas appears slower than the downstream gas if the jet is not in the plane of the sky. The model was initially proposed by \citet{Raga+etal_1990} and a clear sawtooth pattern can be observed in synthetic P-V diagrams from simulations in \citet{Stone+Norman_1993} \citep[see][for a more detailed discussion]{Santiago-Garcia+etal_2009, Tafalla+etal_2017}. This is the most plausible interpretation of the observed structure in IRS2.

A second notable characteristic of the IRS2 jet, unlike IRS1, is the asymmetry in velocity (visible in Fig.\,\ref{fig:irs2_12CO_mapjet}), where the northern jet clearly appears slower than the southern one. Using the P-V diagram, we can define a flux-weighted velocity as follows:


\begin{equation}\label{eq:vmean}
    V_\mathrm{mean} = \frac{\sum_{v_i} v_i F_{tot}(v_i)}{\sum_{v_i} F_{tot}(v_i)}.
\end{equation}

\noindent where $F_{tot}(v_i)$ is the sum of all flux values in each pixel along the jet axis for the velocity $v_i = |V_i + V_\mathrm{sys}|$. Using Equation\,\ref{eq:vmean} for values of flux $> 3\sigma$, we find that the blueshifted jet has a mean velocity V$_\mathrm{b, mean}$ $\sim$ 53~km~s$^{-1}$ while the redshifted jet has a mean velocity V$_\mathrm{r, mean}$ $\sim$ 65~km~s$^{-1}$ (V$_\mathrm{b, mean}$/V$_\mathrm{r, mean} \sim 0.81$). The mean velocities are measured between a distance of 1$\arcsec$ from the protostar (in order to avoid emission too close to the envelope materials) and 15$\arcsec$, and are marked by the horizontal dashed lines in Fig.\,\ref{fig:pvjet}. A similar asymmetry was also reported in the L1157 molecular jet \citep{Podio+etal_2016}.

\section{Discussion} \label{sec:discussion}

\subsection{Brightness asymmetry}\label{sec:discus_asymmetry}
In Section\,\ref{sec:res_cont}, we have shown that the dust continuum component in IRS1 shows a shift in the peak brightness along the minor axis. This feature has also been observed toward Class 0 sources HOPS 124 \citep{Sheehan+etal_2020} and HH 212 \citep{Lee+etal_2017, lin+etal_2021} as well as several eDisk sources \citep[e.g.][]{Lin+edisk_2023,Sharma+edisk_2023,Takakuwa+edisk_2024}. If the 1.3-mm dust continuum emission is geometrically thick and the disk is flared, then the far side of the disk can appear brighter than the near side. This effect was discussed in more details by \citet{SOhashi+etal_2022b} and \citet{Ohashi+edisk_2023}. \citet{Takakuwa+edisk_2024} reproduced the skewed observed emission along the minor axis of the disk around R CrA IRS 7B-a using a dust disk model, and found that a high flaring index ($q \gtrsim 0.25$, where $q$ is defined as $h/r \sim r^q$) is necessary to reproduce the observed asymmetry. This geometrical effect therefore depends on the disk being inclined relative to the line-of-sight. The effect is less prominent towards more evolved disks where the millimeter emission is less flared and the dust is more settled \citep{Pinte+etal_2016, Villenave+etal_2020, Villenave+etal_2022}. 

In the case of IRS1, the object is brighter on the southern part which must therefore be the far side. This is consistent with the outflow velocity structure, which is blueshifted on the southern side (Section\,\ref{sec:res_outflow} and \ref{sec:res_fast}). However, the dust continuum component is far from being edge-on (incl $\sim$40$\arcdeg$), so the disk must have a high flaring index. Testing whether the asymmetry is due to the inclination rather than an azimuthally asymmetric dust distribution requires modeling efforts that will be performed in future works. In IRS2, no such effect is visible, which is likely because IRS2 is not well-resolved. Moreover, given the small derived mass of the mm-component, it is also possible that the mm-component is optically thin, explaining the observed symmetric structure.

 \subsection{Stellar to disk mass ratio}\label{sec:discus_mass}
 
Based on the detailed analysis of the C$^{18}$O P-V diagrams, shown in Sec.~\ref{sec:ana_starmass}, we could estimate upper limits of the stellar mass to be $\sim$ 0.46 M$_\odot$ and $\sim$~0.26~M$_\odot$ for IRS1 and IRS2, respectively.

These stellar mass values raise the questions whether the presumed disks are gravitationally stable. In Section\,\ref{sec:ana_mass} we derived a disk mass (gas+dust) of $\sim$0.1~M$_\odot$ in IRS1 and $\sim$3.9$\times10^{-3}$ M$_\odot$ in IRS2, assuming the 1.3-mm continuum emission at $T_\mathrm{dust} = 20$~K, whereas we derived a mass of $\sim$2.4$\times10^{-2}$~M$_\odot$ and $\sim$1.5$\times10^{-3}$~M$_\odot$ for IRS1 and IRS2, respectively, assuming the average dust temperature (Eq.\,\ref{eq:Tbol}). These set lower limits of the dust masses (for the given temperatures) because the derivations are on the assumption of an optically thin mm-component.

It is expected that a ratio M$_\mathrm{disk}$/M$_\mathrm{\star} > 10^{-2}$ implies that a disk can be considered as self-gravitating and thus very likely unstable \citep{Toomre_1964, Goodman+etal_1993, Adams+etal_2006, Eisner+etal_2008, Kratter+etal_2016}. Assuming $T_\mathrm{dust} = 20$~K, we obtain a ratio of $\sim$0.22 for IRS1, which is one order of magnitude larger than the limit, and we find $\sim$1.5$\times10^{-2}$ for IRS2. In this scenario, the disk is largely gravitationally unstable in IRS1 while it can be marginally unstable in IRS2. On the other hand, assuming the average dust temperatures (76 K and 44 K for IRS1 and IRS2, respectively), we find a ratio of $\sim$5.2$\times10^{-2}$ and $\sim$5.8$\times10^{-3}$ for IRS1 and IRS2, respectively. In this scenario, the disk around IRS1 remains unstable whereas the disk around IRS2 can be considered stable. Note that the derived stellar masses are upper limits and the derived disk masses are lower limits, meaning that the derived ratios of M$_\mathrm{disk}$/M$_\mathrm{\star}$ are lower limits. This suggests that the disks should be both gravitationally unstable. Yet, we do not see spiral arms or signature of fragmentation. The most likely explanation is that these structures are hidden in optically thick emission, if they exist. On the other hand, because instabilities can be at the origin of accretion outbursts and variations in the accretion rate \citep[e.g.][]{Kuffmeier+etal_2018, Vorobyov+etal_2021}, the fact that the ratio M$_\mathrm{disk}$/M$_\mathrm{\star}$ is larger in IRS1 than in IRS2 (assuming that this difference implies that the disk around IRS1 is more unstable than the disk around IRS2) could explain the different morphologies between the jets powered by IRS1 and IRS2.

\section{Summary} \label{sec:summary}

We have presented the observations conducted with ALMA at an angular resolution of $\sim$0\farcs1 toward the binary system BHR 71 as part of the ALMA Large Program eDisk. For this study, we have investigated the ALMA Band 6 dust continuum emission as well as the $^{12}$CO ($J$=2--1), $^{13}$CO ($J$=2--1), C$^{18}$O ($J$=2--1), and SiO ($J$=5--4) lines. The main results can be summarized as follows:
\begin{itemize}
    \item [1.]
    The 1.3-mm emission reveals the dust continuum in the two protostellar sources BHR 71 IRS1 and BHR 71 IRS2. In IRS1, we detect a compact elongated mm-component with a semimajor axis of 278.94 $\pm$ 0.69 mas ($\sim$50~au) at P.A. = $98\fdg15 \pm 0\fdg41$. We derive a total optically thin mass (dust+gas) of $2.40\times10^{-2}$~M$_\odot$. In IRS2, we detect a much smaller mm component of radius $48.72 \pm 0.59$ mas ($\sim$9~au) with the major axis at P.A. $=67\fdg6 \pm 3\fdg7$. The structure also present an elongated shape. The total mass is $1.50\times10^{-3}$~M$_\odot$. 
 
    \item [2.]
     There is no substructure (ring, gap, spiral) detected in the dust continuum in either source. If substructures are present, they may be obscured by the optical depth of the continuum emission and require more in-depth study to be detected or longer wavelength observations. The dust continuum, however, reveals an asymmetry along the minor axis of IRS1, where the southern part (far side) is brighter than the northern part. This could indicate that the disk is geometrically thick and the dust has yet to be settled. IRS2, on the other hand, shows no such asymmetry. However, IRS2 is very compact relative to IRS1 and is not well-resolved enough to detect an asymmetry.

    \item [3.]

    The C$^{18}$O emission exhibits compact structures around both IRS1 and IRS2 and it also exhibits velocity gradients suggestive of disk rotation. The C$^{18}$O P-V diagrams along the continuum major axes show clear evidence for differential rotation, with additional signs of emission arising from infalling and rotating envelopes. Based on detailed analysis of the P-V diagrams, we set upper limits of dynamical stellar masses of 0.46 and 0.26~M$_\odot$ for IRS1 and IRS2, respectively. These masses provide us with lower limits of the disk to stellar mass ratio of $\sim$0.22 and $\sim$1.5$\times10^{-2}$ for IRS1 and IRS2, respectively. These values are both larger than the limit value of 10$^{-2}$, suggesting that these disks can be considered self-gravitating.
    
    \item [4.]
    The $^{12}$CO emission traces a bipolar wide-angle outflow and a high-velocity collimated jet in both sources. The jet in IRS1 shows a striking double helical structure whereas the jet in IRS2 shows a chain of shock regions with signs of episodic accretion, suggesting different mechanisms that carries away the angular momentum. In IRS1, the jet in SiO is much less spatially extended than in $^{12}$CO. IRS2, on the other hand, has a prominent jet in SiO that kinematically overlaps the jet in $^{12}$CO. The jet in IRS2 also presents an asymmetry in velocity: the northern jet has a mean velocity of $\sim$53~km~s$^{-1}$ whereas the southern jet has a mean velocity of $\sim$65~km~s$^{-1}$.

\end{itemize}

For future work it will be important to put the different aspects of the structure of these deeply embedded sources into context of the larger sample of sources observed within eDisk, work that is currently ongoing. Also, additional observations and/or modeling efforts may be needed to shed further light on the link between the dynamics of the infalling envelope and the emergence of the deeply embedded disks. In particular, high angular resolution observations at longer wavelengths may be needed to fully reveal the distribution of the dust in the embedded disks, while detailed dust and line radiative transfer models may help understand the relation between the dynamics of the envelope and the emergence of these disks.

\newpage
\begin{acknowledgments}
This study makes use of the following ALMA data: ADS/JAO.ALMA\#2019.1.00261.L and ADS/JAO.ALMA\#2021.1.00262.S. ALMA is a partnership of ESO (representing its member states), NSF (USA) and NINS (Japan), together with NRC (Canada), MOST and ASIAA (Taiwan), and KASI (Republic of Korea), in cooperation with the Republic of Chile. The Joint ALMA Observatory is operated by ESO, AUI/NRAO and NAOJ. The National Radio Astronomy Observatory is a facility of the National Science Foundation operated
under cooperative agreement by Associated Universities, Inc. 
S.G., J.K.J, and R.S acknowledge support from the Independent Research Fund Denmark (grant No. 0135-00123B). N.O. acknowledges support from National Science and Technology Council (NSTC) in Taiwan through grants NSTC 109-2112-M-001-051, 110- 2112-M-001-031, 110-2124-M-001-007, and 111-2124-M- 001-005. J.J.T. acknowledges support from NASA XRP 80NSSC22K1159.  YLY acknowledges support from a Grant-in-Aid from the Ministry of Education, Culture, Sports, Science, and Technology of Japan (22K20389, 20H05845, 20H05844), and a pioneering project in RIKEN (Evolution of Matter in the Universe). ZYL is supported in part by NASA 80NSSC20K0533 and NSF AST-2307199 and AST-1910106. ZYDL acknowledges support from NASA 80NSSCK1095, the Jefferson Scholars Foundation, the NRAO ALMA Student Observing Support (SOS) SOSPA8-003, the Achievements Rewards for College Scientists (ARCS) Foundation Washington Chapter, the Virginia Space Grant Consortium (VSGC), and UVA research computing (RIVANNA). M.L.R.H acknowledges support from the Michigan Society of Fellows. S.T. is supported by JSPS KAKENHI Grant Numbers 21H00048 and 21H04495. This work was supported by NAOJ ALMA Scientific Research Grant Code 2022-20A.
LWL acknowledges support from NSF AST-2108794.
\end{acknowledgments}

\vspace{5mm}
\facilities{ALMA}


\software{CASA \citep{Mcmullin+etal_2007}, Numpy \citep{Harris_2020}, Astropy \citep{astropy_2013, astropy_2018, astropy_2022}, Matplotlib \citep{Hunter_2007}} \citet{}

\bibliography{ref}{}

\begin{thebibliography}{}
\expandafter\ifx\csname natexlab\endcsname\relax\def\natexlab#1{#1}\fi
\providecommand{\url}[1]{\href{#1}{#1}}
\providecommand{\dodoi}[1]{doi:~\href{http://doi.org/#1}{\nolinkurl{#1}}}
\providecommand{\doeprint}[1]{\href{http://ascl.net/#1}{\nolinkurl{http://ascl.net/#1}}}
\providecommand{\doarXiv}[1]{\href{https://arxiv.org/abs/#1}{\nolinkurl{https://arxiv.org/abs/#1}}}

\bibitem[{{Adams} {et~al.}(2006){Adams}, {Proszkow}, {Fatuzzo}, \&
  {Myers}}]{Adams+etal_2006}
{Adams}, F.~C., {Proszkow}, E.~M., {Fatuzzo}, M., \& {Myers}, P.~C. 2006, \apj,
  641, 504, \dodoi{10.1086/500393}

\bibitem[{{ALMA Partnership} {et~al.}(2015){ALMA Partnership}, {Brogan},
  {P{\'e}rez}, {Hunter}, {Dent}, {Hales}, {Hills}, {Corder}, {Fomalont},
  {Vlahakis}, {Asaki}, {Barkats}, {Hirota}, {Hodge}, {Impellizzeri}, {Kneissl},
  {Liuzzo}, {Lucas}, {Marcelino}, {Matsushita}, {Nakanishi}, {Phillips},
  {Richards}, {Toledo}, {Aladro}, {Broguiere}, {Cortes}, {Cortes}, {Espada},
  {Galarza}, {Garcia-Appadoo}, {Guzman-Ramirez}, {Humphreys}, {Jung}, {Kameno},
  {Laing}, {Leon}, {Marconi}, {Mignano}, {Nikolic}, {Nyman}, {Radiszcz},
  {Remijan}, {Rod{\'o}n}, {Sawada}, {Takahashi}, {Tilanus}, {Vila Vilaro},
  {Watson}, {Wiklind}, {Akiyama}, {Chapillon}, {de Gregorio-Monsalvo}, {Di
  Francesco}, {Gueth}, {Kawamura}, {Lee}, {Nguyen Luong}, {Mangum}, {Pietu},
  {Sanhueza}, {Saigo}, {Takakuwa}, {Ubach}, {van Kempen}, {Wootten},
  {Castro-Carrizo}, {Francke}, {Gallardo}, {Garcia}, {Gonzalez}, {Hill},
  {Kaminski}, {Kurono}, {Liu}, {Lopez}, {Morales}, {Plarre}, {Schieven},
  {Testi}, {Videla}, {Villard}, {Andreani}, {Hibbard}, \&
  {Tatematsu}}]{Partnership+etal_2015}
{ALMA Partnership}, {Brogan}, C.~L., {P{\'e}rez}, L.~M., {et~al.} 2015, \apjl,
  808, L3, \dodoi{10.1088/2041-8205/808/1/L3}

\bibitem[{{Alves} {et~al.}(2017){Alves}, {Girart}, {Caselli}, {Franco}, {Zhao},
  {Vlemmings}, {Evans}, \& {Ricci}}]{Alves+etal_2017}
{Alves}, F.~O., {Girart}, J.~M., {Caselli}, P., {et~al.} 2017, \aap, 603, L3,
  \dodoi{10.1051/0004-6361/201731077}

\bibitem[{{Andrews} \& {Williams}(2005)}]{Andrews+Williams_2005}
{Andrews}, S.~M., \& {Williams}, J.~P. 2005, \apj, 631, 1134,
  \dodoi{10.1086/432712}

\bibitem[{{Andrews} {et~al.}(2018){Andrews}, {Huang}, {P{\'e}rez}, {Isella},
  {Dullemond}, {Kurtovic}, {Guzm{\'a}n}, {Carpenter}, {Wilner}, {Zhang}, {Zhu},
  {Birnstiel}, {Bai}, {Benisty}, {Hughes}, {{\"O}berg}, \&
  {Ricci}}]{Andrews+etal_2018}
{Andrews}, S.~M., {Huang}, J., {P{\'e}rez}, L.~M., {et~al.} 2018, \apjl, 869,
  L41, \dodoi{10.3847/2041-8213/aaf741}

\bibitem[{{Ansdell} {et~al.}(2016){Ansdell}, {Williams}, {van der Marel},
  {Carpenter}, {Guidi}, {Hogerheijde}, {Mathews}, {Manara}, {Miotello},
  {Natta}, {Oliveira}, {Tazzari}, {Testi}, {van Dishoeck}, \& {van
  Terwisga}}]{Ansdell+etal_2016}
{Ansdell}, M., {Williams}, J.~P., {van der Marel}, N., {et~al.} 2016, \apj,
  828, 46, \dodoi{10.3847/0004-637X/828/1/46}

\bibitem[{{Armitage}(2011)}]{Armitage_2011}
{Armitage}, P.~J. 2011, \araa, 49, 195,
  \dodoi{10.1146/annurev-astro-081710-102521}

\bibitem[{Aso \& Sai(2023)}]{Aso_2023}
Aso, Y., \& Sai, J. 2023, jinshisai/SLAM: First Release of SLAM, v1.0.0,
  Zenodo, \dodoi{10.5281/zenodo.7783868}

\bibitem[{{Aso} {et~al.}(2015){Aso}, {Ohashi}, {Saigo}, {Koyamatsu}, {Aikawa},
  {Hayashi}, {Machida}, {Saito}, {Takakuwa}, {Tomida}, {Tomisaka}, \&
  {Yen}}]{Aso+etal_2015}
{Aso}, Y., {Ohashi}, N., {Saigo}, K., {et~al.} 2015, \apj, 812, 27,
  \dodoi{10.1088/0004-637X/812/1/27}

\bibitem[{{Aso} {et~al.}(2017){Aso}, {Ohashi}, {Aikawa}, {Machida}, {Saigo},
  {Saito}, {Takakuwa}, {Tomida}, {Tomisaka}, \& {Yen}}]{Aso+etal_2017}
{Aso}, Y., {Ohashi}, N., {Aikawa}, Y., {et~al.} 2017, \apj, 849, 56,
  \dodoi{10.3847/1538-4357/aa8264}

\bibitem[{{Astropy Collaboration} {et~al.}(2013){Astropy Collaboration},
  {Robitaille}, {Tollerud}, {Greenfield}, {Droettboom}, {Bray}, {Aldcroft},
  {Davis}, {Ginsburg}, {Price-Whelan}, {Kerzendorf}, {Conley}, {Crighton},
  {Barbary}, {Muna}, {Ferguson}, {Grollier}, {Parikh}, {Nair}, {Unther},
  {Deil}, {Woillez}, {Conseil}, {Kramer}, {Turner}, {Singer}, {Fox}, {Weaver},
  {Zabalza}, {Edwards}, {Azalee Bostroem}, {Burke}, {Casey}, {Crawford},
  {Dencheva}, {Ely}, {Jenness}, {Labrie}, {Lim}, {Pierfederici}, {Pontzen},
  {Ptak}, {Refsdal}, {Servillat}, \& {Streicher}}]{astropy_2013}
{Astropy Collaboration}, {Robitaille}, T.~P., {Tollerud}, E.~J., {et~al.} 2013,
  \aap, 558, A33, \dodoi{10.1051/0004-6361/201322068}

\bibitem[{{Astropy Collaboration} {et~al.}(2018){Astropy Collaboration},
  {Price-Whelan}, {Sip{\H{o}}cz}, {G{\"u}nther}, {Lim}, {Crawford}, {Conseil},
  {Shupe}, {Craig}, {Dencheva}, {Ginsburg}, {VanderPlas}, {Bradley},
  {P{\'e}rez-Su{\'a}rez}, {de Val-Borro}, {Aldcroft}, {Cruz}, {Robitaille},
  {Tollerud}, {Ardelean}, {Babej}, {Bach}, {Bachetti}, {Bakanov}, {Bamford},
  {Barentsen}, {Barmby}, {Baumbach}, {Berry}, {Biscani}, {Boquien}, {Bostroem},
  {Bouma}, {Brammer}, {Bray}, {Breytenbach}, {Buddelmeijer}, {Burke},
  {Calderone}, {Cano Rodr{\'\i}guez}, {Cara}, {Cardoso}, {Cheedella}, {Copin},
  {Corrales}, {Crichton}, {D'Avella}, {Deil}, {Depagne}, {Dietrich}, {Donath},
  {Droettboom}, {Earl}, {Erben}, {Fabbro}, {Ferreira}, {Finethy}, {Fox},
  {Garrison}, {Gibbons}, {Goldstein}, {Gommers}, {Greco}, {Greenfield},
  {Groener}, {Grollier}, {Hagen}, {Hirst}, {Homeier}, {Horton}, {Hosseinzadeh},
  {Hu}, {Hunkeler}, {Ivezi{\'c}}, {Jain}, {Jenness}, {Kanarek}, {Kendrew},
  {Kern}, {Kerzendorf}, {Khvalko}, {King}, {Kirkby}, {Kulkarni}, {Kumar},
  {Lee}, {Lenz}, {Littlefair}, {Ma}, {Macleod}, {Mastropietro}, {McCully},
  {Montagnac}, {Morris}, {Mueller}, {Mumford}, {Muna}, {Murphy}, {Nelson},
  {Nguyen}, {Ninan}, {N{\"o}the}, {Ogaz}, {Oh}, {Parejko}, {Parley}, {Pascual},
  {Patil}, {Patil}, {Plunkett}, {Prochaska}, {Rastogi}, {Reddy Janga},
  {Sabater}, {Sakurikar}, {Seifert}, {Sherbert}, {Sherwood-Taylor}, {Shih},
  {Sick}, {Silbiger}, {Singanamalla}, {Singer}, {Sladen}, {Sooley},
  {Sornarajah}, {Streicher}, {Teuben}, {Thomas}, {Tremblay}, {Turner},
  {Terr{\'o}n}, {van Kerkwijk}, {de la Vega}, {Watkins}, {Weaver}, {Whitmore},
  {Woillez}, {Zabalza}, \& {Astropy Contributors}}]{astropy_2018}
{Astropy Collaboration}, {Price-Whelan}, A.~M., {Sip{\H{o}}cz}, B.~M., {et~al.}
  2018, \aj, 156, 123, \dodoi{10.3847/1538-3881/aabc4f}

\bibitem[{{Astropy Collaboration} {et~al.}(2022){Astropy Collaboration},
  {Price-Whelan}, {Lim}, {Earl}, {Starkman}, {Bradley}, {Shupe}, {Patil},
  {Corrales}, {Brasseur}, {N{\"o}the}, {Donath}, {Tollerud}, {Morris},
  {Ginsburg}, {Vaher}, {Weaver}, {Tocknell}, {Jamieson}, {van Kerkwijk},
  {Robitaille}, {Merry}, {Bachetti}, {G{\"u}nther}, {Aldcroft},
  {Alvarado-Montes}, {Archibald}, {B{\'o}di}, {Bapat}, {Barentsen},
  {Baz{\'a}n}, {Biswas}, {Boquien}, {Burke}, {Cara}, {Cara}, {Conroy},
  {Conseil}, {Craig}, {Cross}, {Cruz}, {D'Eugenio}, {Dencheva}, {Devillepoix},
  {Dietrich}, {Eigenbrot}, {Erben}, {Ferreira}, {Foreman-Mackey}, {Fox},
  {Freij}, {Garg}, {Geda}, {Glattly}, {Gondhalekar}, {Gordon}, {Grant},
  {Greenfield}, {Groener}, {Guest}, {Gurovich}, {Handberg}, {Hart},
  {Hatfield-Dodds}, {Homeier}, {Hosseinzadeh}, {Jenness}, {Jones}, {Joseph},
  {Kalmbach}, {Karamehmetoglu}, {Ka{\l}uszy{\'n}ski}, {Kelley}, {Kern},
  {Kerzendorf}, {Koch}, {Kulumani}, {Lee}, {Ly}, {Ma}, {MacBride}, {Maljaars},
  {Muna}, {Murphy}, {Norman}, {O'Steen}, {Oman}, {Pacifici}, {Pascual},
  {Pascual-Granado}, {Patil}, {Perren}, {Pickering}, {Rastogi}, {Roulston},
  {Ryan}, {Rykoff}, {Sabater}, {Sakurikar}, {Salgado}, {Sanghi}, {Saunders},
  {Savchenko}, {Schwardt}, {Seifert-Eckert}, {Shih}, {Jain}, {Shukla}, {Sick},
  {Simpson}, {Singanamalla}, {Singer}, {Singhal}, {Sinha}, {Sip{\H{o}}cz},
  {Spitler}, {Stansby}, {Streicher}, {{\v{S}}umak}, {Swinbank}, {Taranu},
  {Tewary}, {Tremblay}, {de Val-Borro}, {Van Kooten}, {Vasovi{\'c}}, {Verma},
  {de Miranda Cardoso}, {Williams}, {Wilson}, {Winkel}, {Wood-Vasey}, {Xue},
  {Yoachim}, {Zhang}, {Zonca}, \& {Astropy Project
  Contributors}}]{astropy_2022}
{Astropy Collaboration}, {Price-Whelan}, A.~M., {Lim}, P.~L., {et~al.} 2022,
  \apj, 935, 167, \dodoi{10.3847/1538-4357/ac7c74}

\bibitem[{{Beckwith} {et~al.}(1990){Beckwith}, {Sargent}, {Chini}, \&
  {Guesten}}]{Beckwith+etal_1990}
{Beckwith}, S. V.~W., {Sargent}, A.~I., {Chini}, R.~S., \& {Guesten}, R. 1990,
  \aj, 99, 924, \dodoi{10.1086/115385}

\bibitem[{{Benedettini} {et~al.}(2017){Benedettini}, {Gusdorf}, {Nisini},
  {Lefloch}, {Anderl}, {Busquet}, {Ceccarelli}, {Codella}, {Leurini}, \&
  {Podio}}]{Benedettini+etal_2017}
{Benedettini}, M., {Gusdorf}, A., {Nisini}, B., {et~al.} 2017, \aap, 598, A14,
  \dodoi{10.1051/0004-6361/201629451}

\bibitem[{{Bourke}(2001)}]{Bourke_2001}
{Bourke}, T.~L. 2001, \apjl, 554, L91, \dodoi{10.1086/320921}

\bibitem[{{Bourke} {et~al.}(1997){Bourke}, {Garay}, {Lehtinen},
  {K{\"o}hnenkamp}, {Launhardt}, {Nyman}, {May}, {Robinson}, \&
  {Hyland}}]{Bourke+etal_1997}
{Bourke}, T.~L., {Garay}, G., {Lehtinen}, K.~K., {et~al.} 1997, \apj, 476, 781,
  \dodoi{10.1086/303642}

\bibitem[{{Cabrit} {et~al.}(2012){Cabrit}, {Codella}, {Gueth}, \&
  {Gusdorf}}]{Cabrit+etal_2012}
{Cabrit}, S., {Codella}, C., {Gueth}, F., \& {Gusdorf}, A. 2012, \aap, 548, L2,
  \dodoi{10.1051/0004-6361/201219784}

\bibitem[{{Cieza} {et~al.}(2019){Cieza}, {Ru{\'\i}z-Rodr{\'\i}guez}, {Hales},
  {Casassus}, {P{\'e}rez}, {Gonzalez-Ruilova}, {C{\'a}novas}, {Williams},
  {Zurlo}, {Ansdell}, {Avenhaus}, {Bayo}, {Bertrang}, {Christiaens}, {Dent},
  {Ferrero}, {Gamen}, {Olofsson}, {Orcajo}, {Pe{\~n}a Ram{\'\i}rez},
  {Principe}, {Schreiber}, \& {van der Plas}}]{Cieza+etal_2019}
{Cieza}, L.~A., {Ru{\'\i}z-Rodr{\'\i}guez}, D., {Hales}, A., {et~al.} 2019,
  \mnras, 482, 698, \dodoi{10.1093/mnras/sty2653}

\bibitem[{{Codella} {et~al.}(2007){Codella}, {Cabrit}, {Gueth}, {Cesaroni},
  {Bacciotti}, {Lefloch}, \& {McCaughrean}}]{Codella+etal_2007}
{Codella}, C., {Cabrit}, S., {Gueth}, F., {et~al.} 2007, \aap, 462, L53,
  \dodoi{10.1051/0004-6361:20066800}

\bibitem[{{de Valon} {et~al.}(2022){de Valon}, {Dougados}, {Cabrit}, {Louvet},
  {Zapata}, \& {Mardones}}]{deValon+2022}
{de Valon}, A., {Dougados}, C., {Cabrit}, S., {et~al.} 2022, \aap, 668, A78,
  \dodoi{10.1051/0004-6361/202141316}

\bibitem[{{Dong} {et~al.}(2015){Dong}, {Zhu}, {Rafikov}, \&
  {Stone}}]{Dong+etal_2015}
{Dong}, R., {Zhu}, Z., {Rafikov}, R.~R., \& {Stone}, J.~M. 2015, \apjl, 809,
  L5, \dodoi{10.1088/2041-8205/809/1/L5}

\bibitem[{{Eisner} {et~al.}(2008){Eisner}, {Plambeck}, {Carpenter}, {Corder},
  {Qi}, \& {Wilner}}]{Eisner+etal_2008}
{Eisner}, J.~A., {Plambeck}, R.~L., {Carpenter}, J.~M., {et~al.} 2008, \apj,
  683, 304, \dodoi{10.1086/588524}

\bibitem[{{Flores} {et~al.}(2023){Flores}, {Ohashi}, {Tobin}, {J{\o}rgensen},
  {Takakuwa}, {Li}, {Lin}, {van't Hoff}, {Plunkett}, {Yamato}, {Sai (Insa
  Choi)}, {Koch}, {Yen}, {Aikawa}, {Aso}, {de Gregorio-Monsalvo}, {Kido},
  {Kwon}, {Lee}, {Lee}, {Looney}, {Santamar{\'\i}a-Miranda}, {Sharma},
  {Thieme}, {Williams}, {Han}, {Narayanan}, \& {Lai}}]{Flores+edisk_2023}
{Flores}, C., {Ohashi}, N., {Tobin}, J.~J., {et~al.} 2023, \apj, 958, 98,
  \dodoi{10.3847/1538-4357/acf7c1}

\bibitem[{{Frank} {et~al.}(2014){Frank}, {Ray}, {Cabrit}, {Hartigan}, {Arce},
  {Bacciotti}, {Bally}, {Benisty}, {Eisl{\"o}ffel}, {G{\"u}del}, {Lebedev},
  {Nisini}, \& {Raga}}]{Frank+etal_2014}
{Frank}, A., {Ray}, T.~P., {Cabrit}, S., {et~al.} 2014, in Protostars and
  Planets VI, ed. H.~{Beuther}, R.~S. {Klessen}, C.~P. {Dullemond}, \&
  T.~{Henning}, 451--474, \dodoi{10.2458/azu_uapress_9780816531240-ch020}

\bibitem[{{Garay} {et~al.}(1998){Garay}, {K{\"o}hnenkamp}, {Bourke},
  {Rodr{\'\i}guez}, \& {Lehtinen}}]{Garay+etal_1998}
{Garay}, G., {K{\"o}hnenkamp}, I., {Bourke}, T.~L., {Rodr{\'\i}guez}, L.~F., \&
  {Lehtinen}, K.~K. 1998, \apj, 509, 768, \dodoi{10.1086/306534}

\bibitem[{{Goodman} {et~al.}(1993){Goodman}, {Benson}, {Fuller}, \&
  {Myers}}]{Goodman+etal_1993}
{Goodman}, A.~A., {Benson}, P.~J., {Fuller}, G.~A., \& {Myers}, P.~C. 1993,
  \apj, 406, 528, \dodoi{10.1086/172465}

\bibitem[{{Guillet} {et~al.}(2011){Guillet}, {Pineau Des For{\^e}ts}, \&
  {Jones}}]{Guillet+etal_2008}
{Guillet}, V., {Pineau Des For{\^e}ts}, G., \& {Jones}, A.~P. 2011, \aap, 527,
  A123, \dodoi{10.1051/0004-6361/201015973}

\bibitem[{{Gusdorf} {et~al.}(2011){Gusdorf}, {Giannini}, {Flower}, {Parise},
  {G{\"u}sten}, \& {Kristensen}}]{Gusdorf+etal_2011}
{Gusdorf}, A., {Giannini}, T., {Flower}, D.~R., {et~al.} 2011, \aap, 532, A53,
  \dodoi{10.1051/0004-6361/201116758}

\bibitem[{{Gusdorf} {et~al.}(2008){Gusdorf}, {Pineau Des For{\^e}ts}, {Cabrit},
  \& {Flower}}]{Gusdorf+etal_2008}
{Gusdorf}, A., {Pineau Des For{\^e}ts}, G., {Cabrit}, S., \& {Flower}, D.~R.
  2008, \aap, 490, 695, \dodoi{10.1051/0004-6361:200810443}

\bibitem[{{Gusdorf} {et~al.}(2015){Gusdorf}, {Riquelme}, {Anderl},
  {Eisl{\"o}ffel}, {Codella}, {G{\'o}mez-Ruiz}, {Graf}, {Kristensen},
  {Leurini}, {Parise}, {Requena-Torres}, {Ricken}, \&
  {G{\"u}sten}}]{Gusdorf+etal_2015}
{Gusdorf}, A., {Riquelme}, D., {Anderl}, S., {et~al.} 2015, \aap, 575, A98,
  \dodoi{10.1051/0004-6361/201425142}

\bibitem[{Harris {et~al.}(2020)Harris, Millman, van~der Walt, Gommers,
  Virtanen, Cournapeau, Wieser, Taylor, Berg, Smith, Kern, Picus, Hoyer, van
  Kerkwijk, Brett, Haldane, del R{\'{i}}o, Wiebe, Peterson,
  G{\'{e}}rard-Marchant, Sheppard, Reddy, Weckesser, Abbasi, Gohlke, \&
  Oliphant}]{Harris_2020}
Harris, C.~R., Millman, K.~J., van~der Walt, S.~J., {et~al.} 2020, Nature, 585,
  357, \dodoi{10.1038/s41586-020-2649-2}

\bibitem[{{Harsono} {et~al.}(2018){Harsono}, {Bjerkeli}, {van der Wiel},
  {Ramsey}, {Maud}, {Kristensen}, \& {J{\o}rgensen}}]{Harsono+etal_2018}
{Harsono}, D., {Bjerkeli}, P., {van der Wiel}, M. H.~D., {et~al.} 2018, Nature
  Astronomy, 2, 646, \dodoi{10.1038/s41550-018-0497-x}

\bibitem[{Hunter(2007)}]{Hunter_2007}
Hunter, J.~D. 2007, Computing in Science \& Engineering, 9, 90,
  \dodoi{10.1109/MCSE.2007.55}

\bibitem[{{Johansen} \& {Lacerda}(2010)}]{Johansen+etal_2010}
{Johansen}, A., \& {Lacerda}, P. 2010, \mnras, 404, 475,
  \dodoi{10.1111/j.1365-2966.2010.16309.x}

\bibitem[{{Kley} \& {Nelson}(2012)}]{Kley+Nelson_2012}
{Kley}, W., \& {Nelson}, R.~P. 2012, \araa, 50, 211,
  \dodoi{10.1146/annurev-astro-081811-125523}

\bibitem[{{Kratter} \& {Lodato}(2016)}]{Kratter+etal_2016}
{Kratter}, K., \& {Lodato}, G. 2016, \araa, 54, 271,
  \dodoi{10.1146/annurev-astro-081915-023307}

\bibitem[{{Kuffmeier} {et~al.}(2018){Kuffmeier}, {Frimann}, {Jensen}, \&
  {Haugb{\o}lle}}]{Kuffmeier+etal_2018}
{Kuffmeier}, M., {Frimann}, S., {Jensen}, S.~S., \& {Haugb{\o}lle}, T. 2018,
  \mnras, 475, 2642, \dodoi{10.1093/mnras/sty024}

\bibitem[{{Kwon} {et~al.}(2015){Kwon}, {Fern{\'a}ndez-L{\'o}pez}, {Stephens},
  \& {Looney}}]{Kwon+etal_2015}
{Kwon}, W., {Fern{\'a}ndez-L{\'o}pez}, M., {Stephens}, I.~W., \& {Looney},
  L.~W. 2015, \apj, 814, 43, \dodoi{10.1088/0004-637X/814/1/43}

\bibitem[{{Kwon} {et~al.}(2009){Kwon}, {Looney}, {Mundy}, {Chiang}, \&
  {Kemball}}]{Kwon+etal_2009}
{Kwon}, W., {Looney}, L.~W., {Mundy}, L.~G., {Chiang}, H.-F., \& {Kemball},
  A.~J. 2009, \apj, 696, 841, \dodoi{10.1088/0004-637X/696/1/841}

\bibitem[{{Law} {et~al.}(2021){Law}, {Loomis}, {Teague}, {{\"O}berg},
  {Czekala}, {Andrews}, {Huang}, {Aikawa}, {Alarc{\'o}n}, {Bae}, {Bergin},
  {Bergner}, {Boehler}, {Booth}, {Bosman}, {Calahan}, {Cataldi}, {Cleeves},
  {Furuya}, {Guzm{\'a}n}, {Ilee}, {Le Gal}, {Liu}, {Long}, {M{\'e}nard},
  {Nomura}, {Qi}, {Schwarz}, {Sierra}, {Tsukagoshi}, {Yamato}, {van't Hoff},
  {Walsh}, {Wilner}, \& {Zhang}}]{Law+etal_2021}
{Law}, C.~J., {Loomis}, R.~A., {Teague}, R., {et~al.} 2021, \apjs, 257, 3,
  \dodoi{10.3847/1538-4365/ac1434}

\bibitem[{{Lee} {et~al.}(2017){Lee}, {Li}, {Ho}, {Hirano}, {Zhang}, \&
  {Shang}}]{Lee+etal_2017}
{Lee}, C.-F., {Li}, Z.-Y., {Ho}, P. T.~P., {et~al.} 2017, Science Advances, 3,
  e1602935, \dodoi{10.1126/sciadv.1602935}

\bibitem[{{Lee} {et~al.}(2022){Lee}, {Li}, {Shang}, \&
  {Hirano}}]{Lee+etal_2022}
{Lee}, C.-F., {Li}, Z.-Y., {Shang}, H., \& {Hirano}, N. 2022, \apjl, 927, L27,
  \dodoi{10.3847/2041-8213/ac59c0}

\bibitem[{{Lin} {et~al.}(2021){Lin}, {Lee}, {Li}, {Tobin}, \&
  {Turner}}]{lin+etal_2021}
{Lin}, Z.-Y.~D., {Lee}, C.-F., {Li}, Z.-Y., {Tobin}, J.~J., \& {Turner}, N.~J.
  2021, \mnras, 501, 1316, \dodoi{10.1093/mnras/staa3685}

\bibitem[{{Lin} {et~al.}(2023){Lin}, {Li}, {Tobin}, {Ohashi}, {J{\o}rgensen},
  {Looney}, {Aso}, {Takakuwa}, {Aikawa}, {van't Hoff}, {de Gregorio-Monsalvo},
  {Encalada}, {Flores}, {Gavino}, {Han}, {Kido}, {Koch}, {Kwon}, {Lai}, {Lee},
  {Lee}, {Phuong}, {Sai}, {Sharma}, {Sheehan}, {Thieme}, {Williams}, {Yamato},
  \& {Yen}}]{Lin+edisk_2023}
{Lin}, Z.-Y.~D., {Li}, Z.-Y., {Tobin}, J.~J., {et~al.} 2023, \apj, 951, 9,
  \dodoi{10.3847/1538-4357/acd5c9}

\bibitem[{{Long} {et~al.}(2018){Long}, {Pinilla}, {Herczeg}, {Harsono},
  {Dipierro}, {Pascucci}, {Hendler}, {Tazzari}, {Ragusa}, {Salyk}, {Edwards},
  {Lodato}, {van de Plas}, {Johnstone}, {Liu}, {Boehler}, {Cabrit}, {Manara},
  {Menard}, {Mulders}, {Nisini}, {Fischer}, {Rigliaco}, {Banzatti}, {Avenhaus},
  \& {Gully-Santiago}}]{Long+etal_2018}
{Long}, F., {Pinilla}, P., {Herczeg}, G.~J., {et~al.} 2018, \apj, 869, 17,
  \dodoi{10.3847/1538-4357/aae8e1}

\bibitem[{{L{\'o}pez-V{\'a}zquez} {et~al.}(2023){L{\'o}pez-V{\'a}zquez},
  {Zapata}, \& {Lee}}]{lopez+2023}
{L{\'o}pez-V{\'a}zquez}, J.~A., {Zapata}, L.~A., \& {Lee}, C.-F. 2023, \apj,
  944, 63, \dodoi{10.3847/1538-4357/acb439}

\bibitem[{{Machida} {et~al.}(2007){Machida}, {Inutsuka}, \&
  {Matsumoto}}]{machida+etal_2007}
{Machida}, M.~N., {Inutsuka}, S.-i., \& {Matsumoto}, T. 2007, \apj, 670, 1198,
  \dodoi{10.1086/521779}

\bibitem[{{Maret} {et~al.}(2020){Maret}, {Maury}, {Belloche}, {Gaudel},
  {Andr{\'e}}, {Cabrit}, {Codella}, {Lef{\'e}vre}, {Podio}, {Anderl}, {Gueth},
  \& {Hennebelle}}]{Maret+etal_2020}
{Maret}, S., {Maury}, A.~J., {Belloche}, A., {et~al.} 2020, \aap, 635, A15,
  \dodoi{10.1051/0004-6361/201936798}

\bibitem[{{Masciadri} \& {Raga}(2002)}]{Masciadri+etal_2002}
{Masciadri}, E., \& {Raga}, A.~C. 2002, \apj, 568, 733, \dodoi{10.1086/338767}

\bibitem[{{McMullin} {et~al.}(2007){McMullin}, {Waters}, {Schiebel}, {Young},
  \& {Golap}}]{Mcmullin+etal_2007}
{McMullin}, J.~P., {Waters}, B., {Schiebel}, D., {Young}, W., \& {Golap}, K.
  2007, in Astronomical Society of the Pacific Conference Series, Vol. 376,
  Astronomical Data Analysis Software and Systems XVI, ed. R.~A. {Shaw},
  F.~{Hill}, \& D.~J. {Bell}, 127

\bibitem[{{Momose} {et~al.}(1998){Momose}, {Ohashi}, {Kawabe}, {Nakano}, \&
  {Hayashi}}]{Momose+etal_1998}
{Momose}, M., {Ohashi}, N., {Kawabe}, R., {Nakano}, T., \& {Hayashi}, M. 1998,
  \apj, 504, 314, \dodoi{10.1086/306061}

\bibitem[{{Montgomery}(2009)}]{Montgomery_2009}
{Montgomery}, M.~M. 2009, \apj, 705, 603, \dodoi{10.1088/0004-637X/705/1/603}

\bibitem[{{Mottram} {et~al.}(2014){Mottram}, {Kristensen}, {van Dishoeck},
  {Bruderer}, {San Jos{\'e}-Garc{\'\i}a}, {Karska}, {Visser}, {Santangelo},
  {Benz}, {Bergin}, {Caselli}, {Herpin}, {Hogerheijde}, {Johnstone}, {van
  Kempen}, {Liseau}, {Nisini}, {Tafalla}, {van der Tak}, \&
  {Wyrowski}}]{Mottram+etal_2014}
{Mottram}, J.~C., {Kristensen}, L.~E., {van Dishoeck}, E.~F., {et~al.} 2014,
  \aap, 572, A21, \dodoi{10.1051/0004-6361/201424267}

\bibitem[{{Neufeld} {et~al.}(2009){Neufeld}, {Nisini}, {Giannini}, {Melnick},
  {Bergin}, {Yuan}, {Maret}, {Tolls}, {G{\"u}sten}, \&
  {Kaufman}}]{Neufeld+etal_2009}
{Neufeld}, D.~A., {Nisini}, B., {Giannini}, T., {et~al.} 2009, \apj, 706, 170,
  \dodoi{10.1088/0004-637X/706/1/170}

\bibitem[{{Ohashi} {et~al.}(1997){Ohashi}, {Hayashi}, {Ho}, \&
  {Momose}}]{Ohashi+etal_1997}
{Ohashi}, N., {Hayashi}, M., {Ho}, P. T.~P., \& {Momose}, M. 1997, \apj, 475,
  211, \dodoi{10.1086/303533}

\bibitem[{{Ohashi} {et~al.}(2014){Ohashi}, {Saigo}, {Aso}, {Aikawa},
  {Koyamatsu}, {Machida}, {Saito}, {Takahashi}, {Takakuwa}, {Tomida},
  {Tomisaka}, \& {Yen}}]{Ohashi+etal_2014}
{Ohashi}, N., {Saigo}, K., {Aso}, Y., {et~al.} 2014, \apj, 796, 131,
  \dodoi{10.1088/0004-637X/796/2/131}

\bibitem[{{Ohashi} {et~al.}(2023){Ohashi}, {Tobin}, {J{\o}rgensen}, {Takakuwa},
  {Sheehan}, {Aikawa}, {Li}, {Looney}, {Williams}, {Aso}, {Sharma}, {Sai},
  {Yamato}, {Lee}, {Tomida}, {Yen}, {Encalada}, {Flores}, {Gavino}, {Kido},
  {Han}, {Lin}, {Narayanan}, {Phuong}, {Santamar{\'\i}a-Miranda}, {Thieme},
  {van't Hoff}, {de Gregorio-Monsalvo}, {Koch}, {Kwon}, {Lai}, {Lee},
  {Plunkett}, {Saigo}, {Hirano}, {Lam}, \& {Mori}}]{Ohashi+edisk_2023}
{Ohashi}, N., {Tobin}, J.~J., {J{\o}rgensen}, J.~K., {et~al.} 2023, \apj, 951,
  8, \dodoi{10.3847/1538-4357/acd384}

\bibitem[{{Ohashi} {et~al.}(2022){Ohashi}, {Nakatani}, {Liu}, {Kobayashi},
  {Zhang}, {Hanawa}, \& {Sakai}}]{SOhashi+etal_2022b}
{Ohashi}, S., {Nakatani}, R., {Liu}, H.~B., {et~al.} 2022, \apj, 934, 163,
  \dodoi{10.3847/1538-4357/ac794e}

\bibitem[{{Ossenkopf} \& {Henning}(1994)}]{Ossenkopf+Henning_1994}
{Ossenkopf}, V., \& {Henning}, T. 1994, \aap, 291, 943

\bibitem[{{Parise} {et~al.}(2006){Parise}, {Belloche}, {Leurini}, {Schilke},
  {Wyrowski}, \& {G{\"u}sten}}]{Parise+etal_2006}
{Parise}, B., {Belloche}, A., {Leurini}, S., {et~al.} 2006, \aap, 454, L79,
  \dodoi{10.1051/0004-6361:20065363}

\bibitem[{{Pinte} {et~al.}(2016){Pinte}, {Dent}, {M{\'e}nard}, {Hales}, {Hill},
  {Cortes}, \& {de Gregorio-Monsalvo}}]{Pinte+etal_2016}
{Pinte}, C., {Dent}, W.~R.~F., {M{\'e}nard}, F., {et~al.} 2016, \apj, 816, 25,
  \dodoi{10.3847/0004-637X/816/1/25}

\bibitem[{{Podio} {et~al.}(2016){Podio}, {Codella}, {Gueth}, {Cabrit}, {Maury},
  {Tabone}, {Lef{\`e}vre}, {Anderl}, {Andr{\'e}}, {Belloche}, {Bontemps},
  {Hennebelle}, {Lefloch}, {Maret}, \& {Testi}}]{Podio+etal_2016}
{Podio}, L., {Codella}, C., {Gueth}, F., {et~al.} 2016, \aap, 593, L4,
  \dodoi{10.1051/0004-6361/201628876}

\bibitem[{{Raga} {et~al.}(1990){Raga}, {Canto}, {Binette}, \&
  {Calvet}}]{Raga+etal_1990}
{Raga}, A.~C., {Canto}, J., {Binette}, L., \& {Calvet}, N. 1990, \apj, 364,
  601, \dodoi{10.1086/169443}

\bibitem[{{Reynolds} {et~al.}(2021){Reynolds}, {Tobin}, {Sheehan}, {Sadavoy},
  {Kratter}, {Li}, {Chandler}, {Segura-Cox}, {Looney}, \&
  {Dunham}}]{Reynolds+etal_2021}
{Reynolds}, N.~K., {Tobin}, J.~J., {Sheehan}, P., {et~al.} 2021, \apjl, 907,
  L10, \dodoi{10.3847/2041-8213/abcc02}

\bibitem[{{Sai} {et~al.}(2020){Sai}, {Ohashi}, {Saigo}, {Matsumoto}, {Aso},
  {Takakuwa}, {Aikawa}, {Kurose}, {Yen}, {Tomisaka}, {Tomida}, \&
  {Machida}}]{Sai+etal_2020}
{Sai}, J., {Ohashi}, N., {Saigo}, K., {et~al.} 2020, \apj, 893, 51,
  \dodoi{10.3847/1538-4357/ab8065}

\bibitem[{{Santiago-Garc{\'\i}a} {et~al.}(2009){Santiago-Garc{\'\i}a},
  {Tafalla}, {Johnstone}, \& {Bachiller}}]{Santiago-Garcia+etal_2009}
{Santiago-Garc{\'\i}a}, J., {Tafalla}, M., {Johnstone}, D., \& {Bachiller}, R.
  2009, \aap, 495, 169, \dodoi{10.1051/0004-6361:200810739}

\bibitem[{{Schilke} {et~al.}(1997){Schilke}, {Walmsley}, {Pineau des Forets},
  \& {Flower}}]{Schilke+etal_1997}
{Schilke}, P., {Walmsley}, C.~M., {Pineau des Forets}, G., \& {Flower}, D.~R.
  1997, \aap, 321, 293

\bibitem[{{Seifried} {et~al.}(2016){Seifried}, {S{\'a}nchez-Monge}, {Walch}, \&
  {Banerjee}}]{Seifried+etal_2016}
{Seifried}, D., {S{\'a}nchez-Monge}, {\'A}., {Walch}, S., \& {Banerjee}, R.
  2016, \mnras, 459, 1892, \dodoi{10.1093/mnras/stw785}

\bibitem[{{Sharma} {et~al.}(2023){Sharma}, {J{\o}rgensen}, {Gavino}, {Ohashi},
  {Tobin}, {Lin}, {Li}, {Takakuwa}, {Lee}, {Sai (Insa Choi)}, {Kwon}, {de
  Gregorio-Monsalvo}, {Santamar{\'\i}a-Miranda}, {Yen}, {Aikawa}, {Aso}, {Lai},
  {Lee}, {Looney}, {Phuong}, {Thieme}, \& {Williams}}]{Sharma+edisk_2023}
{Sharma}, R., {J{\o}rgensen}, J.~K., {Gavino}, S., {et~al.} 2023, \apj, 954,
  69, \dodoi{10.3847/1538-4357/ace35c}

\bibitem[{{Sheehan} {et~al.}(2020){Sheehan}, {Tobin}, {Federman}, {Megeath}, \&
  {Looney}}]{Sheehan+etal_2020}
{Sheehan}, P.~D., {Tobin}, J.~J., {Federman}, S., {Megeath}, S.~T., \&
  {Looney}, L.~W. 2020, \apj, 902, 141, \dodoi{10.3847/1538-4357/abbad5}

\bibitem[{{Sheehan} {et~al.}(2022){Sheehan}, {Tobin}, {Li}, {van't Hoff},
  {J{\o}rgensen}, {Kwon}, {Looney}, {Ohashi}, {Takakuwa}, {Williams}, {Aso},
  {Gavino}, {Gregorio-Monsalvo}, {Han}, {Lee}, {Plunkett}, {Sharma}, {Aikawa},
  {Lai}, {Lee}, {Lin}, {Saigo}, {Tomida}, \& {Yen}}]{Sheehan+etal_2022}
{Sheehan}, P.~D., {Tobin}, J.~J., {Li}, Z.-Y., {et~al.} 2022, \apj, 934, 95,
  \dodoi{10.3847/1538-4357/ac7a3b}

\bibitem[{{Shu} {et~al.}(1987){Shu}, {Adams}, \& {Lizano}}]{Shu+Adam_1987}
{Shu}, F.~H., {Adams}, F.~C., \& {Lizano}, S. 1987, \araa, 25, 23,
  \dodoi{10.1146/annurev.aa.25.090187.000323}

\bibitem[{{Stone} \& {Norman}(1993)}]{Stone+Norman_1993}
{Stone}, J.~M., \& {Norman}, M.~L. 1993, \apj, 413, 210, \dodoi{10.1086/172989}

\bibitem[{{Tafalla} {et~al.}(2015){Tafalla}, {Bachiller}, {Lefloch},
  {Rodr{\'\i}guez-Fern{\'a}ndez}, {Codella}, {L{\'o}pez-Sepulcre}, \&
  {Podio}}]{Tafalla+etal_2015}
{Tafalla}, M., {Bachiller}, R., {Lefloch}, B., {et~al.} 2015, \aap, 573, L2,
  \dodoi{10.1051/0004-6361/201425255}

\bibitem[{{Tafalla} {et~al.}(2017){Tafalla}, {Su}, {Shang}, {Johnstone},
  {Zhang}, {Santiago-Garc{\'\i}a}, {Lee}, {Hirano}, \&
  {Wang}}]{Tafalla+etal_2017}
{Tafalla}, M., {Su}, Y.~N., {Shang}, H., {et~al.} 2017, \aap, 597, A119,
  \dodoi{10.1051/0004-6361/201629493}

\bibitem[{{Takakuwa} {et~al.}(2024){Takakuwa}, {Saigo}, {Kido}, {Ohashi},
  {Tobin}, {J{\o}rgensen}, {Aikawa}, {Aso}, {Gavino}, {Han}, {Koch}, {Kwon},
  {Lee}, {Lee}, {Li}, {Lin}, {Looney}, {Mori}, {Sai}, {Sharma}, {Sheehan},
  {Tomida}, {Williams}, {Yamato}, \& {Yen}}]{Takakuwa+edisk_2024}
{Takakuwa}, S., {Saigo}, K., {Kido}, M., {et~al.} 2024, arXiv e-prints,
  arXiv:2401.08722, \dodoi{10.48550/arXiv.2401.08722}

\bibitem[{{Terebey} {et~al.}(1984){Terebey}, {Shu}, \&
  {Cassen}}]{Terebey+etal_1984}
{Terebey}, S., {Shu}, F.~H., \& {Cassen}, P. 1984, \apj, 286, 529,
  \dodoi{10.1086/162628}

\bibitem[{{Terquem} {et~al.}(1999){Terquem}, {Eisl{\"o}ffel}, {Papaloizou}, \&
  {Nelson}}]{Terquem+etal_1999}
{Terquem}, C., {Eisl{\"o}ffel}, J., {Papaloizou}, J.~C.~B., \& {Nelson}, R.~P.
  1999, \apjl, 512, L131, \dodoi{10.1086/311880}

\bibitem[{{Testi} {et~al.}(2014){Testi}, {Birnstiel}, {Ricci}, {Andrews},
  {Blum}, {Carpenter}, {Dominik}, {Isella}, {Natta}, {Williams}, \&
  {Wilner}}]{Testi+etal_2014}
{Testi}, L., {Birnstiel}, T., {Ricci}, L., {et~al.} 2014, in Protostars and
  Planets VI, ed. H.~{Beuther}, R.~S. {Klessen}, C.~P. {Dullemond}, \&
  T.~{Henning}, 339, \dodoi{10.2458/azu_uapress_9780816531240-ch015}

\bibitem[{Tobin(2023)}]{john_tobin_2023_7986682}
Tobin, J. 2023, eDisk data reduction scripts, 1.0.0,  Zenodo,
  \dodoi{10.5281/zenodo.7986682}

\bibitem[{{Tobin} {et~al.}(2015){Tobin}, {Looney}, {Wilner}, {Kwon},
  {Chandler}, {Bourke}, {Loinard}, {Chiang}, {Schnee}, \&
  {Chen}}]{Tobin+etal_2015}
{Tobin}, J.~J., {Looney}, L.~W., {Wilner}, D.~J., {et~al.} 2015, \apj, 805,
  125, \dodoi{10.1088/0004-637X/805/2/125}

\bibitem[{{Tobin} {et~al.}(2019){Tobin}, {Bourke}, {Mader}, {Kristensen},
  {Arce}, {Gueth}, {Gusdorf}, {Codella}, {Leurini}, \&
  {Chen}}]{Tobin+etal_2019}
{Tobin}, J.~J., {Bourke}, T.~L., {Mader}, S., {et~al.} 2019, \apj, 870, 81,
  \dodoi{10.3847/1538-4357/aaef87}

\bibitem[{{Tobin} {et~al.}(2020){Tobin}, {Sheehan}, {Megeath},
  {D{\'\i}az-Rodr{\'\i}guez}, {Offner}, {Murillo}, {van 't Hoff}, {van
  Dishoeck}, {Osorio}, {Anglada}, {Furlan}, {Stutz}, {Reynolds}, {Karnath},
  {Fischer}, {Persson}, {Looney}, {Li}, {Stephens}, {Chandler}, {Cox},
  {Dunham}, {Tychoniec}, {Kama}, {Kratter}, {Kounkel}, {Mazur}, {Maud},
  {Patel}, {Perez}, {Sadavoy}, {Segura-Cox}, {Sharma}, {Stephenson}, {Watson},
  \& {Wyrowski}}]{Tobin+etal_2020}
{Tobin}, J.~J., {Sheehan}, P.~D., {Megeath}, S.~T., {et~al.} 2020, \apj, 890,
  130, \dodoi{10.3847/1538-4357/ab6f64}

\bibitem[{{Toomre}(1964)}]{Toomre_1964}
{Toomre}, A. 1964, \apj, 139, 1217, \dodoi{10.1086/147861}

\bibitem[{{van't Hoff} {et~al.}(2023){van't Hoff}, {Tobin}, {Li}, {Ohashi},
  {J{\o}rgensen}, {Lin}, {Aikawa}, {Aso}, {de Gregorio-Monsalvo}, {Gavino},
  {Han}, {Koch}, {Kwon}, {Lee}, {Lee}, {Looney}, {Narayanan}, {Plunkett},
  {Sai}, {Santamar{\'\i}a-Miranda}, {Sharma}, {Sheehan}, {Takakuwa}, {Thieme},
  {Williams}, {Lai}, {Phuong}, \& {Yen}}]{vanHoff+edisk_2023}
{van't Hoff}, M. L.~R., {Tobin}, J.~J., {Li}, Z.-Y., {et~al.} 2023, \apj, 951,
  10, \dodoi{10.3847/1538-4357/accf87}

\bibitem[{{Villenave} {et~al.}(2020){Villenave}, {M{\'e}nard}, {Dent},
  {Duch{\^e}ne}, {Stapelfeldt}, {Benisty}, {Boehler}, {van der Plas}, {Pinte},
  {Telkamp}, {Wolff}, {Flores}, {Lesur}, {Louvet}, {Riols}, {Dougados},
  {Williams}, \& {Padgett}}]{Villenave+etal_2020}
{Villenave}, M., {M{\'e}nard}, F., {Dent}, W.~R.~F., {et~al.} 2020, \aap, 642,
  A164, \dodoi{10.1051/0004-6361/202038087}

\bibitem[{{Villenave} {et~al.}(2022){Villenave}, {Stapelfeldt}, {Duch{\^e}ne},
  {M{\'e}nard}, {Lambrechts}, {Sierra}, {Flores}, {Dent}, {Wolff}, {Ribas},
  {Benisty}, {Cuello}, \& {Pinte}}]{Villenave+etal_2022}
{Villenave}, M., {Stapelfeldt}, K.~R., {Duch{\^e}ne}, G., {et~al.} 2022, \apj,
  930, 11, \dodoi{10.3847/1538-4357/ac5fae}

\bibitem[{{Voirin} {et~al.}(2018){Voirin}, {Manara}, \&
  {Prusti}}]{Voirin+etal_2018}
{Voirin}, J., {Manara}, C.~F., \& {Prusti}, T. 2018, \aap, 610, A64,
  \dodoi{10.1051/0004-6361/201731153}

\bibitem[{{Vorobyov} {et~al.}(2021){Vorobyov}, {Elbakyan}, {Liu}, \&
  {Takami}}]{Vorobyov+etal_2021}
{Vorobyov}, E.~I., {Elbakyan}, V.~G., {Liu}, H.~B., \& {Takami}, M. 2021, \aap,
  647, A44, \dodoi{10.1051/0004-6361/202039391}

\bibitem[{{Yang} {et~al.}(2017){Yang}, {Evans}, {Green}, {Dunham}, \&
  {J{\o}rgensen}}]{Yang+etal_2017}
{Yang}, Y.-L., {Evans}, Neal~J., I., {Green}, J.~D., {Dunham}, M.~M., \&
  {J{\o}rgensen}, J.~K. 2017, \apj, 835, 259,
  \dodoi{10.3847/1538-4357/835/2/259}

\bibitem[{{Yang} {et~al.}(2020){Yang}, {Evans}, {Smith}, {Lee}, {Tobin},
  {Terebey}, {Calcutt}, {J{\o}rgensen}, {Green}, \& {Bourke}}]{Yang+etal_2020}
{Yang}, Y.-L., {Evans}, Neal~J., I., {Smith}, A., {et~al.} 2020, \apj, 891, 61,
  \dodoi{10.3847/1538-4357/ab7201}

\bibitem[{{Yen} {et~al.}(2017){Yen}, {Koch}, {Takakuwa}, {Krasnopolsky},
  {Ohashi}, \& {Aso}}]{Yen+etal_2017}
{Yen}, H.-W., {Koch}, P.~M., {Takakuwa}, S., {et~al.} 2017, \apj, 834, 178,
  \dodoi{10.3847/1538-4357/834/2/178}

\bibitem[{{Yen} {et~al.}(2016){Yen}, {Liu}, {Gu}, {Hirano}, {Lee},
  {Puspitaningrum}, \& {Takakuwa}}]{Yen+etal_2016}
{Yen}, H.-W., {Liu}, H.~B., {Gu}, P.-G., {et~al.} 2016, \apjl, 820, L25,
  \dodoi{10.3847/2041-8205/820/2/L25}

\bibitem[{{Zhang} {et~al.}(2021){Zhang}, {Booth}, {Law}, {Bosman}, {Schwarz},
  {Bergin}, {{\"O}berg}, {Andrews}, {Guzm{\'a}n}, {Walsh}, {Qi}, {van't Hoff},
  {Long}, {Wilner}, {Huang}, {Czekala}, {Ilee}, {Cataldi}, {Bergner}, {Aikawa},
  {Teague}, {Bae}, {Loomis}, {Calahan}, {Alarc{\'o}n}, {M{\'e}nard}, {Le Gal},
  {Sierra}, {Yamato}, {Nomura}, {Tsukagoshi}, {P{\'e}rez}, {Trapman}, {Liu}, \&
  {Furuya}}]{Zhang+etal_2021}
{Zhang}, K., {Booth}, A.~S., {Law}, C.~J., {et~al.} 2021, \apjs, 257, 5,
  \dodoi{10.3847/1538-4365/ac1580}

\bibitem[{{Zhang} {et~al.}(2018){Zhang}, {Zhu}, {Huang}, {Guzm{\'a}n},
  {Andrews}, {Birnstiel}, {Dullemond}, {Carpenter}, {Isella}, {P{\'e}rez},
  {Benisty}, {Wilner}, {Baruteau}, {Bai}, \& {Ricci}}]{Zhang+etal_2018}
{Zhang}, S., {Zhu}, Z., {Huang}, J., {et~al.} 2018, \apjl, 869, L47,
  \dodoi{10.3847/2041-8213/aaf744}

\end{thebibliography}
\bibliographystyle{aasjournal}

\appendix

\section{Outflows in IRS1: channel maps}\label{app:chanmap_IRS1}

\begin{figure*}
\plotone{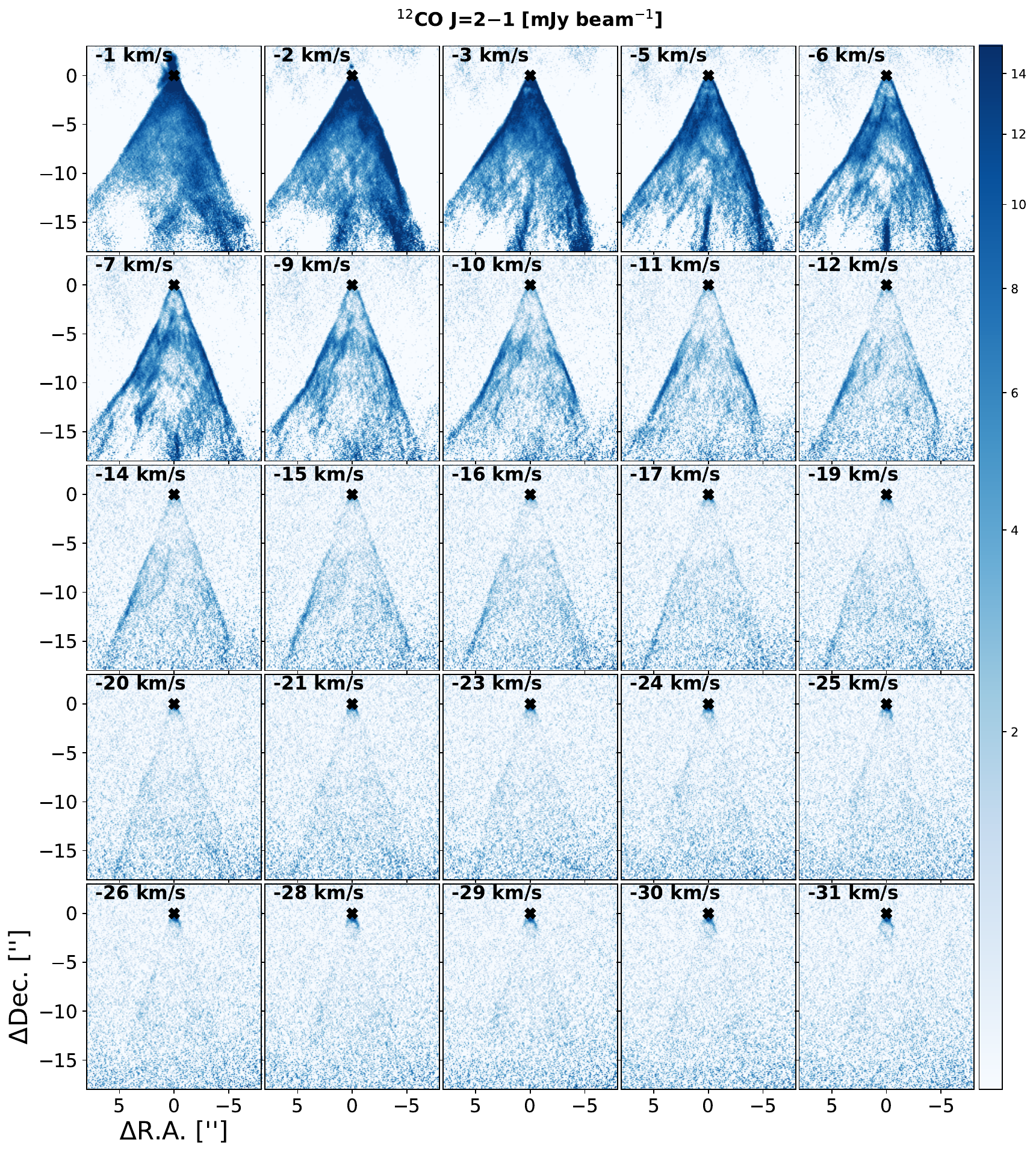}
 \caption{Channel maps of the blueshifted emission of $^{12}$CO in the SHV regime ($|V| < 30$ km~s$^{-1}$) in IRS1, shown by decreasing velocity values relative to the systemic velocity from -1 km~s$^{-1}$ to -31 km~s$^{-1}$.}
\label{fig:chan_irs1_shv_12co_blue}
\end{figure*}

\begin{figure*}
\plotone{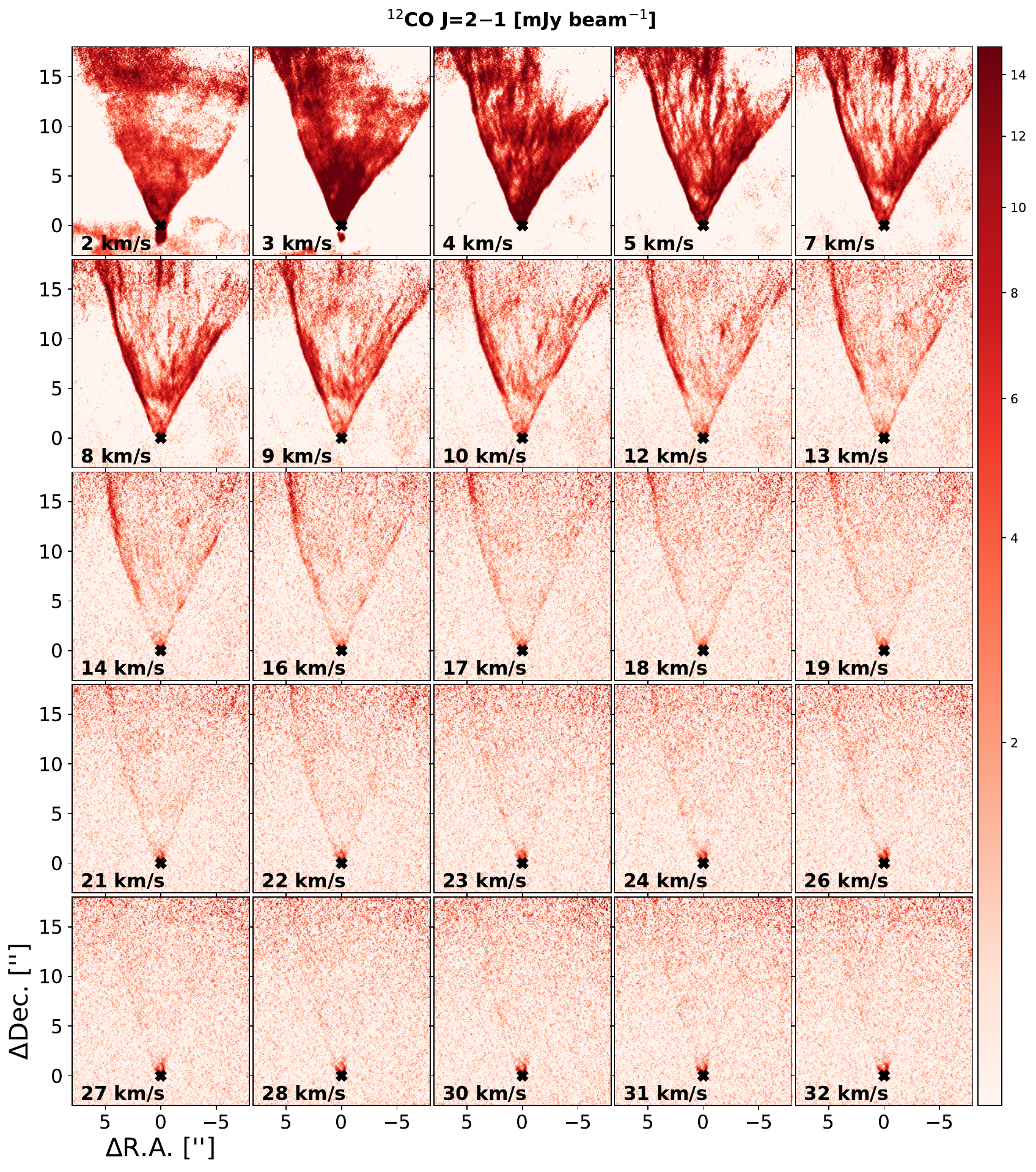}
 \caption{Channel maps of the redshifted emission of $^{12}$CO in the SHV regime ($|V| < 30$ km~s$^{-1}$) in IRS1, shown by increasing velocity values relative to the systemic velocity from 2 km~s$^{-1}$ to 32 km~s$^{-1}$.}
\label{fig:chan_irs1_shv_12co_red}
\end{figure*}

\begin{figure*}
\plotone{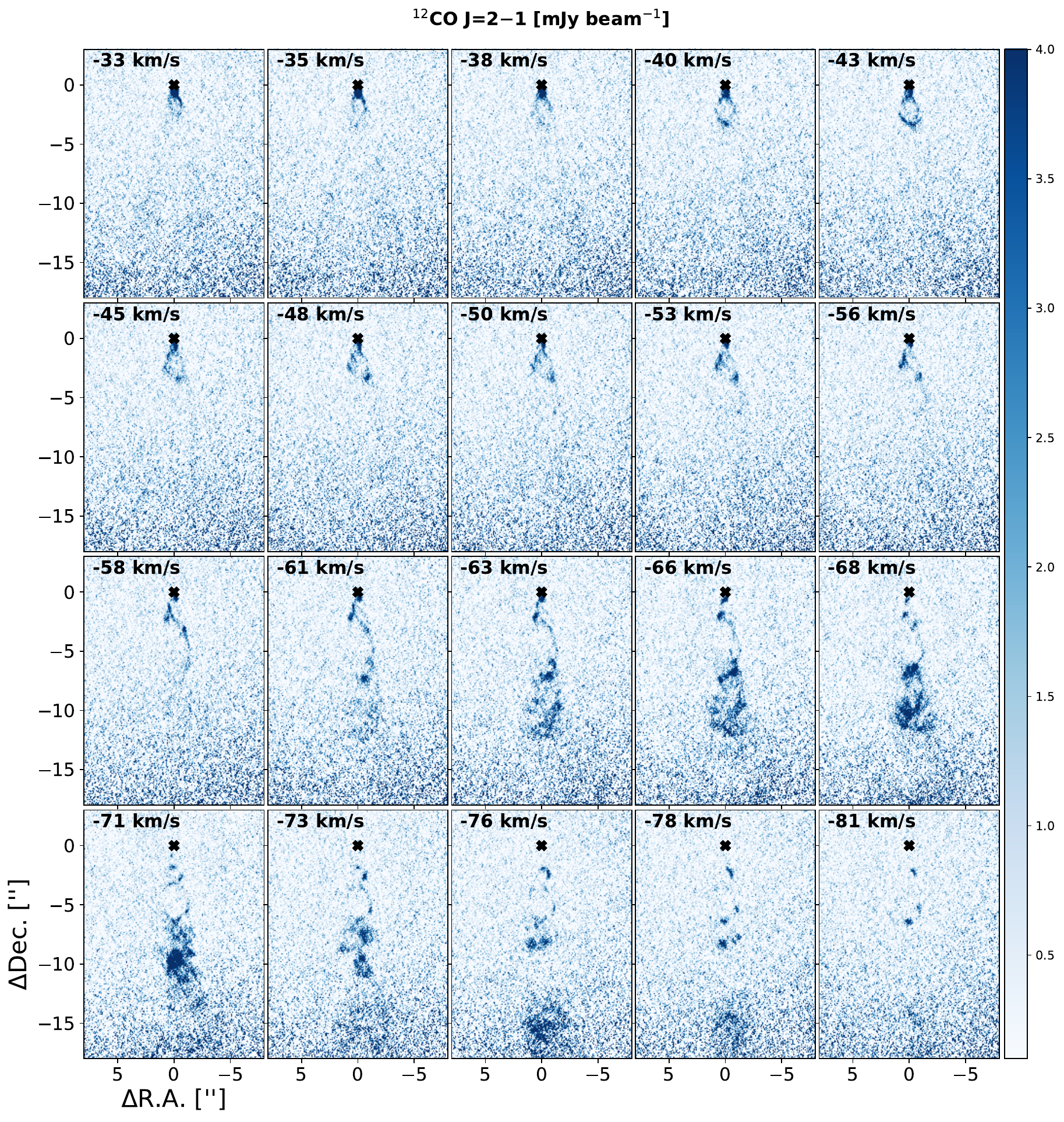}
 \caption{Channel maps of the blueshifted emission of $^{12}$CO in the EHV regime ($|V| > 30$ km~s$^{-1}$) in IRS1, shown by decreasing velocity values relative to the systemic velocity from -33 km~s$^{-1}$ to -81 km~s$^{-1}$.}
\label{fig:chan_irs1_vhv_12co_blue}
\end{figure*}

\begin{figure*}
\plotone{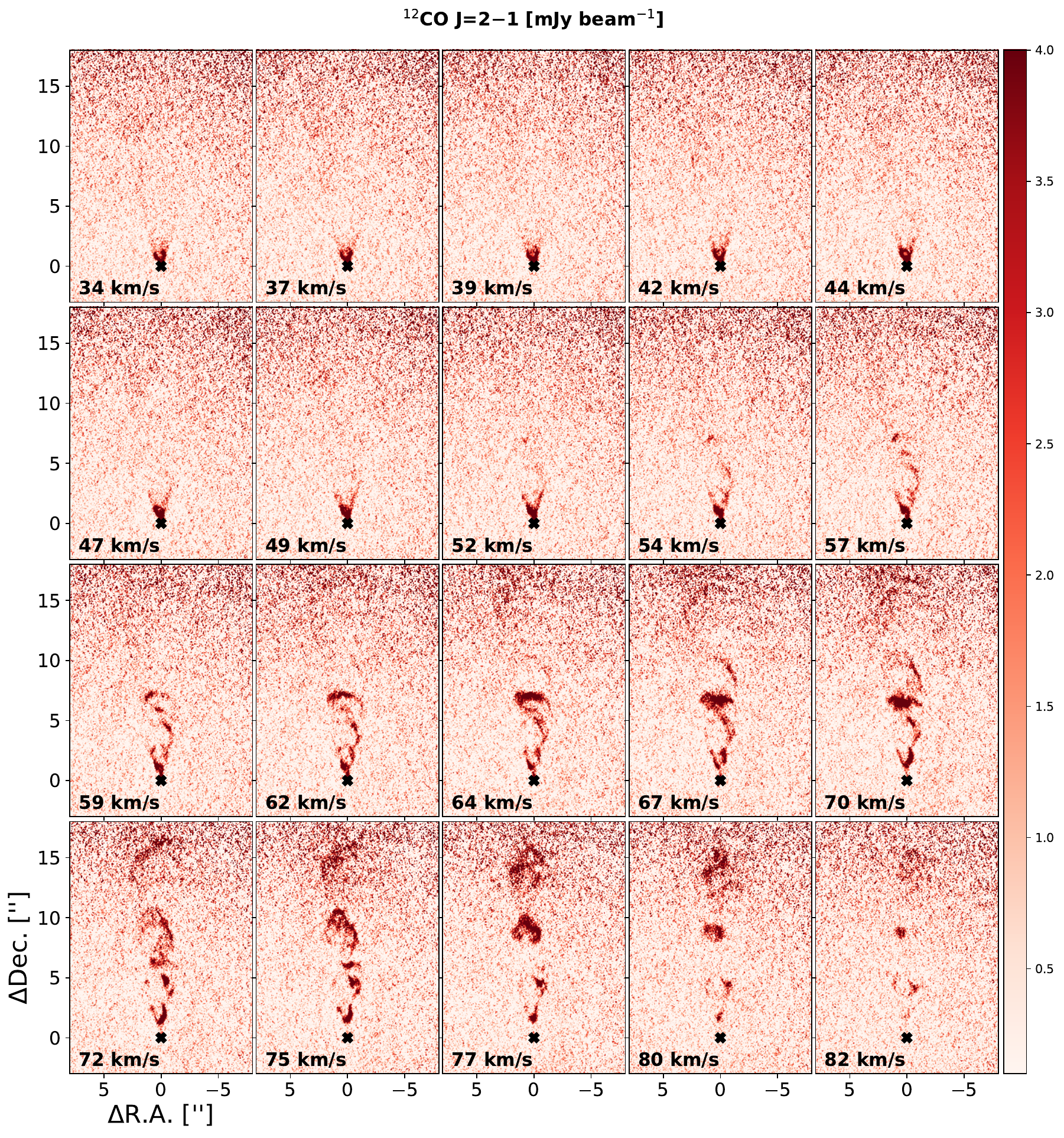}
 \caption{Channel maps of the redshifted emission of $^{12}$CO in the EHV regime ($|V| > 30$ km~s$^{-1}$) in IRS1, shown by increasing velocity values relative to the systemic velocity from 34 km~s$^{-1}$ to 82 km~s$^{-1}$.}
\label{fig:chan_irs1_vhv_12co_red}
\end{figure*}

\begin{figure*}
\plotone{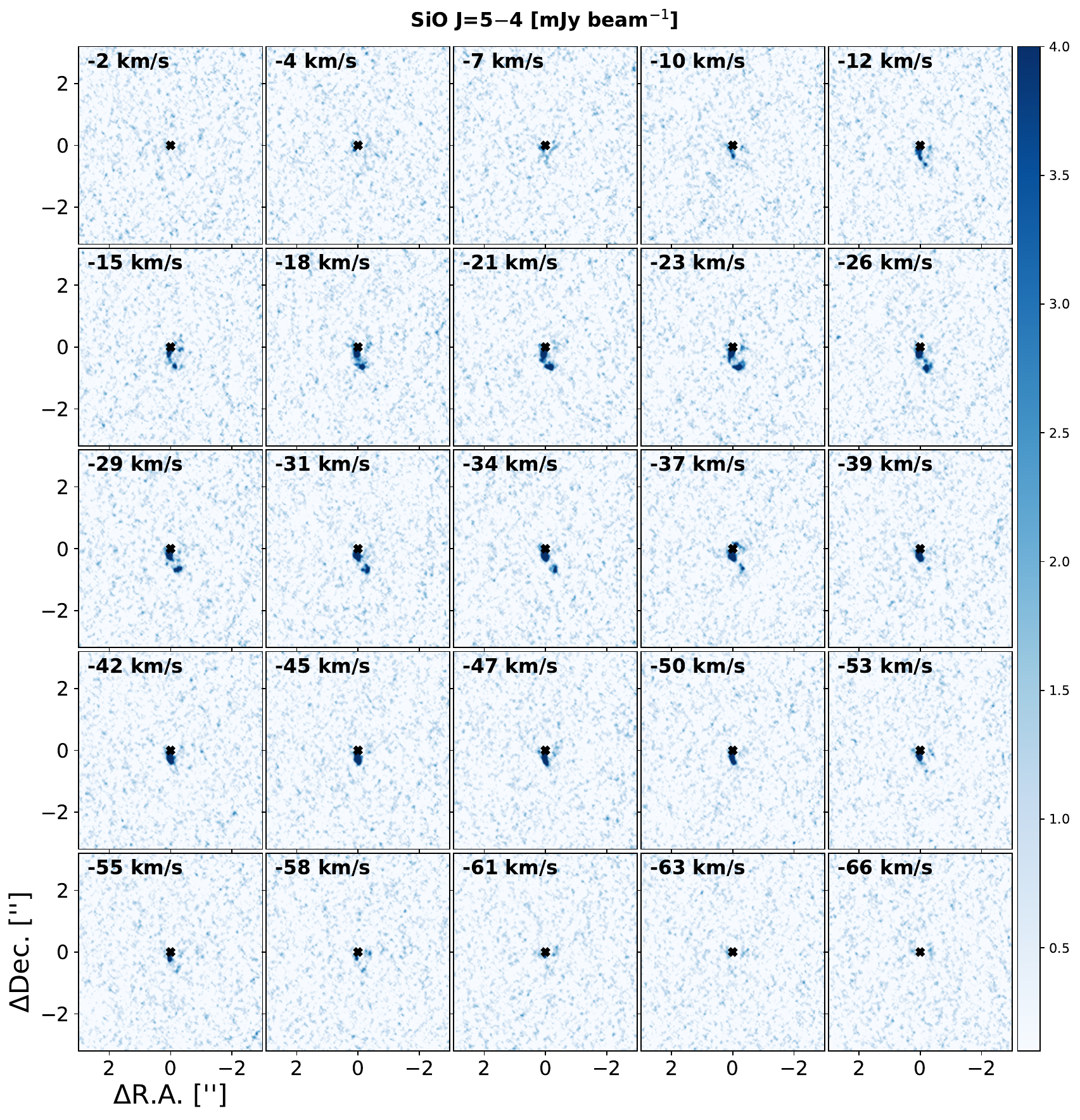}
 \caption{Channel maps of the blueshifted emission of SiO in IRS1, shown by decreasing velocity values relative to the systemic velocity from -2 km~s$^{-1}$ to -66 km~s$^{-1}$.}
\label{fig:chan_irs1_shv_sio_blue}
\end{figure*}

\begin{figure*}
\plotone{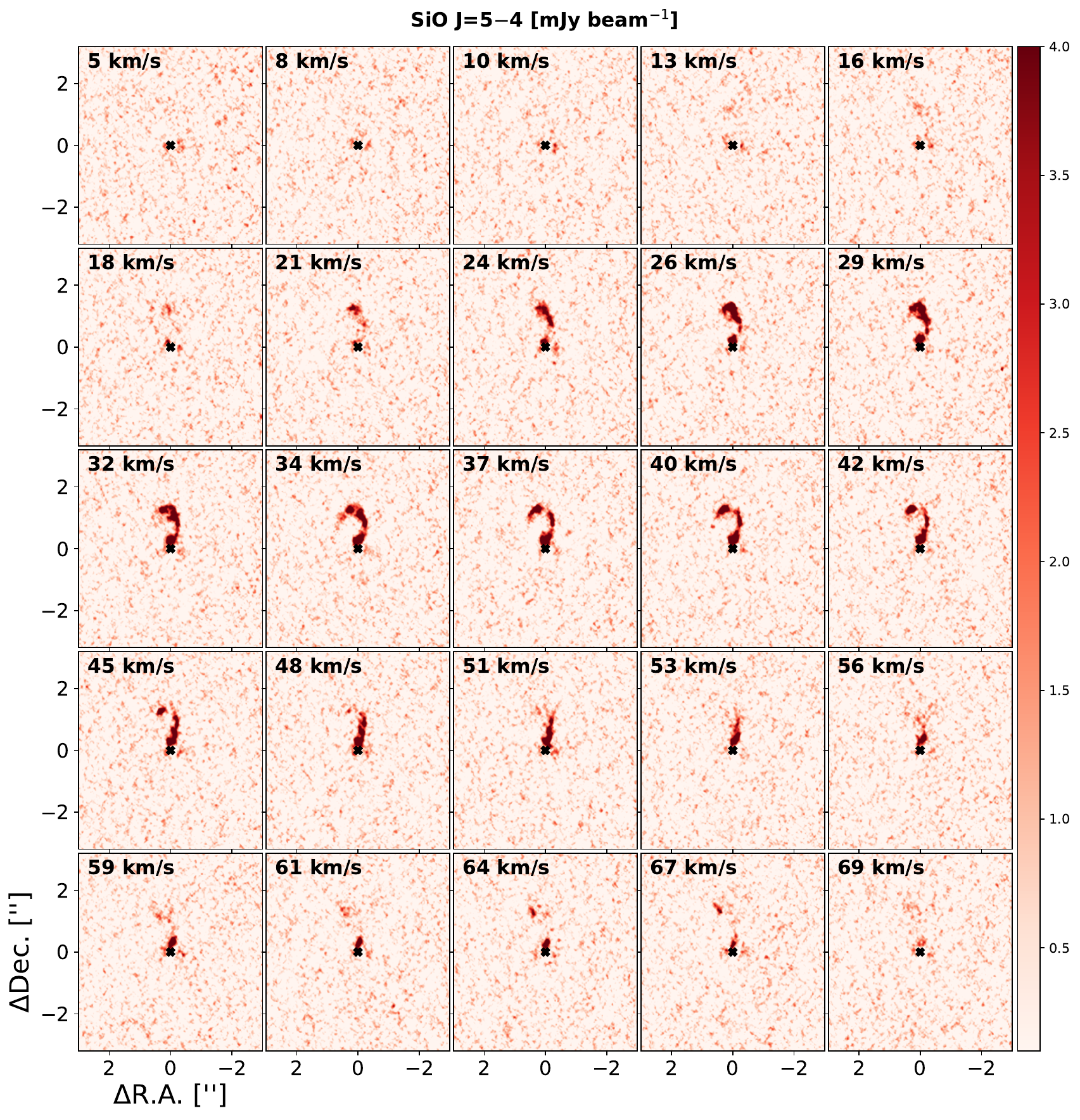}
 \caption{Channel maps of the redshifted emission of SiO in IRS1, shown by increasing velocity values relative to the systemic velocity from 5 km~s$^{-1}$ to 69 km~s$^{-1}$.}
\label{fig:chan_irs1_shv_sio_red}
\end{figure*}

We present a selected series of channel maps in each of the two velocity regimes based on the $^{12}$CO kinematics (see Sect.\,\ref{sec:results}), SHV ($< 30$ km~s$^{-1}$) and EHV ($> 30$ km~s$^{-1}$), in the case of BHR 71 IRS1. Figures \ref{fig:chan_irs1_shv_12co_blue} and \ref{fig:chan_irs1_shv_12co_red} show the channel maps of $^{12}$CO in the SHV regime for the blueshifted and redshifted emission, respectively, whereas Figures \ref{fig:chan_irs1_vhv_12co_blue} and \ref{fig:chan_irs1_vhv_12co_red} show the channel maps of $^{12}$CO in the EHV regime for the blueshifted and redshifted emission, respectively. The SiO emission is showed in Figures \ref{fig:chan_irs1_shv_sio_blue} and \ref{fig:chan_irs1_shv_sio_red}.


\section{Outflows in IRS2: channel maps}\label{app:chanmap_IRS2}

\begin{figure*}
\plotone{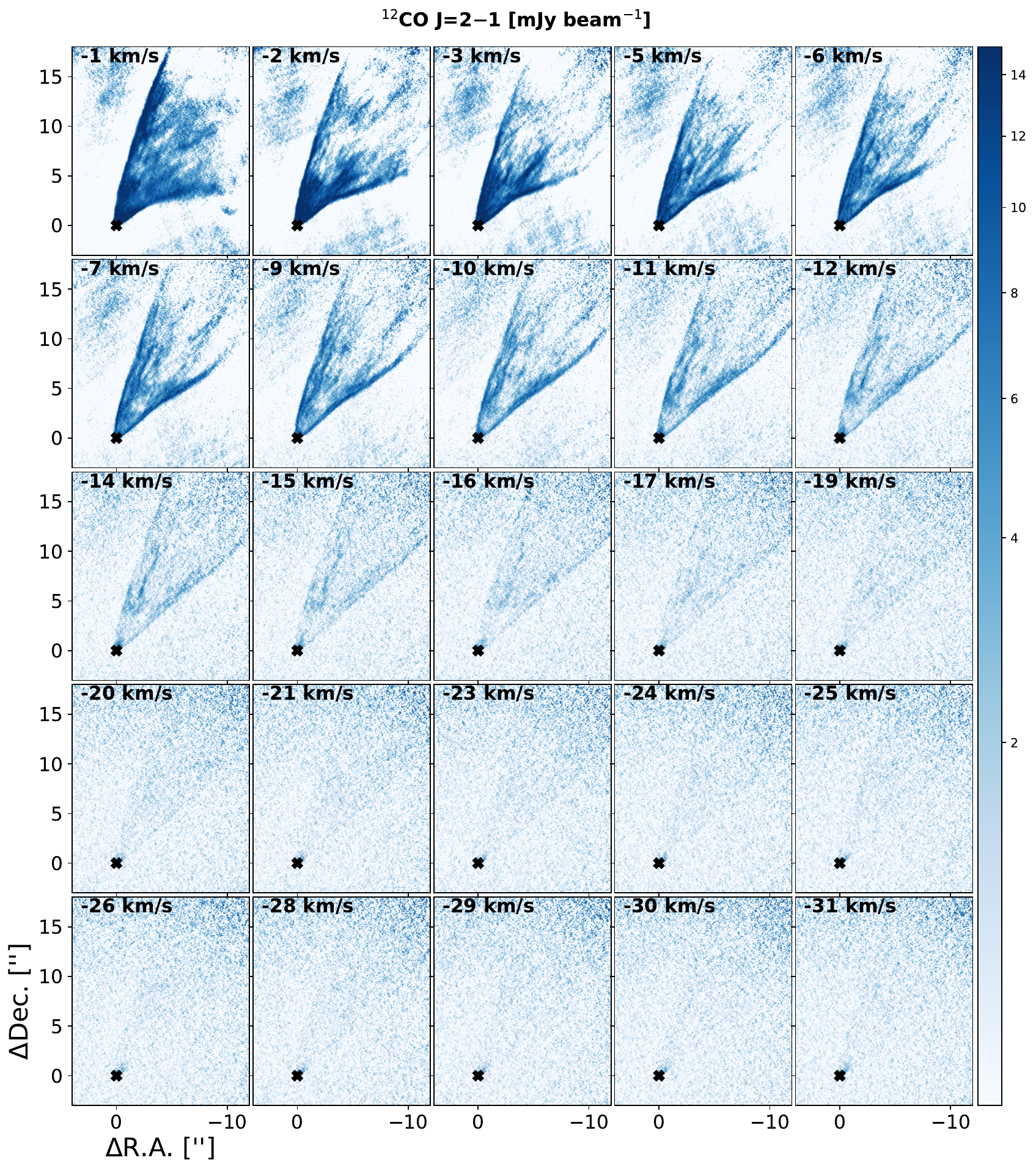}
 \caption{Channel maps of the blueshifted emission of $^{12}$CO in the SHV regime ($|V| < 30$ km~s$^{-1}$) in IRS2, shown by decreasing velocity values relative to the systemic velocity from -1 km~s$^{-1}$ to -31 km~s$^{-1}$.}
\label{fig:chan_irs2_shv_12co_blue}
\end{figure*}

\begin{figure*}
\plotone{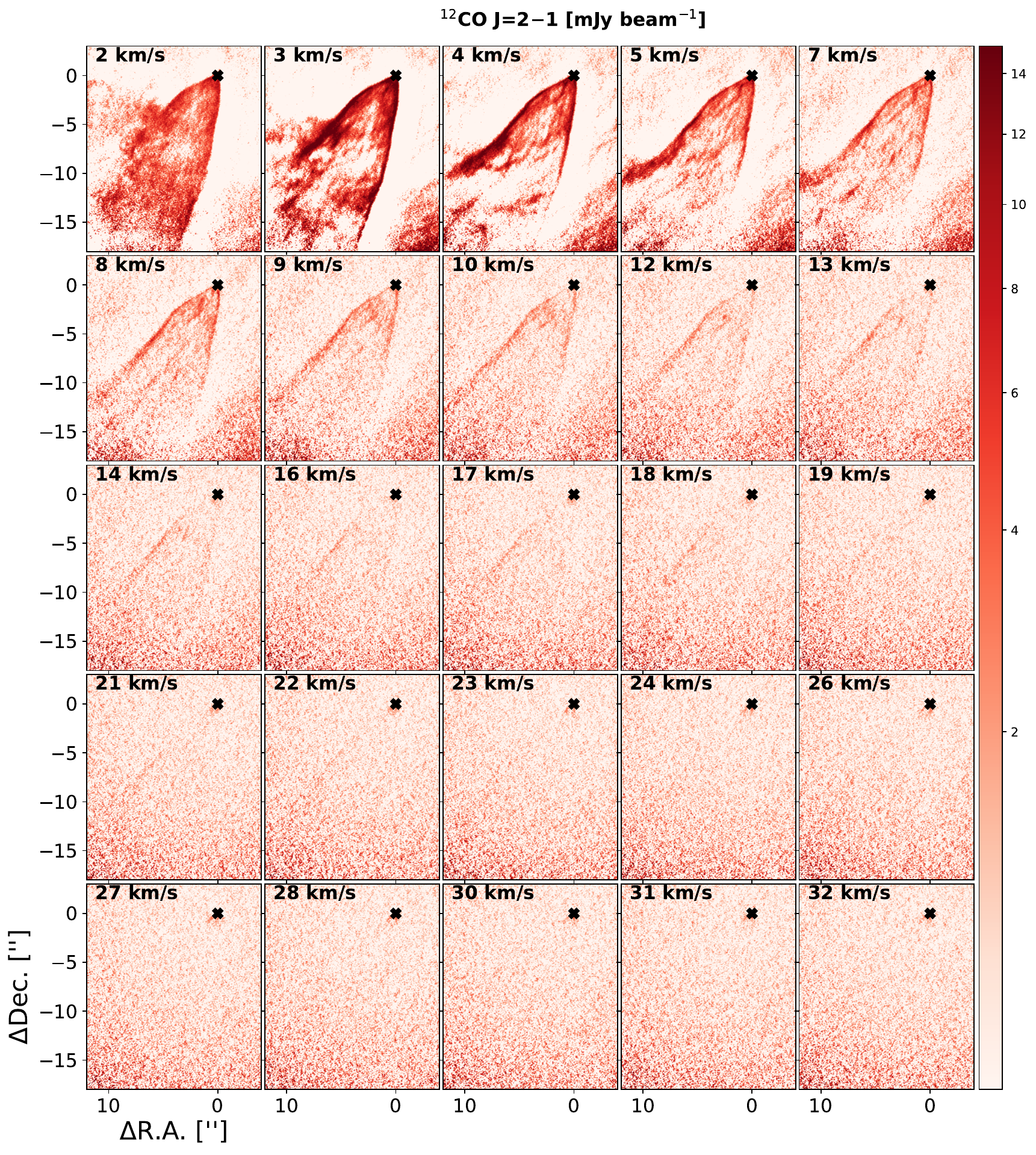}
 \caption{Channel maps of the redshifted emission of $^{12}$CO in the SHV regime ($|V| < 30$ km~s$^{-1}$) in IRS2, shown by increasing velocity values relative to the systemic velocity from 2 km~s$^{-1}$ to 32 km~s$^{-1}$.}
\label{fig:chan_irs2_shv_12co_red}
\end{figure*}

\begin{figure*}
\plotone{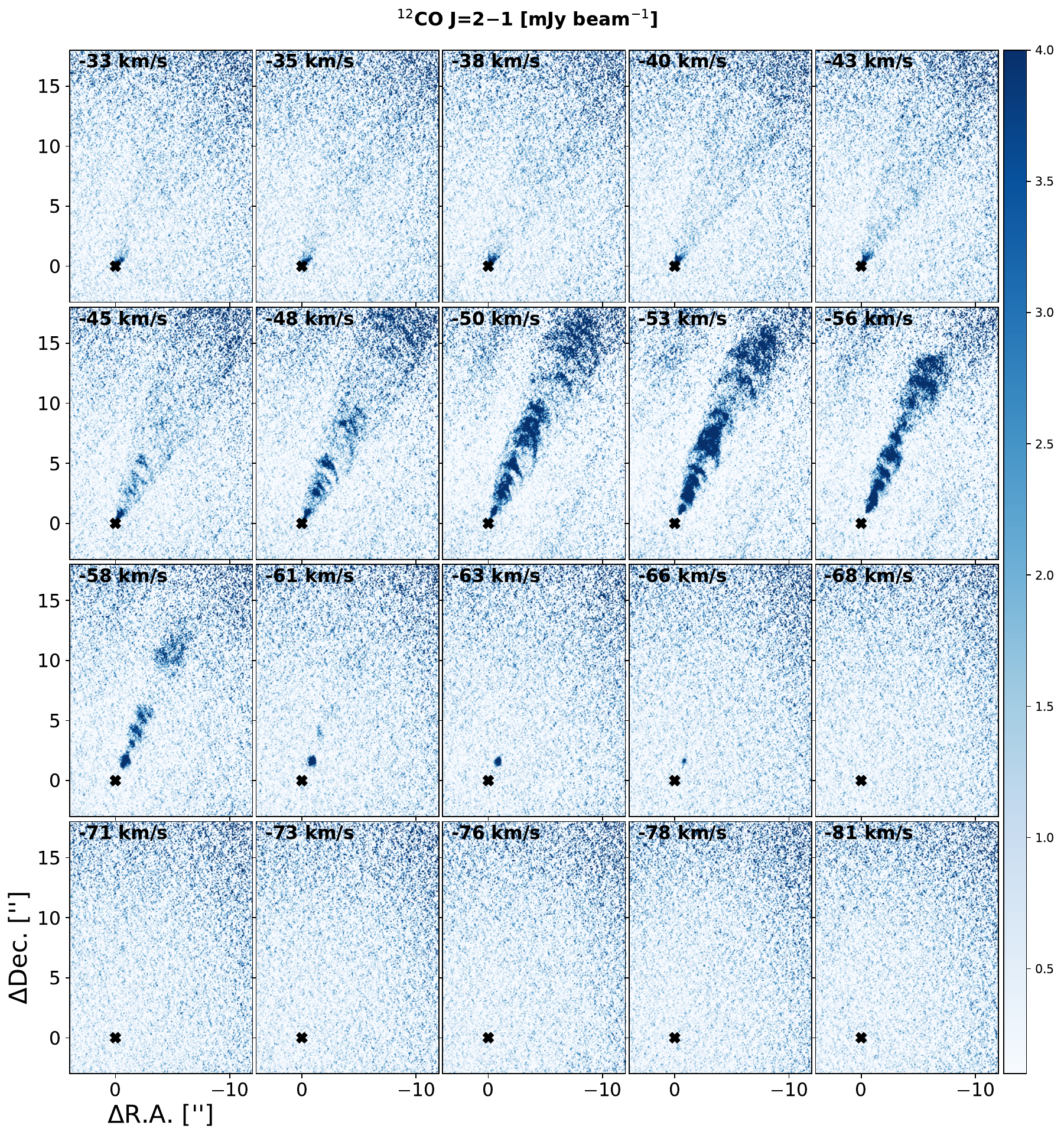}
 \caption{Channel maps of the blueshifted emission of $^{12}$CO in the EHV regime ($|V| > 30$ km~s$^{-1}$) in IRS2, shown by decreasing velocity values relative to the systemic velocity from -33 km~s$^{-1}$ to -81 km~s$^{-1}$.}
\label{fig:chan_irs2_vhv_12co_blue}
\end{figure*}

\begin{figure*}
\plotone{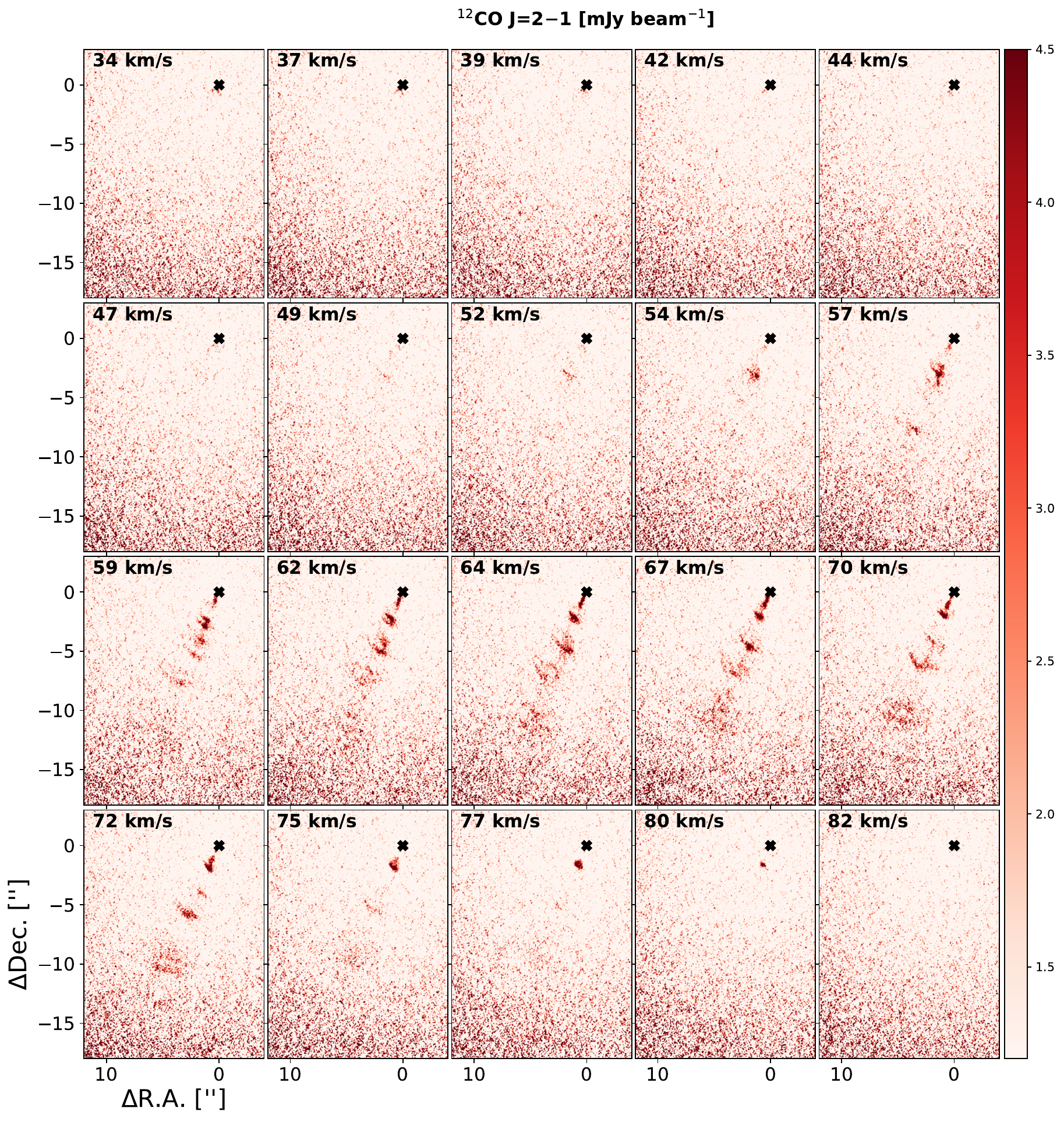}
 \caption{Channel maps of the redshifted emission of $^{12}$CO in the EHV regime ($|V| > 30$ km~s$^{-1}$) in IRS2, shown by increasing velocity values relative to the systemic velocity from 34 km~s$^{-1}$ to 82 km~s$^{-1}$.}
\label{fig:chan_irs2_vhv_12co_red}
\end{figure*}

\begin{figure*}
\plotone{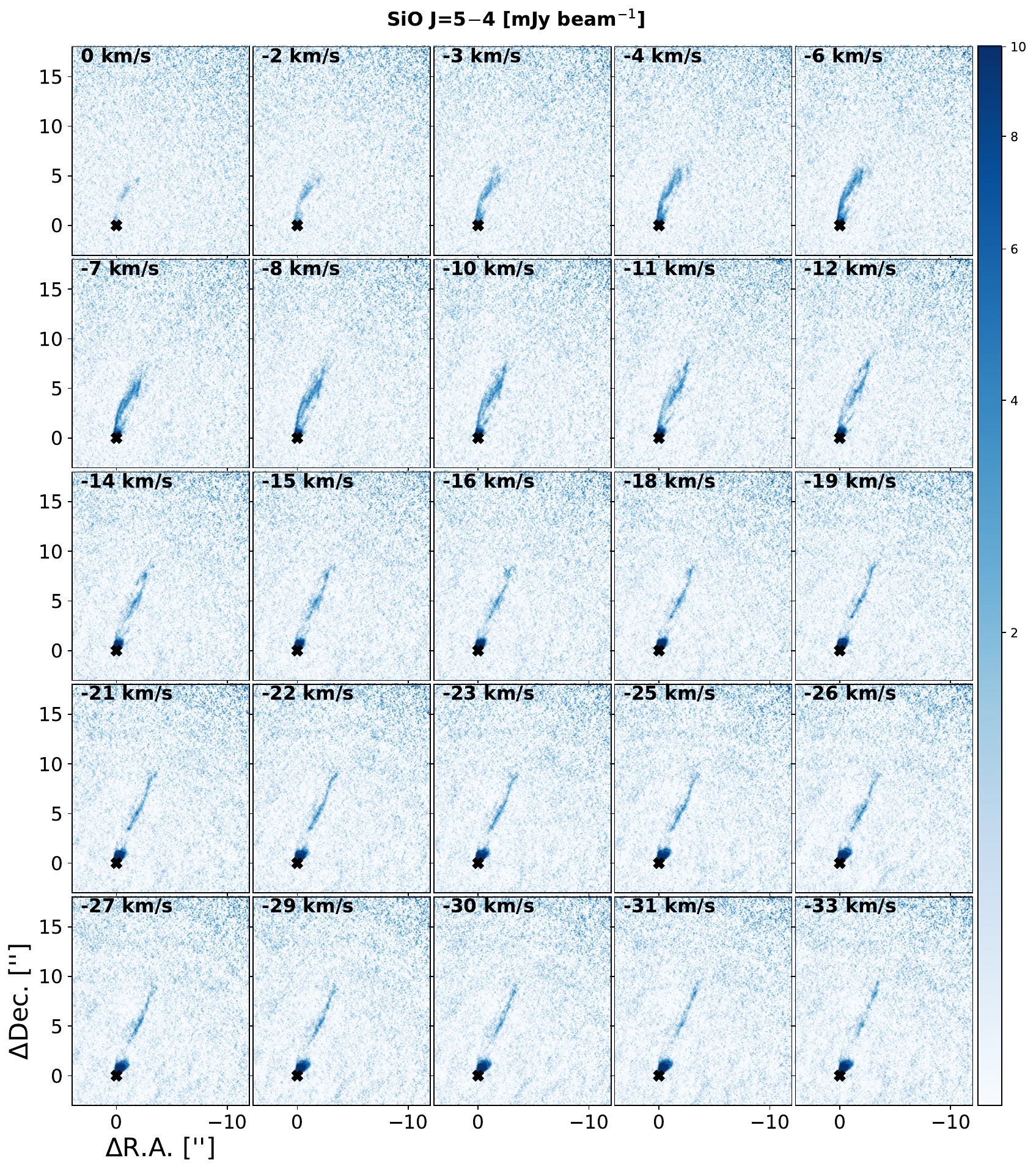}
 \caption{Channel maps of the blueshifted emission of SiO in the SHV regime ($|V| < 30$ km~s$^{-1}$) in IRS2, shown by decreasing velocity values relative to the systemic velocity from 0 km~s$^{-1}$ to -33 km~s$^{-1}$.}
\label{fig:chan_irs2_shv_sio_blue}
\end{figure*}

\begin{figure*}
\plotone{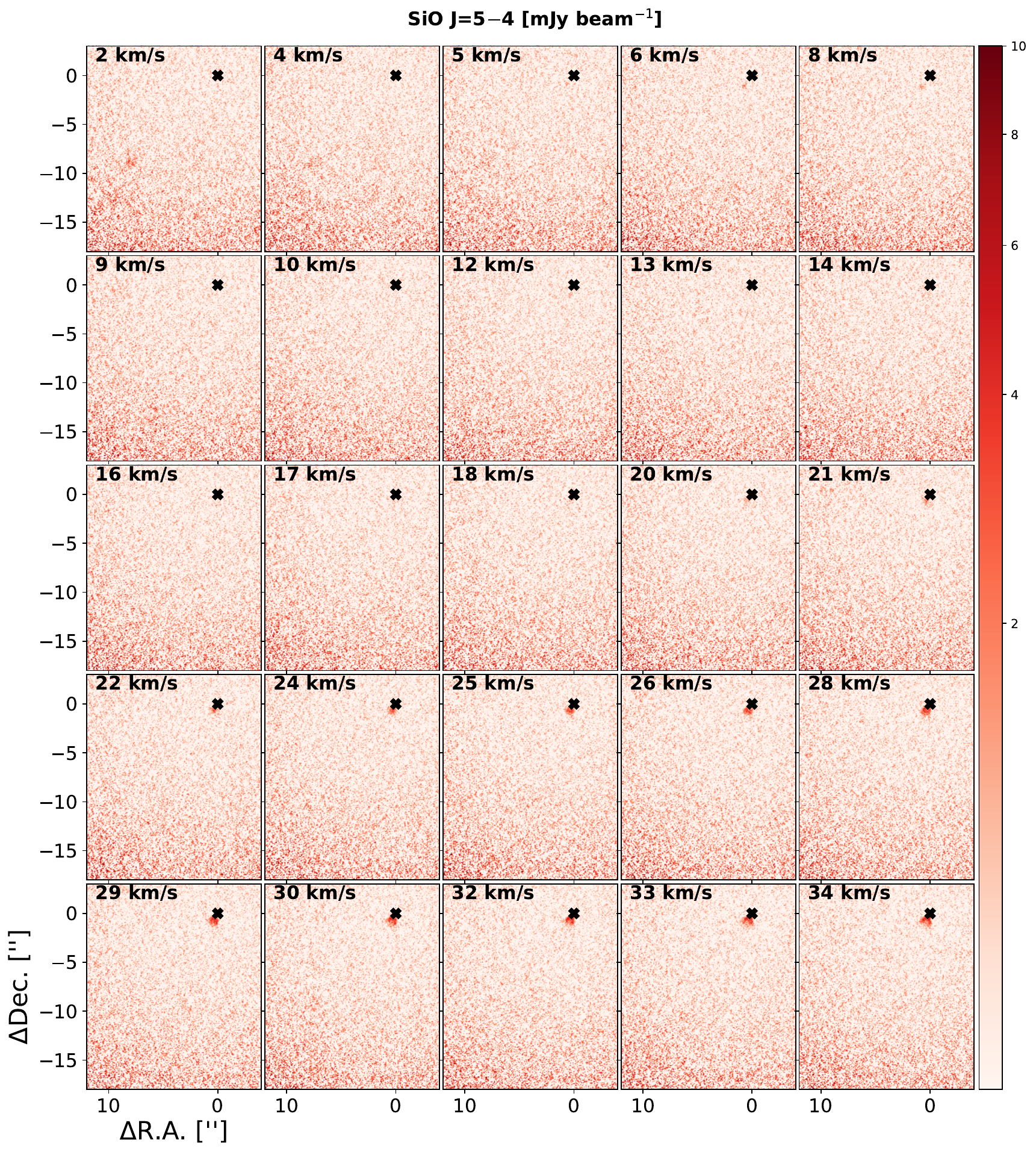}
 \caption{Channel maps of the redshifted emission of SiO in the SHV regime ($|V| < 30$ km~s$^{-1}$) in IRS2, shown by increasing velocity values relative to the systemic velocity from 2 km~s$^{-1}$ to 34 km~s$^{-1}$.}
\label{fig:chan_irs2_shv_sio_red}
\end{figure*}

\begin{figure*}
\plotone{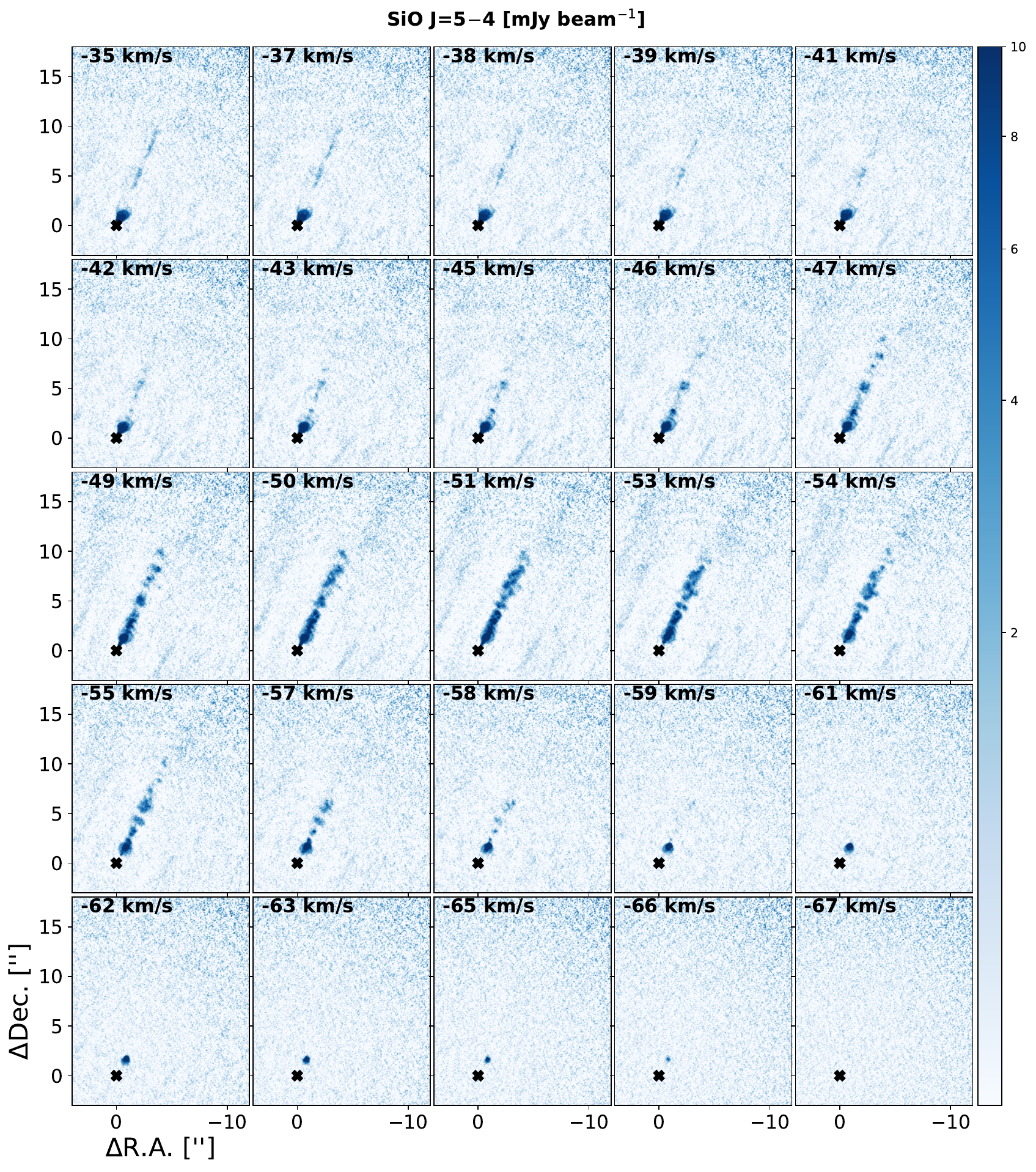}
 \caption{Channel maps of the blueshifted emission of SiO in the EHV regime ($|V| > 30$ km~s$^{-1}$) in IRS2, shown by decreasing velocity values relative to the systemic velocity from -35 km~s$^{-1}$ to -67 km~s$^{-1}$.}
\label{fig:chan_irs2_vhv_sio_blue}
\end{figure*}

\begin{figure*}
\plotone{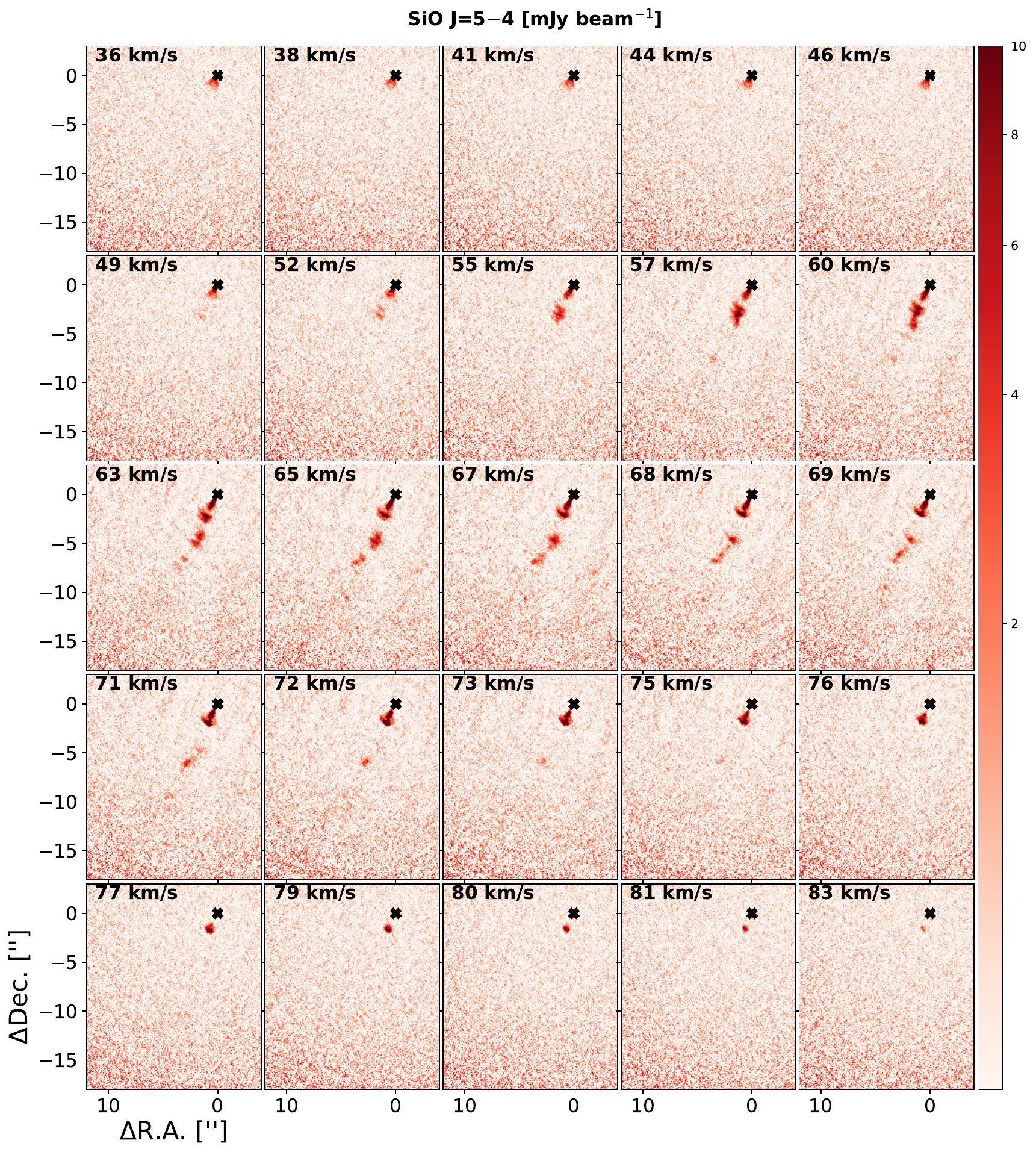}
 \caption{Channel maps of the redshifted emission of SiO in the EHV regime ($|V| > 30$ km~s$^{-1}$) in IRS2, shown by increasing velocity values relative to the systemic velocity from 36 km~s$^{-1}$ to 83 km~s$^{-1}$.}
\label{fig:chan_irs2_vhv_sio_red}
\end{figure*}

Similarly to Appendix~\ref{app:chanmap_IRS1}, we present a selected series of channel maps in each of the two velocity regimes, SHV ($< 30$ km~s$^{-1}$) and EHV ($> 30$ km~s$^{-1}$), in the case of BHR 71 IRS2. Figures \ref{fig:chan_irs2_shv_12co_blue} and \ref{fig:chan_irs2_shv_12co_red} show the channel maps of $^{12}$CO in the SHV regime for the blueshifted and redshifted emission, respectively, whereas Figures \ref{fig:chan_irs2_vhv_12co_blue} and \ref{fig:chan_irs2_vhv_12co_red} show the channel maps of $^{12}$CO in the EHV regime for the blueshifted and redshifted emission, respectively. The SiO emission is showed in Figures~\ref{fig:chan_irs2_shv_sio_blue} and \ref{fig:chan_irs2_shv_sio_red} in the SHV regime, and in Figures~\ref{fig:chan_irs2_vhv_sio_blue} and \ref{fig:chan_irs2_vhv_sio_red} in the EHV regime.


\end{document}